\documentclass[
	a4paper, 
	12pt, 
	review,
	twoside, 
]{LTJournalArticle_mod}

\usepackage{soul}
\usepackage{subfig}
\usepackage{amsthm}
\usepackage{amsfonts}
\usepackage{amsmath}
\usepackage{comment}
\usepackage{amssymb}
\usepackage{enumitem}
\usepackage{float}
\usepackage{makecell}
\usepackage{multirow}
\usepackage{xcolor}
\usepackage{algorithm}
\usepackage{algpseudocode}
\usepackage{bm}

\title{Generating Synthetic Rainfall Fields by R-vine Copulas Applied to Seamless Probabilistic Predictions}

\author{Peter Schaumann$^1$, Martin Rempel$^2$, Ulrich Blahak$^2$, Volker Schmidt$^1$}

\date{\footnotesize$^1$Institute of Stochastics, Ulm University, Ulm, Germany\\ $^2$Deutscher Wetterdienst, Offenbach, Germany}

\renewcommand{\maketitlehookd}{%
	\begin{abstract}
		\noindent Many post-processing methods improve forecasts at individual locations but remove their correlation structure, which is crucial for predicting larger-scale events like total precipitation amount over areas such as river catchments that are relevant for weather warnings and flood predictions. We propose a method to reintroduce spatial correlation into a post-processed forecast using an R-vine copula fitted to historical observations. This method works similarly to related approaches like the Schaake shuffle and ensemble copula coupling, i.e., by rearranging predictions at individual locations and reintroducing spatial correlation while maintaining the post-processed marginal distribution. Here, the copula measures how well an arrangement compares with the historical distribution of precipitation. No close relationship is needed between the post-processed marginal distributions and the spatial correlation source. This is an advantage compared to Schaake shuffle and ensemble copula coupling, which rely on a ranking with no ties at each considered location in their source for spatial correlations. However, weather variables such as the precipitation amount, whose distribution has an atom at zero, have rankings with ties. To evaluate the proposed method, it is applied to a precipitation forecast produced by a combination model with two input forecasts that deliver calibrated marginal distributions but without spatial correlations. The obtained results indicate that the calibration of the combination model carries over to the output of the proposed model, i.e., the evaluation of area predictions shows a similar improvement in forecast quality as the predictions for individual locations. Additionally, the spatial correlation of the forecast is evaluated with the help of object-based metrics, for which the proposed model also shows an improvement compared to both input forecasts.\\
        \textbf{Keywords:} precipitation forecast, post-processing, spatial correlation, R-vine Copula
	\end{abstract}
}

\newcommand{\R}{\mathbb{R}}

\DeclareMathOperator*{\argmax}{arg\,max}

\begin{document}

\maketitle


\section{Introduction}
\label{sec.introduction}

In operational weather forecasting, forecasters are supported by various forecast models to issue targeted warnings of potentially hazardous weather phenomena. The longer such warnings are accurate and reliable, the more time decision-makers in hydrological and civil protection have to diminish possible harm to life and property. Usually, these warnings rely on nowcasting systems and numerical weather prediction (NWP). Both give valuable guidance, however, for different lead times \citep{lit:hess2020,lit:Ruti2020}. Thus, both systems represent different sources of predictability that can be combined to give the best forecast at any time. The so-called seamless prediction is part of a whole value cycle as described in \cite{lit:Ruti2020}, reaching from information generation toward the outcomes and values. The project Seamless Integrated Forecasting System (SINFONY) of Deutscher Wetterdienst (DWD) focuses on the seamless prediction of precipitation within the short-term range up to 12\,h ahead. As a first step to achieve this combination, both forecasting techniques---nowcasting and NWP---are enhanced individually such that the gap between both of them is narrowed in terms of a verification metric. Then, based on the improved forecast systems, developing and implementing custom combination methods leads to a unique user-oriented forecast, including the best information for both individual forecasts. Within this project, among other techniques, we have developed an approach to seamlessly combine 6\,h forecasts of an observation-based precipitation nowcasting scheme and the NWP model ICON-D2.

The main component of the approach mentioned above is a model we proposed in \cite{Rempel2022}. It is based on neural networks for combining and post-processing two ensemble forecasts (briefly called C$^3$-model in the following, where C$^3$ stands for \textit{combined, calibrated, consistent}). Forecasts of this model could provide a data basis for customized warnings. The output of the C$^3$-model consists of calibrated probabilities for the exceedance of a set of precipitation thresholds at each grid point of a regular horizontal grid, see also \citep{lit:Schaumann2021}. However, a drawback of the model is the loss of spatial correlations of both input ensembles during the calibration process.

This loss is a well-known problem often discussed in the literature about post-processing of ensemble forecasts \citep{ bellier2018generating,jobst2023d,moller2013multivariate,wu2018comparative}. Several approaches have been developed to overcome this situation that reintroduces spatial information to post-processed marginal forecasts. Typically, these approaches are based on the ``Schaake shuffle'' \citep{clark2004schaake} or on ``ensemble copula coupling'' \citep{schefzik2013uncertainty}, where sample values drawn from the post-processed marginal distributions for individual locations are rearranged into synthetic ensembles according to the rank correlation of a provided ensemble forecast or a period of past observations. The resulting synthetic ensemble follows the post-processed marginal distributions while preserving the spatial correlations of the raw (i.e. unprocessed) data.

To obtain suitable spatial correlations from a specific set of ensemble members, a close correspondence between the provided ensemble and the post-processed marginal distributions is required \citep{schefzik2016similarity}, since spatial correlations are usually neither stationary in space nor in time. Additionally, the ranks are not unambiguous at locations where ensemble members take the same value, which is common for precipitation forecasts.

The marginal distributions provided by the C$^3$-model \citep{Rempel2022} are based on combining two, sometimes vastly different, ensemble forecasts. Thus, no corresponding ensemble forecasts exist that could be used to transfer its spatial correlation structure.

Therefore,  in the present paper, we propose a multivariate stochastic model for generating synthetic ensemble members based on the idea of the ``Schaake shuffle''. However, in our approach, the spatial correlations are not derived from a given ensemble nor directly from historical observations but are modelled through an R-vine copula. Vine copulas are a powerful tool for parametric modelling of multivariate probability distributions \citep{joe2011dependence} and offer the flexibility to describe their tail behaviour adequately \citep{lit:Czado2021}. Thus, the need for having an ensemble forecast available in advance that follows the post-processed marginal distributions is circumvented. Furthermore, the spatial correlation structure can be reconstructed in any area without a readjustment of the model components for the new area considered. As a result, we offer the possibility to provide forecast information on user-oriented customized areas, which can be a basis for flexible warnings to end-users. Thus, with additional impact data, we can support the transition process of the current operational warning system towards an impact-oriented one \citep{ lit:Kaltenberger2020,lit:Kox2018, lit:Potter2021}. Finally, we estimate the distribution of the total precipitation amount within given regions (e.g. river catchments or municipal areas) using synthetic ensemble members drawn from the multivariate stochastic model.

Vine copulas are recently used in the field of energy meteorology to model the spatial interrelation of errors in probabilistic forecasts to assess the uncertainties in the power generation of photovoltaic systems \citep{aigner2023robust,lit:SchinkeNendza2021}, or for the inclusion of spatial dependencies between multiple wind farms in wind energy scenarios \citep{lit:Li2022,lit:Tu2023}. Another field of application is within so-called weather generators. Since high-resolution time series of atmospheric variables are only available for a limited period (e.g. 30 years), it is necessary to extend these time series when assessing extreme events \citep{lit:VanDeVelde2023} or designing hydrological applications \citep{lit:Callau2018}. In \cite{lit:Brunner2019}, several copula models have been evaluated to reproduce the spatial dependencies of gauging stations in a river catchment. Furthermore, in \cite{lit:Ehrhardt2015},  a spatial model based on an R-vine copula has been introduced to predict time series of the daily mean temperature at unobserved locations, and,  in \cite{lit:Tahroudi2022}, various  C-, D-, and R-vine copulas have been examined to estimate rainfall deficiency structures in an Iranian river basin.

It should be noted that in the meteorological literature, the term ``spatial correlation'' sometimes refers to the spatial correlation of forecast uncertainty, which is then called ``spatial error correlation'', see, e.g.  \citep{feldmann2015spatial},  as opposed to the spatial correlation of the precipitation amount itself. The reason for this distinction is that the (actually observed) precipitation amount can be mathematically modelled either as a deterministic quantity or as a random variable.

The present paper is structured as follows. First, the utilized datasets are introduced in Section~\ref{sec.data}. This is followed in Section~\ref{sec.methods} by a description of the adjustment of the C$^3$-model by a quantile regression, the generation of synthetic ensemble members, and the prediction of the total precipitation amount in a given area. Then, Section~\ref{sec.implementation} discusses the proposed model's implementation details. Afterwards, the validation results are presented in Section~\ref{sec.verification}. Finally, conclusions are drawn in Section~\ref{sec.conclusion}.

\section{Data}
\label{sec.data}
For adding spatial correlations to the C$^3$-model forecasts, we use the same dataset as in \cite{Rempel2022}. That set consists of, on the one hand, precipitation extrapolations of STEPS-DWD, a nowcasting scheme. On the other hand, forecasts of an experimental version of ICON-D2 are used, a high-resolution short-term NWP model. Both systems provide ensemble forecasts. In order to keep the present paper largely self-contained, we provide a brief description of the dataset, which covers three time periods from the years 2016, 2019 and 2020 (05/26/2016 - 06/26/2016, 06/01/2019 - 06/23/2019, 06/03/2020 - 07/16/2020). 

\subsection{STEPS-DWD}

STEPS-DWD has been developed within SINFONY as an adaption of STEPS (Short-Term Ensemble Prediction System), see \cite{lit:bowler2006, lit:foresti2016,lit:seed2003, lit:seed2013}, where DWD's radar network provides radar reflectivities based on which precipitation rates are estimated by a hydrometeor-dependent Z-R relation optimised for the radar stations utilised by DWD \citep{lit:Steinert2021}.

For the present study, STEPS-DWD is configured to consist of a cascade of first-order autoregressive processes on twelve spatial scales and to apply a new localisation approach \citep{lit:Pulkkinen2020, lit:Reinoso2021} for the estimation of the autoregressive parameters on each individual scale. The spatially correlated noise field is estimated globally but is imprinted only in regions with precipitation due to the localised autoregressive parameters.

We use further 30-member STEPS-DWD extrapolations that are generated every 30\,minutes running 6\,h ahead on a $1\times1\,\mathrm{km}^2$ grid with a temporal resolution of 5 minutes. For our purposes, all 5-minute forecasts within a given hour are aggregated into one hourly forecast. In the same way, radar observations are aggregated to obtain the ground truth for the C$^3$-model and the estimated precipitation amounts for validating the synthetic rainfall fields.

\subsection{ICON-D2-EPS}

Furthermore, we use forecasts of an experimental version of ICON-D2-EPS \citep{lit:zaengl2015} that runs in limited area mode (LAM) with a horizontal grid spacing of $\Delta x \approx 2.2\,\mathrm{km}$. The domain of the 20-members ensemble is centred on Central Europe. In addition to conventional observation data and MODE-S aircraft measurements, radar reflectivities and radial winds of the 3D volume scans are assimilated by DWD's kilometre-scale ensemble data assimilation system KENDA. It implements a localised ensemble transform Kalman filter \citep{lit:Bick2016,lit:schraff2016}. Only the first 20 members serve as initial conditions for the forecasts, while 40 members are used for the assimilation. ICON-EU ensemble forecasts (larger trans-European domain, grid spacing 6.5\,km, parameterised deep convection) provide lateral and upper boundary conditions. The operational conventional one-moment cloud microphysics scheme is used.

Our experimental setting of ICON-D2-EPS generates hourly forecasts running $12\,\mathrm{h}$ ahead. The native ICON output is provided on an irregular triangular grid. Hence, forecasts of both---STEPS-DWD and ICON-D2-EPS---as well as the observations are interpolated onto a common regular $2.2\times2.2\,\mathrm{km}^2$ grid that was established for the former operational NWP model COSMO-D2.

\section{Methods}
\label{sec.methods}
The modelling approach proposed in this paper consists of several parts. First, a machine learning model called the $ C^3$ model is used to predict calibrated quantiles of the precipitation amount distribution for each location under consideration. However, these pointwise predictions do not take spatial correlation into account. Therefore, in the next step, an R-vine copula is fitted to precipitation observation data for modelling the multivariate precipitation distribution at multiple locations. Finally, the copula model is used with a hill-climbing algorithm to order the predicted quantiles at each considered location into synthetic ensemble members. This step is conceptually similar to Schaake shuffle and ensemble copula coupling in that spatial correlation is imposed on samples of marginal distributions by rearranging the samples.
However, for rearranging the predicted precipitation values, the fitted copula
serves as a measure of how well a given arrangement compares with the observed distribution
of precipitation in the historical data, where no close relationship is required between the post-processed marginal distributions and the spatial correlation source. This is an advantage in
comparison to Schaake shuffle and ensemble copula coupling, which rely on the existence of a
ranking with no ties at each considered location in their source for spatial correlations.
The ensemble members obtained by our modelling approach exhibit not only the calibrated marginal distributions predicted by the $ C^3$ model but also the spatial correlation provided by the copula model. We refer to Section~\ref{sec.implementation} below regarding various implementation aspects.

\subsection{Quantile regression}

Recall that the output of the C$^3$-model, as proposed in \cite{Rempel2022} and    \cite{lit:Schaumann2021}, consists of probabilities for the exceedance of an appropriately selected family of thresholds.
For reliable fitting of probability densities to these datasets of exceedance probabilities, a relatively large number of thresholds would be required to cover the entire spectrum of possible precipitation amounts. In addition, picking a suitable family of parametric densities is not trivial to model the distribution predicted by the C$^3$-model. Thus, we switch from considering threshold exceedance probabilities to a quantile regression approach to limit the number of necessary data points generated by the C$^3$-model.
For this, we replace the softmax layer of the C$^3$-model with a dense layer with a linear activation function and replace the ``categorical cross-entropy'' loss function with ``pinball'' loss functions, see \cite{steinwart2011estimating} for details. This allows us to predict a set of evenly spaced quantiles on the $[0,1]$-interval. By sampling values directly from this set of quantiles, we obtain data which can be seen as realizations of the precipitation amount distribution predicted by the C$^3$-model without the necessity to fit a suitable probability density function.

\subsection{Sampling of predicted precipitation amounts for single 
locations}\label{sec.mar.pre}

For some integer, $m>1$, let $V=\{v_1,\ldots,v_m\}\subset\R^2$ denote a set of locations, for each location $v_i$ of which the C$^3$-model produces a calibrated distribution of precipitation amounts, as described above. For some sample size $N>0$ and for each $i=1,\ldots,m$, we draw $N$ sample values $x^1_i, \ldots, x^N_i>0$ for a specified hour $t > t_c$ from a certain random variable $X_{i}$, whose probability distribution describes the predicted precipitation amount at location $v_i$. Here and in the following, $t_c\in\R$ refers to the present point in time dividing the datasets into past and future, i.e., the data available for training and validation, respectively. Note that these values are independently sampled for each individual location, i.e., the random vector $(X_1,\ldots,X_m)$ has independent components because the C$^3$-model predicts the univariate (marginal)  distributions of precipitation amounts without taking spatial correlations into account. The sample values  $x^1_i, \ldots, x^N_i$ are used later on as building blocks for the construction of synthetic ensemble members, which follow the distribution of some $m$-dimensional random vector $(X'_{1},\ldots,X'_{m})$ that describes the precipitation amounts predicted at locations $v_1, \ldots, v_m \in V$ for the same hour $t$, i.e., the (univariate) distributions of $X'_i$ and $X_i$ coincide for each $i=1,\ldots,m$,  where the spatial correlations observed in historical precipitation data are taken into consideration in  $(X'_{1},\ldots,X'_{m})$.

\subsection{Modeling  spatial dependencies by measured precipitation amounts}\label{sec.spa.dep}

This section explains how copula models can capture spatial dependencies based on measured precipitation amounts. In particular, we use Sklar's fundamental theorem of copula theory, see \cite{joe2011dependence}, which states that the cumulative distribution function of a multivariate probability distribution can be decomposed into several distinct functions. They consist of univariate (marginal) cumulative distribution functions and a so-called copula which models the underlying correlation structure. While many different copula families are used to model different multivariate correlation structures, most are restricted to the two-dimensional case. Thus, so-called vine copulas have been developed to use copula models for higher dimensions. These are based on a graph structure of two-dimensional copulas to construct  $m$-dimensional copulas for some integer $m>2$. For further details regarding the construction of vine copulas, we refer to \cite{joe2011dependence}.

In the present paper, we are interested in the multivariate distribution of $m$-dimensional vectors of precipitation amounts. For technical reasons, however, the implementation explained in Section~\ref{sec.implementation} below only allows for continuous or discrete distributions but not for mixtures of both distribution types. Since the distribution of precipitation amounts has an atom at $0$\;mm, it is a mixture distribution, which necessitates that we consider discretized marginal distributions instead. However, the method presented in this paper can also be applied to meteorological quantities with continuous distributions, such as temperature. Therefore, the continuous and the discrete cases will be discussed in the following.

To determine the joint distribution of (historical) precipitation amounts at a set of locations $v_1,\ldots,v_m\in V$ modelled by an $m$-dimensional random vector $(Y_1,\ldots,Y_m)$, we use historical datasets $(y_{1,t_1},\ldots,y_{m,t_1}),\ldots,(y_{1,t_k},\ldots,y_{m,t_k})$, which are considered as realizations of $(Y_1,\ldots,Y_m)$ being available from measurements of precipitation amounts for certain past times $t_1,\ldots,t_k \leq t_c$. Moreover, we use  Sklar's representation formula \citep{joe2011dependence} for the joint cumulative distribution function $G:\R^m\to [0,1]$ of $(Y_1,\ldots,Y_m)$, which states that
\begin{equation}\label{sklar}
G(y_1, \ldots, y_m) = C(G_1(y_1), \ldots, G_m(y_m)) \qquad\mbox{for all $y_1,\ldots,y_m\in\R$,}
\end{equation}
where  $G_i:\R\to[0,1]$ denotes the univariate  cumulative distribution function  of $Y_i$ for  $i=1,\ldots,m$, and $C:[0,1]^m\to[0,1]$  is an $m$-variate  copula, i.e.,   $C:[0,1]^m\to[0,1]$ is the restriction of
the $m$-variate cumulative distribution function of an $m$-dimensional random vector to the cube $[0,1]^m$ such that its components are uniformly distributed on the unit interval $[0,1]$ for each $i=1,\ldots,m$. Thus, to determine $G$, it is sufficient to determine the univariate distribution functions $G_1,\ldots,G_m$ and the copula $C$, where we assume that $C$ is an R-vine copula. 

Note that Sklar's representation formula given in Equation~\eqref{sklar} can be 
specified, if the joint distribution of $(Y_1,\ldots,Y_m)$ is (purely) continuous or discrete. In the discrete case, it is sufficient to determine the values of the copula $C:[0,1]^m\to[0,1]$  for the joint support $ R_{U_1} \times \ldots \times R_{U_m}\subset[0,1]^m$ of the vector $(U_1,\ldots,U_m)$ of transformed random variables  $U_i = G_i(Y_i)$,   $i=1,\ldots,m$,  instead of considering the values of $C$ for the entire cube $[0,1]^m$, where $R_{U_i}$ denotes the support of $U_i$ for $i=1,\ldots,m$.

In this paper, we are mainly interested in the joint probability density (or, in the discrete case, the probability mass function)
$g:\R^m\to[0,\infty)$
of $(Y_1,\ldots,Y_m)$, instead of determining the joint cumulative distribution function $G:\R^m\to [0,1]$ of $(Y_1,\ldots,Y_m)$.  Then, the following representation formulas for $g$ can be derived from Equation~\eqref{sklar}, see
\citep{joe2011dependence}:  
In the continuous case, the density $g$ of $(Y_1,\ldots,Y_n)$ is given by
\begin{equation}\label{density_cont}
g(y_1,\ldots,y_m)=c(G_1(y_1),\ldots,G_m(y_m))\prod_{i=1}^m g_i(y_i)\qquad\mbox{for all $y_1,\ldots,y_m\in\R$,}
\end{equation}
where $g_i:\R\to[0,\infty)$ denotes the density of $Y_i$ for $i=1,\ldots,m$,
and $c:[0,1]^m\to[0,\infty)$ is the density of $C$.
In the discrete case,

the probability mass function $g:\R^m\to[0,1]$ of $(Y_1,\ldots,Y_m)$ is given by
\begin{equation}\label{density_disc}
g(y_1,\ldots,y_m)=c(G_1(y_1),\ldots,G_m(y_m)) \qquad\mbox{for all $y_1\in R_{Y_1}, \ldots, y_m\in R_{Y_m}$,}
\end{equation}
where  $R_{Y_i}$ denotes the support of $Y_i$ for $i=1,\ldots,m$. Here, 
$c:R_{U_1} \times \ldots \times R_{U_m} \to[0,\infty)$ is the probability mass function of the transformed random vector $(U_1, \ldots, U_m)$ which is given by
\begin{equation} \label{sieve}
c(u_1, \ldots, u_m) = \sum_{j_1 \in \{0, 1\}} \ldots \sum_{j_m \in \{0, 1\}} (-1)^{j_1 + \ldots + j_m} C(u_1^{(j_1)}, \ldots, u_m^{(j_m)})
\end{equation}
for all $(u_1, \ldots, u_m)\in  R_{U_1} \times \ldots \times R_{U_m}$,
where $u_i^{(0)} = u_i$ and $u_i^{(1)} = \lim_{x \uparrow G^{-1}_i(u_i)}G_i(x)$ for $i=1,\ldots,m$.

Thus, in order to obtain $g$ for a dataset $(y_{1,t_1},\ldots,y_{m,t_1}),\ldots,(y_{1,t_k},\ldots,y_{m,t_k})\in\R^m$, we first determine the univariate cumulative distribution functions $G_1, \ldots, G_m:\R\to[0,1]$ for each location in $v_1, \ldots, v_m \in V$. Then, in a second step, an R-vine copula $C$ (or its density resp. probability mass function $c$) is fitted to the transformed dataset $(u_{1,t_1},\ldots,u_{m,t_1}),\ldots,(u_{1,t_k},\ldots,u_{m,t_k})\in[0,1]^m$. For a more detailed description regarding the fitting procedure of R-vine copulas to data, we refer to \cite{joe2011dependence}, see also \cite{aigner2023robust}.

\subsection{Generation of synthetic ensemble members}\label{sec.gen.syn}

From now on, we assume that all random variables, i.e., $X_1,\ldots,X_m$,  $Y_1,\ldots,Y_m$, and  $X'_1,\ldots,X'_m$, as well as the transformed random variables  $U_1,\ldots,U_m$, considered in the following, have discrete distributions.

We  show how the  samples $(x^1_i, \ldots, x^N_i)$ for $i=1,\ldots,m$, which have been independently drawn from the components of the random vector $(X_1,\ldots,X_m)$ as stated above,
can be rearranged into $N$ synthetic ensemble members $(x^{(j)}_1, \ldots, x^{(j)}_m)$ for $j \in \{1, \ldots, N\}$, such that these ensemble members can be considered as realizations of some $m$-dimensional random vector $(X'_1,\ldots,X'_m)$, whose univariate marginal distributions coincide with those of $(X_{1},\ldots,X_{m})$ and, in addition, which exhibits the spatial correlations observed in historical precipitation data.

For this purpose, i.e., to rearrange the samples $\{(x^1_i, \ldots, x^N_i), i=1,\ldots,m\}$ into  $N$ synthetic ensemble members $\{(x^{(j)}_1, \ldots, x^{(j)}_m), j =1, \ldots, N\}$,  we determine a permutation $\sigma_i:\{1,\ldots,N\}\to\{1,\ldots,N\}$ for each  $i=1,\ldots,m$ with $(x^{\sigma^{-1}_i(1)}_i, \ldots, x^{\sigma^{-1}_i(N)}_i) = (x^{(1)}_i, \ldots, x^{(N)}_i)$ such that the sample of ensemble members $\{(x^{(j)}_1, \ldots, x^{(j)}_m), j=1, \ldots, N\}$ exhibits the spatial correlation of precipitation amounts modeled by the copula $C:R_{U_1}\times\ldots\times R_{U_M}\to[0,1]$ introduced above.
For this, we define the likelihood $L(\sigma)$ for a set of permutations $\sigma = \{\sigma_i, i=1,\ldots,m\}$ as 
\begin{equation}\label{lik.lih.ood}
L(\sigma) = \prod_{j=1}^N g(x^{(j)}_1, \ldots, x^{(j)}_m)\,,
\end{equation}
where $g:R_{Y_1}\times\ldots\times R_{Y_m}\to[0,1]$ is the probability mass function  of the discrete random vector $(Y_1,\ldots,Y_m)$ fitted to historical precipitation data.

Thus, a set of permutations $\sigma$, which maximizes the likelihood $L(\sigma)$, rearranges the samples 
$\{(x^1_i, \ldots, x^N_i), i=1,\ldots,m\}$ into  $N$ synthetic ensemble members $\{(x^{(j)}_1, \ldots, x^{(j)}_m), j =1, \ldots, N\}$  such that they match the spatial correlation of precipitation amounts in the best possible way.

\subsection{Predicting the total precipitation amount in a given area}
\label{sec.methods.area_pred}

Recall that the $i$-th component $X_i$ of the random vector $(X_1,\ldots,X_m)$ introduced above has the same (univariate) distribution as the $i$-th component  $X'_i$ of  $(X'_1,\ldots,X'_m)$ for each $i\in\{1,\ldots,m\}$. However, in addition to this, the joint distribution of the random vector $(X'_1,\ldots,X'_m)$ obtained by maximizing the likelihood $L(\sigma)$ given in Equation~\eqref{lik.lih.ood},
also captures the spatial correlation of precipitation amounts. Thus, for any subset $S\subset \{1,\ldots,m\}$, the random sum
$\sum_{i\in S} X'_i$ can be considered as an appropriate prediction model for the total precipitation amount in an area which is represented by the set
$\{v_i,i\in S\}\subset V$ of locations. In particular, the probability $\mathbb{P}(\sum_{i\in S} X'_i \geq z)$ that the total precipitation amount $\sum_{i\in S} X'_i$ exceeds a certain (critical) threshold $z>0$ can be estimated by the relative frequency  $\#\{j: 1\le j\le N,\sum_{v_i \in S} x^{(j)}_i >z\}/N$, where $\#$ denotes cardinality.

\section{Implementation of the copula-based model}
\label{sec.implementation}

We now discuss some implementation details of the copula-based model introduced in Section~\ref{sec.methods}. First, we explain for which areas the proposed model generates synthetic ensembles, which are validated in Section~\ref{sec.verification} below. Next, the procedure for fitting an R-vine copula to historical data is explained in detail. Then, a hill-climbing algorithm is presented for optimizing the likelihood function $L(\sigma)$  introduced in Section~\ref{sec.methods}. Finally, we discuss the validation scheme utilized for evaluating the results stated in Section~\ref{sec.verification}. 

Note that so far in this paper, the term "location" has been used because, in general, the model introduced in Section~\ref{sec.methods}  does not assume that the underlying data is given on a regular grid, i.e., the set  $V=\{v_1,\ldots,v_m\}\subset\R^2$   introduced in Section~\ref{sec.methods}  can be arbitrarily shaped. However, in our case,  the data described in Section~\ref{sec.data} is arranged on a regular grid, and, therefore, we will refer to grid points (or grid boxes) in the following instead of calling them locations. Moreover, instead of considering one single set $V$ of grid points, in the following, we apply the copula-based model introduced in Section~\ref{sec.methods} to several sets of grid points simultaneously, i.e., we assume that the historical observations of precipitation amounts are statistically invariant in space and time. This means, in particular, that we fit one single R-vine copula to multiple precipitation observations from across the entire considered period and from non-overlapping subsets of some sampling window $W \subset \R^2$, respectively.

\subsection{Areas, where synthetic ensembles are generated}

In order to evaluate the copula-based model proposed in the present paper, it is applied to the combined forecast produced by the C$^3$-model, which receives forecasts of STEPS-DWD and ICON-D2-EPS as input and predicts calibrated marginal distributions for all considered grid points without taking spatial correlations into account. This combined forecast is available for a  sampling window $W\subset\R^2$, consisting of $350\times 450$ grid points and a rectangular subset of the regular COSMO-D2 grid, enclosing Germany and parts of neighbouring countries.

However, due to computational complexity, applying the copula-based model to the whole sampling window $W$ in a single step is impossible. So instead, for each hour $t$ available in the dataset, which belongs to the time intervals listed in the introduction of Section~\ref{sec.data}, we successively choose $n_t>1$ non-overlapping quadratic subsets $V_1^t, V_2^t, \ldots, V^t_{n_t} \subset W$, consisting of $9 \times 9$ grid points each and being positioned at random within $W$, until no further non-overlapping $9 \times 9$  subset can be found in $W$. 

\subsection{Fitting  R-vine copulas to historical observations}

For a given hour $t_c$ representing the ``current time'', we fit an R-vine copula to the historical observations made within the areas $V_1^t, V_2^t, \ldots, V^t_{n_t}\subset W$ and for the $k$ past hours  $t\in\{t_1,\ldots,t_k\}$
introduced in Section~\ref{sec.methods} with $t \leq t_c$, i.e., the R-vine copula is fitted to the vector data of precipitation amounts observed within the sets of $9 \times 9$ grid points described above, without taking local peculiarities into account. In other words, since the areas $V_1^t, V_2^t, \ldots, V^t_{n_t}$ are selected at random from all parts of the sampling window $W$, the fitted R-vine copula does not model local correlations which might be specific to a particular $9 \times 9$ area. This has the advantage that rare events, which might occur only at a few grid points within the entire dataset, do not influence how spatial correlation is modelled globally in $W$.

Note that the dataset on which the R-vine copula is fitted consists of observations for different past hours $t\in\{t_1,\ldots,t_k\}$ and different $9\times 9$ areas $V_1^t, \ldots, V^t_{n_t}\subset W$. However, for fitting the copula, this dataset is considered as realizations of one single random vector $(Y_1, \ldots, Y_m)$ for one single (abstract) $9\times 9$ area $V = \{v_1, \ldots, v_m \}$ with $m=81$, as introduced in Section~\ref{sec.methods}. Therefore, in the following, $V = \{v_1, \ldots, v_m \}$ does not refer to one specific area within $W$, but to an unspecified quadratic set of $m=81$ grid points and their relative positions to each other for which the random vector $(Y_1, \ldots, Y_m)$ is defined.

To fit the R-vine copula, the library \texttt{pyvinecopulib} \cite{vinecopulib} is used, where \texttt{pyvinecopulib} requires that the univariate (marginal) distributions of $Y_i$ for $i\in\{1,\ldots,m\}$ fitted to the historical precipitation data are either continuous or discrete. Thus, since the distribution of precipitation amounts has an atom at $0$\,mm, we consider a discretized marginal distribution for each  $i\in\{1,\ldots,m\}$, represented by its cumulative distribution function $G_i:R_{Y_i}\to[0,1]$. Next, using $G_i$, we transform the  precipitation amount $y_i$ observed at  $v_i \in V$  to obtain $u_i = G_i(y_i) \in [0, 1]$.

The R-vine copula is now fitted to the vectors $(u_1, \ldots, u_m)$ of transformed precipitation amounts, where the fact is used that an R-vine copula is built by a set of bivariate copulas. Thus, in the fitting process, \texttt{pyvinecopulib} determines the most suitable copula family for each of these bivariate copulas with the help of statistical tests. For the results derived in this paper, the bivariate copulas are chosen from all available parametric copula families provided by \texttt{pyvinecopulib}. However, to reduce computational costs, we truncate the R-vine copula considering only five trees within the vine structure, i.e., for each of the $m=81$ arguments of the R-vine copula, the correlation structure with five other suitably chosen arguments is considered. For more details regarding truncated vine copulas and vine copulas in general, we refer to \cite{joe2011dependence}.

\subsection{Generating samples of predicted precipitation amounts}

In the following, we describe how to generate a sample $(x_i^1,\ldots,x_i^N)$ of predicted precipitation amounts for an hour $t > t_c$ and for each of the $m=81$ grid points of each $9\times 9$ area $V_1^t, \ldots, V^t_{n_t}\subset W$, where $i\in\{1,\ldots,m\}$.

As described in Section~\ref{sec.methods}, the C$^3$-model has been modified to generate quantiles of the predicted distributions of precipitation amounts for each grid point in $W$. For our purpose, the C$^3$-model has been trained to generate vectors $q=(q_1,\ldots,q_{100})$ of $100$ $\alpha$-quantiles, where the values of $\alpha$ are evenly spaced between $0.0001$ and $0.9999$. 
Then, for each grid point in $V_1^t \cup \ldots \cup V^t_{n_t}$, we select $N=20$ values at random among the components of the corresponding vector $q = (q_1, \ldots, q_{100})$. To ensure that these values are spread out across the entire support of the predicted precipitation distribution, the components of $q=(q_1, \ldots, q_{100})$ are divided into $N=20$ consecutive groups $(q_1, \ldots, q_5), (q_6, \ldots, q_{10}), \ldots, (q_{91}, \ldots, q_{95}), (q_{96}, \ldots, q_{100})$, each consisting of five quantiles from which one quantile is selected at random. Note that this procedure is similar to the stratified sampling approach discussed in \cite{hu2016stratified}. In the following, the matrix of the $N$ quantiles drawn from $q$ for each of the $m=81$ grid points within a given $9 \times 9$ area will be denoted by $x = (x^j_i, i \in \{1, \ldots, m \}, j \in \{1, \ldots, N \}) \in \mathbb{R}^{m \times N}$.

\subsection{Hill climbing algorithm for ensemble generation}

To find a set of permutations  $\sigma = \{\sigma_i, i=1,\ldots,m\}$ that maximizes the likelihood $L(\sigma)$ introduced in Equation~\eqref{lik.lih.ood}, a hill climbing algorithm \citep{skiena1998algorithm} is applied. This algorithm starts with a set of random permutations $\sigma = \{\sigma_i, i=1,\ldots,m\}$, i.e., for each grid point $i \in \{1, \ldots, m \}$ the quantiles in $(x_i^1, \ldots, x_i^N)$ are rearranged at random.
These arrangements are iteratively changed such that the value of the evaluation function $L(\sigma)$ increases with each step, where the algorithm is structured as follows, 
 see also Algorithm \ref{alg.hill_climbing} below
:
\begin{enumerate}
    \item The inputs of the algorithm are the matrix $x = (x^j_i, i \in \{1, \ldots, m \}, j \in \{1, \ldots, N \}) \in \mathbb{R}^{m \times N}$, which contains $N$ predictions for each of the $m$ grid points, and the  $m$-variate density function $g:R_Y\to[0,\infty)$ introduced in Section~\ref{sec.methods}.
    \item As already mentioned above, the algorithm starts with a set of random permutations $\sigma = \{\sigma_i, i=1,\ldots,m\}$,   i.e., there is no (spatial) correlation between the rows $(x_i^1, \ldots, x_i^N)$ of $x$ after applying the set $\sigma$ of random permutations.
    \item Iterate over each synthetic ensemble member $(x_1^j, \ldots, x_m^j)$ for $j \in \{1, \ldots, N \}$:
    (a) Find $k \in \{j, \ldots, N \}$ such that $g(x^j_1, \ldots, x^k_p, \ldots, x^j_m)$ is maximized for each grid point $p \in \{1, \ldots, m \}$. If $k \neq j$ switch $x^j_p$ and $x^k_p$ in order to improve $g(x^j)$.
        (b) Repeat step (a) until no values in $x$ have been switched.
    \item Return $x$. Note that at this step of the algorithm, the variable $x$ contains the ordered values that were referred to as $\{(x^{(j)}_1, \ldots, x^{(j)}_m), j =1, \ldots, N\}$.
\end{enumerate}

\begin{algorithm}
\caption{Synthetic Ensemble Optimization with Hill Climbing}\label{alg.hill_climbing}
\begin{algorithmic}[1]
\Procedure{Hill\_Climbing}{$x, g$}\Comment{$x \in \mathbb{R}^{m \times N}$ contains $N$ predictions for $m$ grid points}
\State \Comment{$g$ is the density function fitted in Section \ref{sec.methods}}
\For{$j \in \{1, \ldots, N \}$} \Comment{Iterate over all synthetic ensemble members}
    \Repeat
        \State $c \gets \texttt{False}$
        \For{$p \in \{1, \ldots, m \}$} \Comment{Iterate over all grid points (in random order)}
            \State $k \gets \argmax_{k \in \{j, \ldots, N \}} g(x^j_1, \ldots, x^k_p ,\ldots, x^j_m)$ \Comment{Find best prediction for $p$}
            \If{$k > j$}\Comment{true, if switching $x^k_p$ with $x^j_p$ improves $g(x^j)$}
                \State $x^j_p \leftrightarrow x^k_p$ \Comment{Predictions in members $j$ and $k$ switch places}
                \State $c \gets \texttt{True}$ \Comment{$c=\texttt{True}$ indicates that a change has been made to $x$}
            \EndIf
        \EndFor
    \Until{$\neg c$} \Comment{Stop if no improvement has been made for any grid point in $x^j$}
\EndFor
\State \Return $x$ \Comment{$x$ contains the ordered values referred to as $\{(x^{(j)}_1, \ldots, x^{(j)}_m), j =1, \ldots, N\}$}
\EndProcedure
\end{algorithmic}
\end{algorithm}

The idea of reordering samples drawn from marginal distributions in order to obtain realistic ensemble members is also used in the approaches of the Schaake shuffle \citep{clark2004schaake} and ensemble copula coupling \citep{Wilks2006}.
However, in those approaches, the permutations for each grid point are provided either by a given set of ensemble members or directly by a set of historical observations. In contrast, the approach considered in the present paper is based on a fitted copula model.

\begin{figure}[H]
	\centering
            \includegraphics[width=0.66\linewidth]{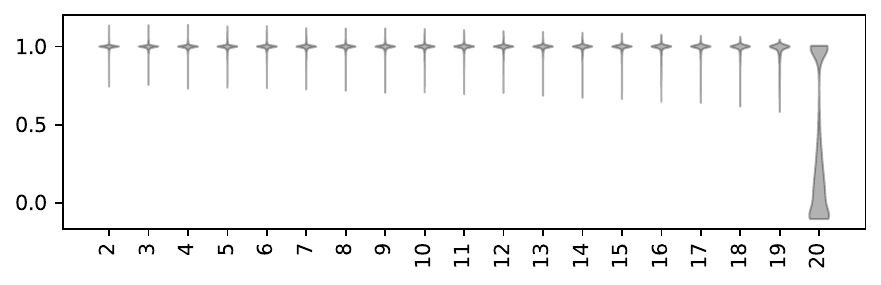}
		\caption{Relative quality of all synthetic ensemble members $(x_1^{(i)}, \ldots, x_m^{(i)})$ evaluated in the dataset ($\sim 50000$ in total) compared to the first member $(x_1^{(1)}, \ldots, x_m^{(1)})$ within the same $9 \times 9$ area for the lead time $+3$\,h. The relative quality is expressed by the distribution of the quotient $\text{log}(g(x_1^{(i)}, \ldots, x_m^{(i)})) / \text{log}(g(x_1^{(1)}, \ldots, x_m^{(1)}))$ for $i=2, \ldots, 20$ with $90\%$ winsorization. Note that the distributions of the relative qualities looks essentially the same for other lead times.}
		\label{fig.loglik}
\end{figure}

For the results stated in this paper, $N$ is chosen to be $20$; however, the last member is omitted since the number of values from which the algorithm can choose at a grid point decreases as $j\in\{1,\ldots,N\}$ increases, the quality of the synthetic members, therefore, decreases for higher values of $j$. The decrease in quality is depicted in Figure~\ref{fig.loglik}, where it is shown that this effect mainly affects the last member $j=N$.

\subsection{Validation scheme}

The copula model proposed in this paper is validated with the help of a rolling-origin scheme \citep{armstrong1972comparative}. In this scheme, the available dataset is split by the ``current time'' $t_{\rm c}$, into a ``past'' and a ``future''.
Throughout the validation, the "current time" $t_{\rm c}$ is incrementally shifted from the start to the end of the dataset in chronological order. In each step, the model is updated on the ''past'' data while the ``future'' data is used to validate model predictions. This approach is especially suitable for datasets with a time axis and also because it closely simulates operational conditions.

Furthermore, in each step, the model is applied to randomly chosen, non-overlapping areas of size $9 \times 9$, as described at the beginning of this section. Each application results in 19 synthetic ensemble members. Based on these synthetic ensemble members, the probability is estimated that the total precipitation amount within the area exceeds a given threshold. The resulting validation scores are discussed in the following section.

\section{Validation of synthetic rainfall fields}
\label{sec.verification}
We want to assess how well the copula-based technique proposed in this paper models realistic spatial dependencies and to which extent the original requirements of a consistent and calibrated combination forecast are retained. For this, we compare the synthetic rainfall fields with two extreme cases of spatial correlation. A minimum of spatial correlation is realized by randomly arranging the $x$-values at each grid point, which have been drawn from the $C^3$-model as described in Section~\ref{sec.implementation}. On the other hand, the maximum possible spatial correlation is achieved by sorting these sample values by size, i.e., the first synthetic ensemble member contains the largest values for each grid point, the second ensemble member the second largest ones, etc. However, since the random arrangement reveals very poor results for all considered metrics,  we only discuss the results of the comparative validation against the sorted arrangement. In the following, forecasts, for which the spatial dependencies are restored by the model proposed in this paper, are referred to as ``COPULA'', whereas the sorted scenario is named ``COPULA sorted''. Furthermore, the sorting arrangement is also made for both input forecasts named ``STEPS sorted'' and ``ICON sorted''. Note that while we append "sorted" to each of the three model names for consistency, the ``COPULA sorted'' forecast consists only of the sorted predictions of the C$^3$-model and does not require the application of a copula model or the hill climbing algorithm.

For the evaluation of the core requirements, we first computed traditional metrics such as bias, Brier skill score and reliability for approximately $50000$ (sub-) regions $V\subset W$ that we have been drawn at random for each considered hour within our three months period and for all lead times, see Section~\ref{sec.data}. Thus, the dataset consists of $1761$ forecast hours for each lead time, where the average number of evaluated sub-regions per forecast hour is equal to $28,4$. These traditional metrics mentioned above have been computed for threshold exceedance probabilities to get easily interpretable information about the systematic model error (bias), the forecast quality regarding both model and forecast error (Brier skill score), and the conditional frequency bias (reliability).

Moreover, to get a better insight into the spatial structure of the synthetic rainfall fields and their realistic appearance, we identified objects as sets of connected grid points, in which precipitation occurs, by a standard segmentation method using 4-connectivity and, furthermore, evaluated the parameters of a fitted exponential variogram model. Based on the identified objects, we computed their number, overall area, scaled volume, and weighted centre distance. Note that the latter two metrics were initially introduced in \cite{wernli2008sal}. In addition, we assessed the model performance by the aggregation metrics $D_0$ and $D_1$, which were introduced in \cite{tobin2012observational}. For a discussion of these object-based metrics, see also \cite{lit:Rempel2017}.

\begin{figure}[H]
	\centering
		\includegraphics[width=0.7\linewidth]{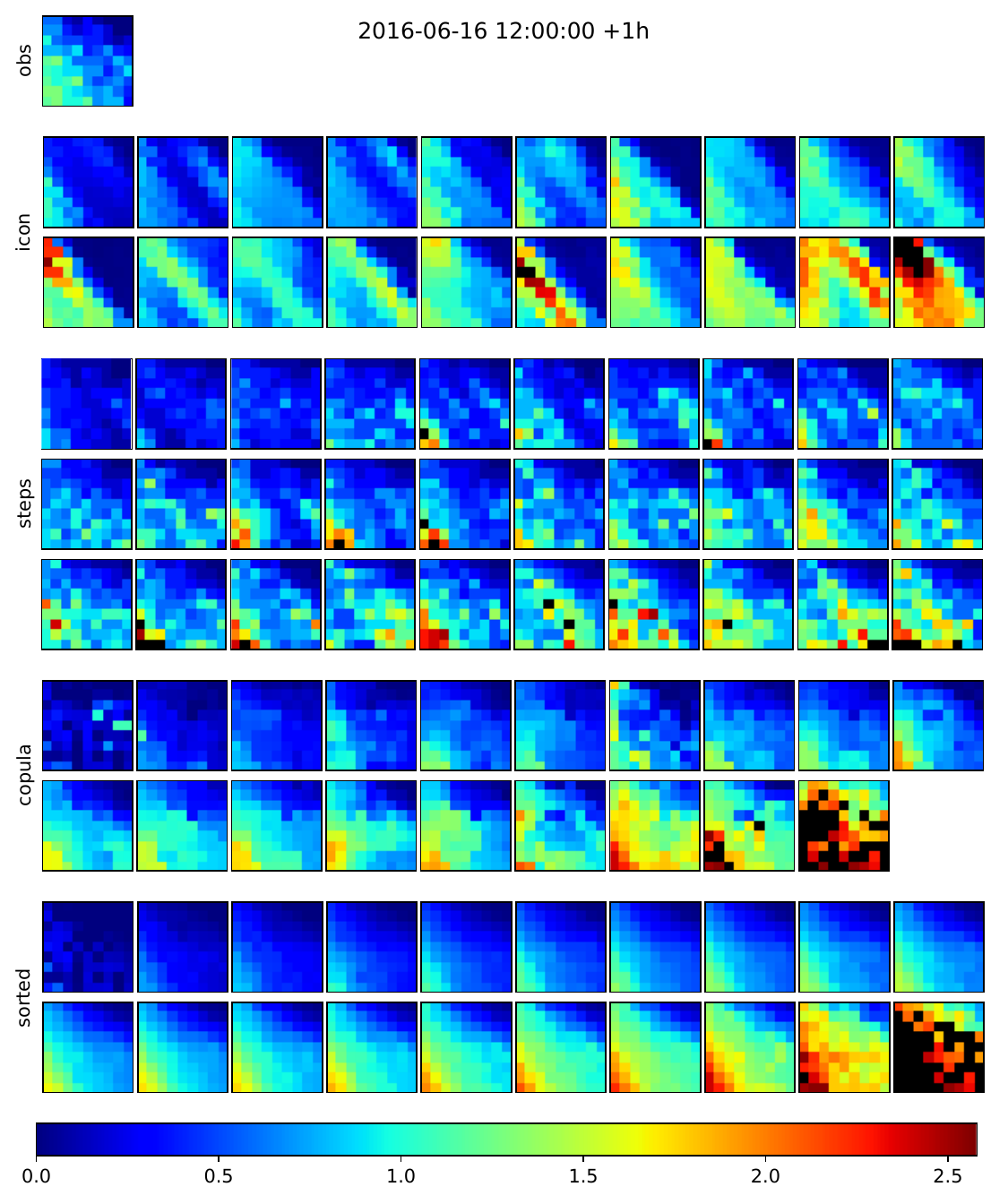}
		\caption{Observed precipitation amounts (obs) and realizations of 1\,h forecasts of the four considered models (ICON, STEPS, COPULA, and the sorted predictions of the C$^3$-model) for a randomly chosen subset  of 9$\times$9 grid points. The colour scale is given in millimetres. Precipitation amounts above the right-hand end of the colour scale are shown in black.}
\label{fig.examples}
\end{figure}

\subsection{Visual Inspection}

Figure~\ref{fig.examples} depicts an exemplary observation (obs) at the top for a randomly chosen sub-region $V\subset W$ and forecast hour, together with the ensemble members of the NWP forecast (ICON) and the precipitation nowcasting (STEPS). Note that ICON provides 20 ensemble members, while STEPS has 30. The synthetic ensemble members resulting from the copula model (COPULA) with restored spatial dependencies are illustrated as a fourth category. At the bottom, the randomly drawn and sorted marginal forecasts (COPULA sorted) are depicted. As discussed in the previous section and illustrated in Figure~\ref{fig.loglik}, the quality of the last synthetic ensemble member generated by the hill climbing algorithm differs vastly from previous iterations and is thus omitted. Therefore,  COPULA shows only 19 ensemble members. For each ensemble, the members are depicted in Figure~\ref{fig.examples}
in ascending order of the total amount of precipitation within the considered area.

It can be seen that each ensemble forecast exhibits different spatial correlation patterns. ICON ensemble members appear relatively smooth, while the ensemble members of STEPS show a high level of variation between neighbouring grid points. This is because the effective resolution of ICON is lower than the resolution of the considered grid, while the native grid of STEPS has an even higher resolution, as discussed in Section \ref{sec.data}. When considering the ensemble members of  COPULA, we see that the proposed model reintroduced spatial features, i.e., values at neighbouring grid points are more similar compared to grid points further apart. When comparing the COPULA ensemble and its sorted version, we see that the sorted version is much smoother since it is the arrangement with the highest possible correlation between grid points. Note that the precipitation amounts for the ensemble members generated by COPULA and its sorted version are provided by the C$^3$-model and differ only in their arrangement to synthetic ensemble members.

\subsection{Traditional  metrics}

\begin{figure}[H]
	\centering
		\includegraphics[width=0.8\linewidth]{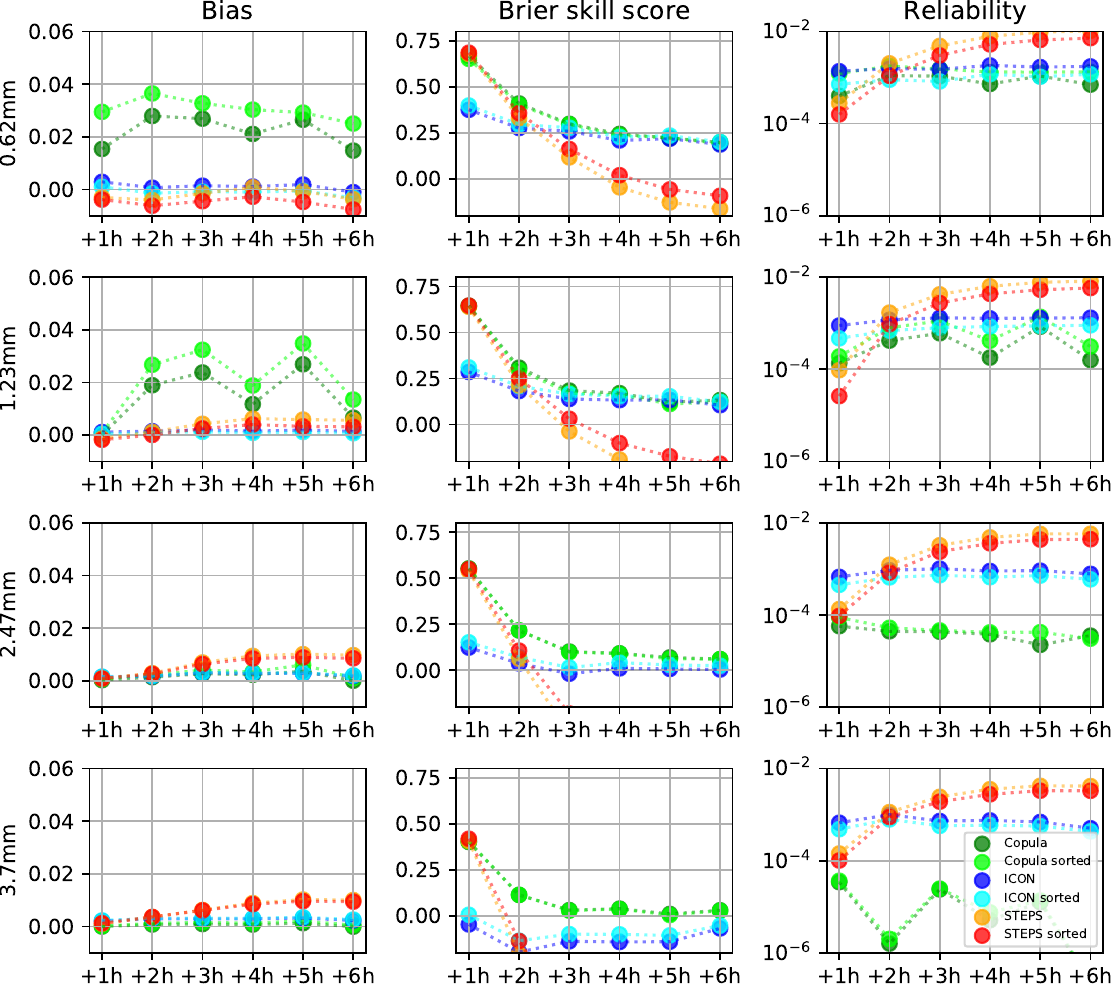}
		\caption{Bias, Brier skill score and reliability for threshold exceedance probabilities predicted by the models STEPS (orange), STEPS sorted (red), ICON (blue), ICON sorted (cyan), COPULA (green) and Copula sorted (light green).}
		\label{fig.scores}
\end{figure}

The results depicted in Figure~\ref{fig.scores} for bias (left), Brier skill score (centre), and reliability (right) for thresholds from 0.62\,mm up to 3.7\,mm are shown for STEPS (orange), STEPS sorted (red), ICON (blue), ICON sorted (cyan), COPULA (green), and COPULA sorted (light green). For lower thresholds (0.62\,mm and 1.23\,mm), the systematic error of area probabilities based on COPULA is decreased compared to the sorted arrangement of C$^3$ samples. Nevertheless, for both thresholds, COPULA and COPULA sorted are systematically overestimating the exceedance probabilities.

However, the forecast quality in terms of the Brier skill score depicted in the centre column of Figure~\ref{fig.scores} indicates no loss compared to the input models, albeit improvements over ICON are unremarkable for lower thresholds at later lead times. The reliability depicted in the right column of Figure~\ref{fig.scores} reveals that the copula model has an improved relative frequency bias compared to the input models except for a lead time of 1\,h for the lower thresholds.

Comparing these results for area probabilities with those from the grid point perspective in \cite{Rempel2022}, it can be seen that there is no large decline in the results based on a traditional forecast verification. That indicates that our core requirements of a calibrated and consistent combination are still fulfilled and that the calibration of marginal distributions also improves the prediction of area probabilities.

\begin{figure}[H]
	\centering
		\includegraphics[width=1\linewidth]{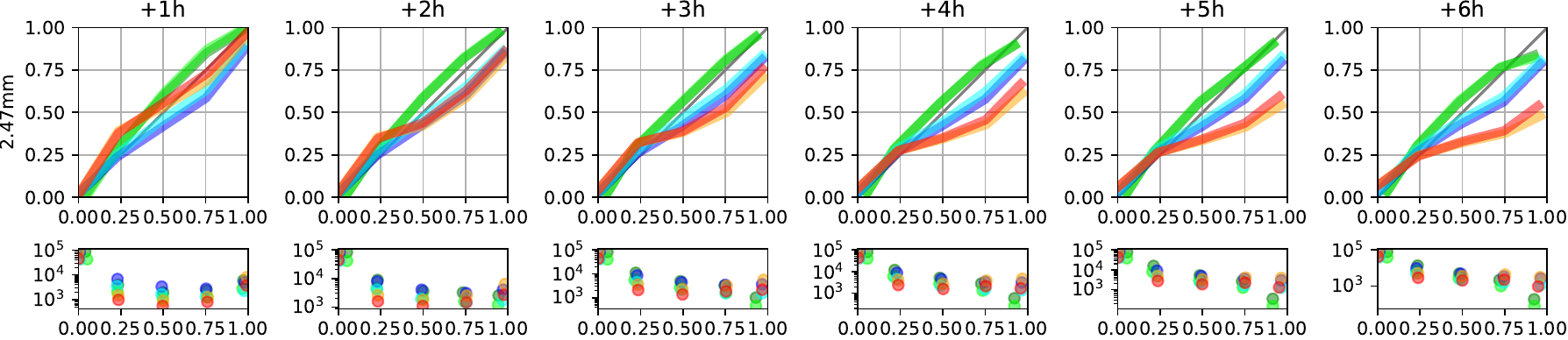}
		\caption{Reliability (top) and frequency diagrams (bottom) for threshold exceedance probabilities ($2.47$\,mm) predicted by the models STEPS (orange), STEPS sorted (red), ICON (blue), ICON sorted (cyan),  COPULA (green) and COPULA sorted (light green).}
		\label{fig.reldiag}
\end{figure}

To further demonstrate the ensemble calibration of area probabilities, Figure~\ref{fig.reldiag} depicts reliability diagrams for every forecast lead time for a threshold of 2.47\,mm. These diagrams allow for a more detailed insight into the reliability as the scores depicted in Figure~\ref{fig.scores}. Here, the curves for COPULA are close to the leading diagonal, indicating that the calibration performed by the C$^3$-model (see \citep{Rempel2022}) also improves the calibration of area forecasts. Furthermore, the occurrence of probabilities close to 1 decreases with increasing forecast uncertainty, i.e., for increasing thresholds and longer lead times. This decrease is indicated by an increasing truncation of the curves. Compared with this, the input forecasts of ICON and STEPS reveal a lead-time invariant and increasing overforecast, respectively.

\begin{figure}[H]
	\center
    \begin{tabular}{c|cccc}
    & \footnotesize \makecell{ICON} & \footnotesize \makecell{STEPS} & \footnotesize \makecell{COPULA} & \footnotesize \makecell{COPULA sorted}\\ \hline
    \footnotesize \makecell{$+1$\,h} &
    \makecell{\includegraphics[width=0.1\textwidth]{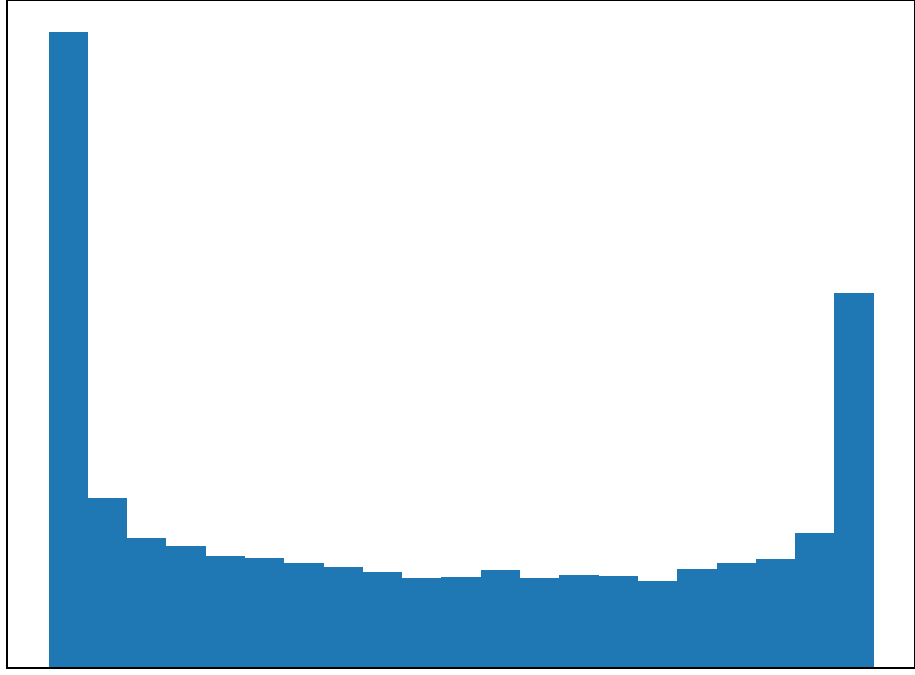}} &
    \makecell{\includegraphics[width=0.1\textwidth]{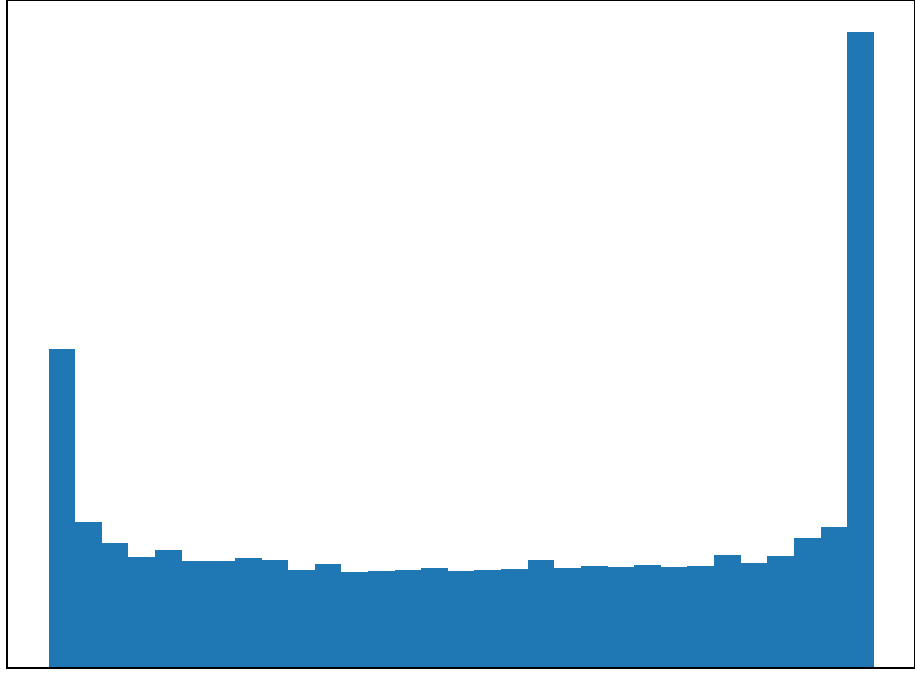}} &
    \makecell{\includegraphics[width=0.1\textwidth]{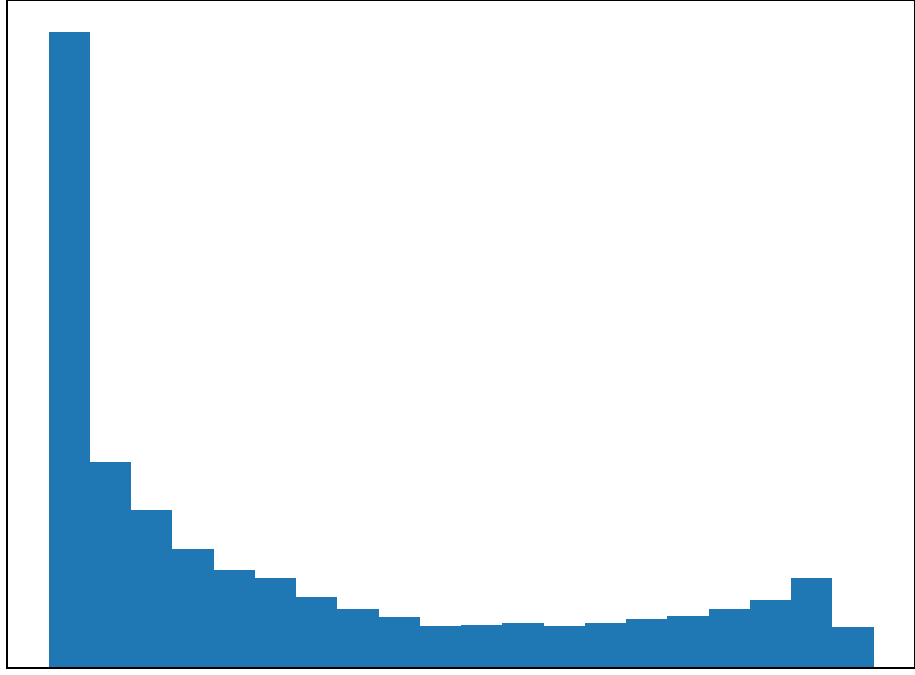}} &
    \makecell{\includegraphics[width=0.1\textwidth]{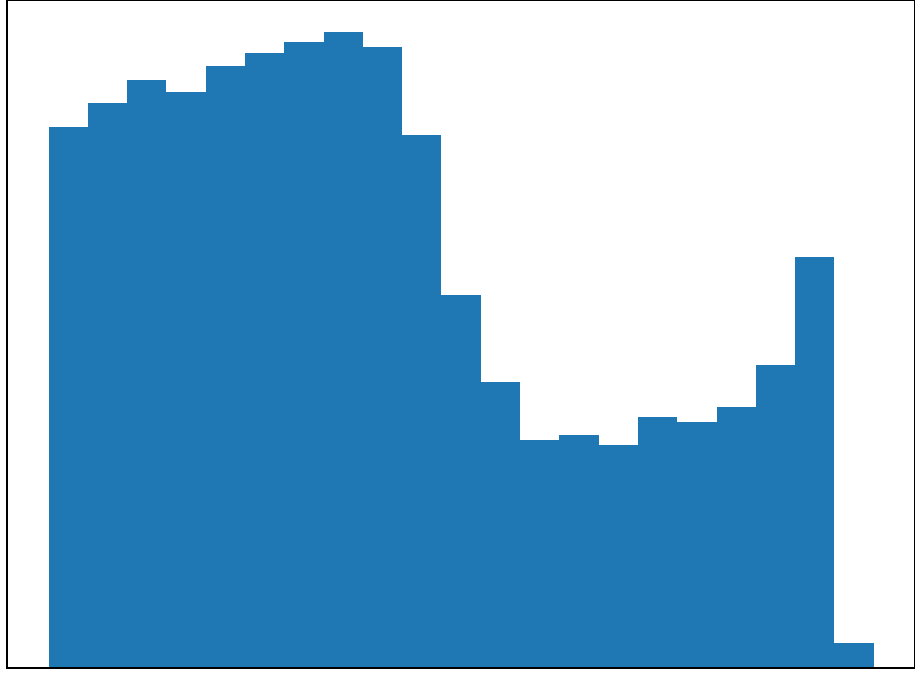}}\\
    \footnotesize \makecell{$+3$\,h} &
    \makecell{\includegraphics[width=0.1\textwidth]{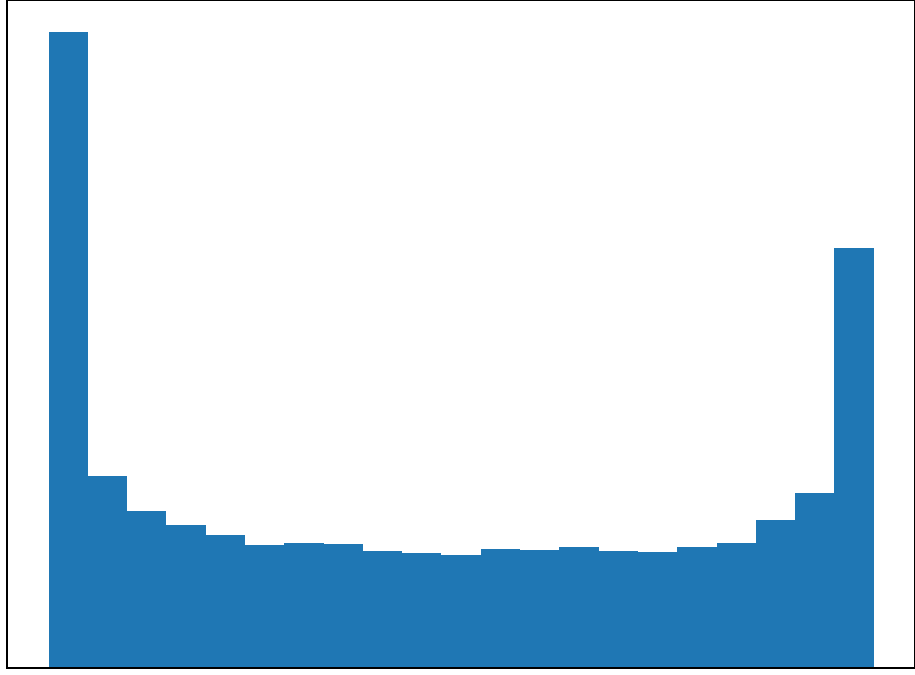}} &
    \makecell{\includegraphics[width=0.1\textwidth]{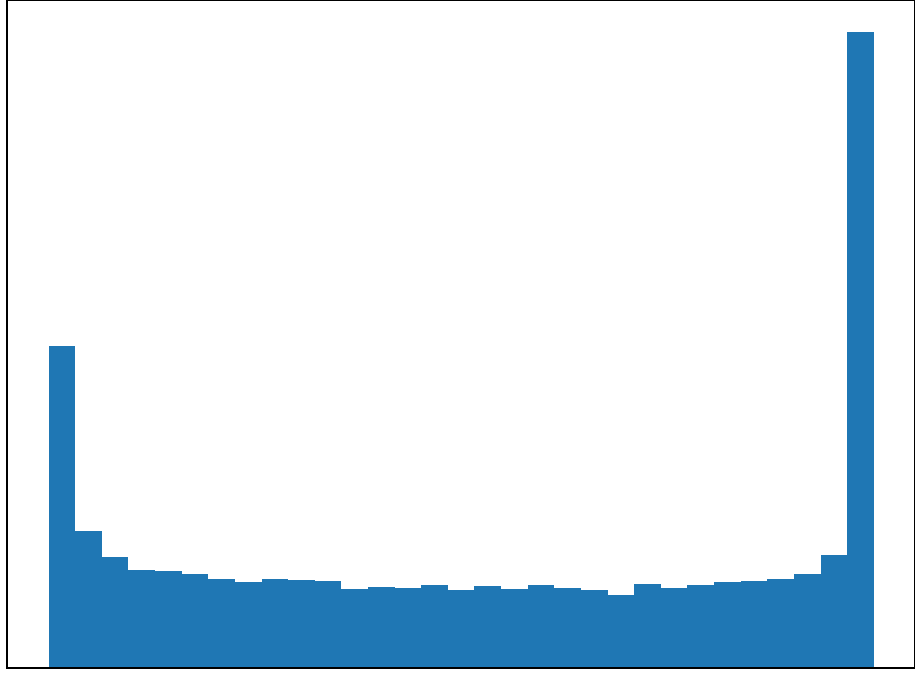}} &
    \makecell{\includegraphics[width=0.1\textwidth]{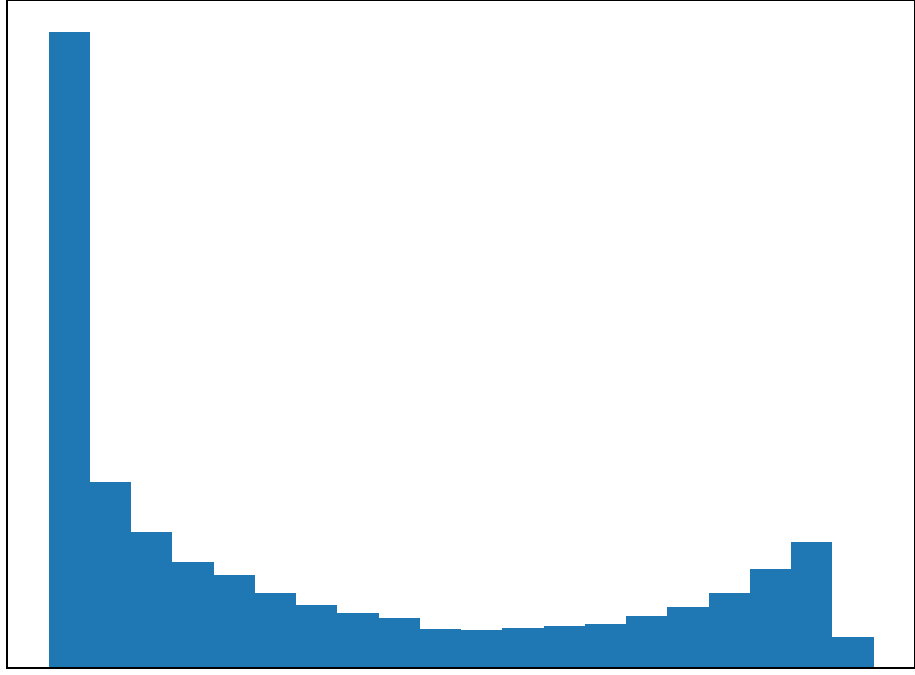}} &
    \makecell{\includegraphics[width=0.1\textwidth]{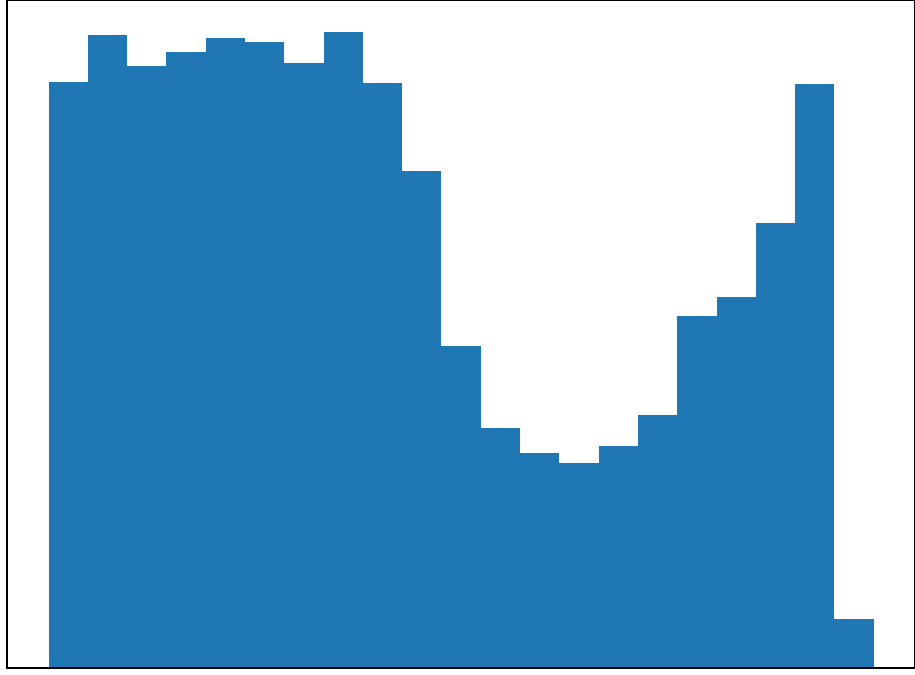}}\\
    \footnotesize \makecell{$+6$\,h} &
    \makecell{\includegraphics[width=0.1\textwidth]{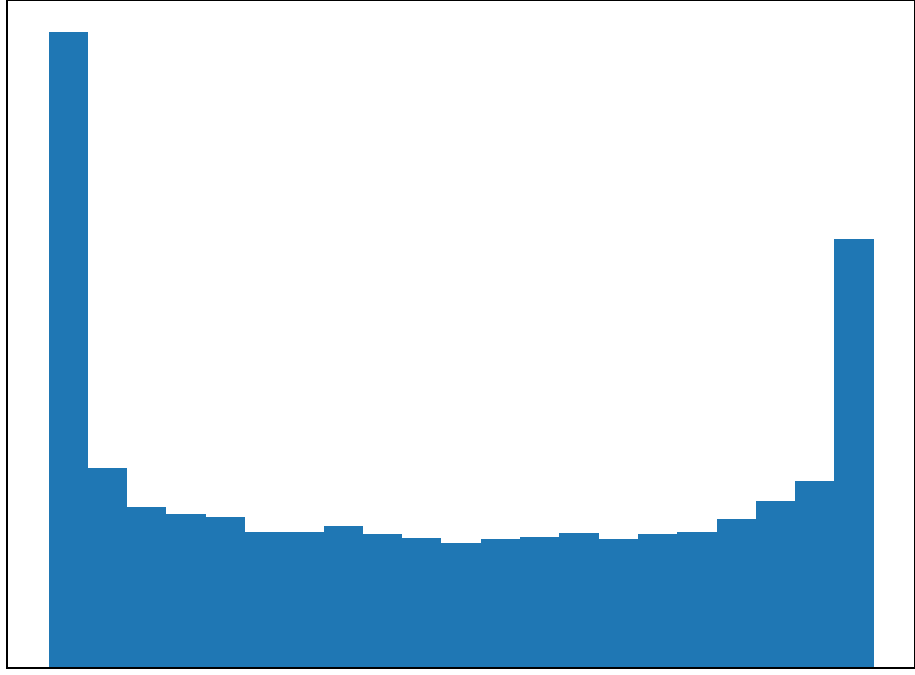}} &
    \makecell{\includegraphics[width=0.1\textwidth]{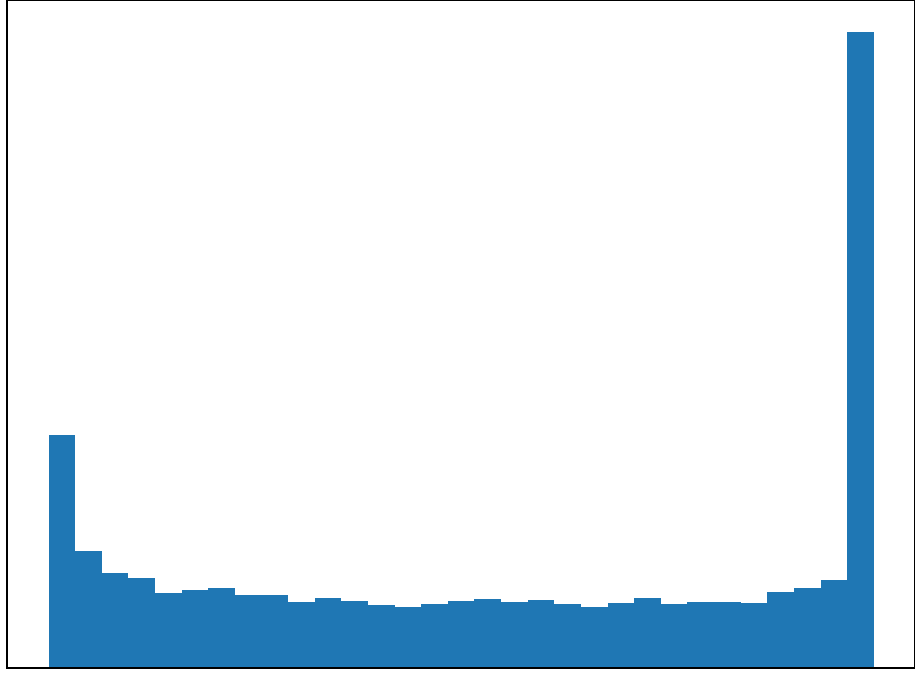}} &
    \makecell{\includegraphics[width=0.1\textwidth]{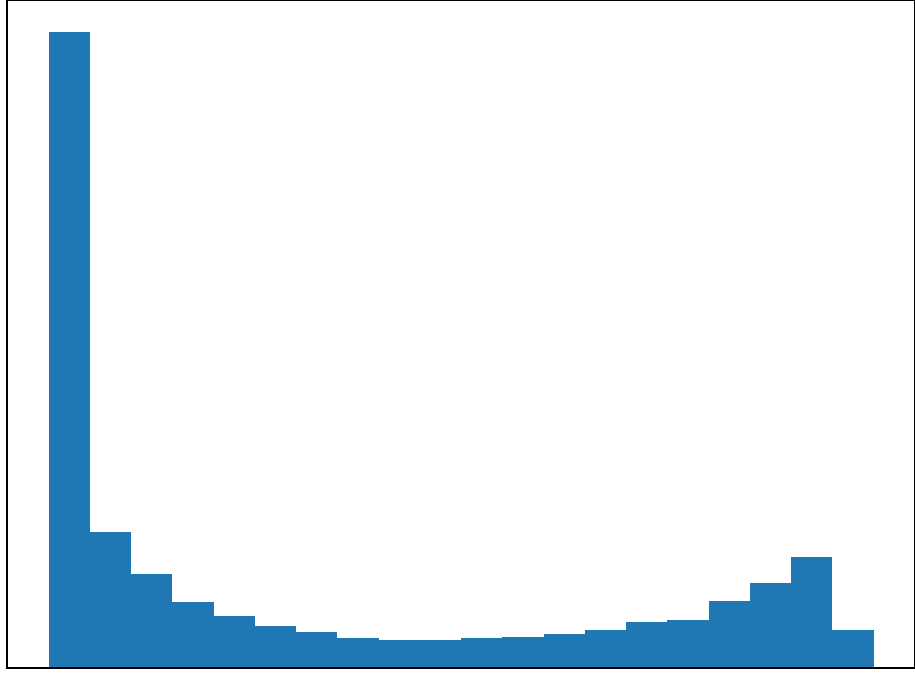}} &
    \makecell{\includegraphics[width=0.1\textwidth]{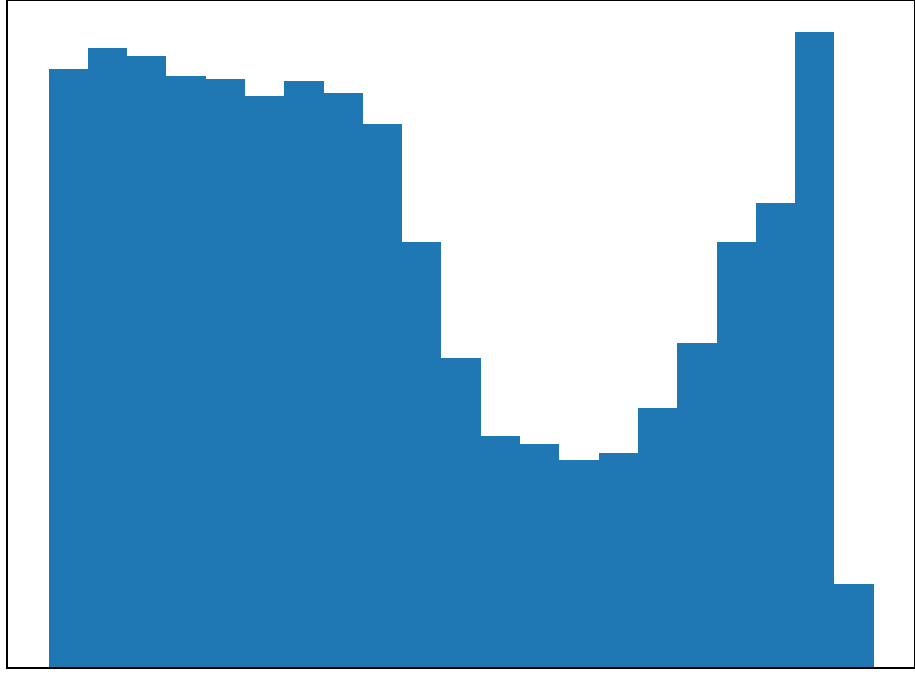}}
    \end{tabular}
    \caption{Probability integral transform (PIT) diagrams for area precipitation amounts predicted by the four models STEPS, ICON, COPULA as well as COPULA sorted for lead times of +1\,h (top), +3\,h (centre), and +6\,h (bottom). Here, the values of $0$ and $1$ along the horizontal axes correspond to the lowest and highest values, respectively, predicted by the corresponding ensemble forecast, whereas the binned events indicate whether the observed precipitation amounts are below or equal to the rainfall amount of the respective ensemble member. Thus, events in the first (last) bin represent observed precipitation amounts below (above) the lowest (highest) forecast value.}
    \label{fig.pit_precipitation}
\end{figure}

We want to assess not only the forecast quality of exceedance probabilities in terms of bias, Brier skill score, and reliability, but also the statistical behavior of predicted precipitation amounts. For this, probability integral transform (PIT) diagrams \citep{czado2009predictive} for areal precipitation amounts predicted by the four models STEPS, ICON, COPULA, as well as COPULA sorted for different lead times are depicted in Figure~\ref{fig.pit_precipitation}. In such diagrams, the horizontal axes are defined as the normalized value ranges of the forecasts, whereas the numbers of bins are equal to the ensemble sizes. The actually occurred precipitation amounts are sorted into the respective ensemble member's bin with an equal or higher rainfall amount forecast. Thus, events in the first (last) bin represent observed rainfall amounts below (above) the forecast value range.

Note that each bin of a PIT diagram corresponds to an interval between two ensemble members for a given metric. In cases where the considered metric is identical for two or more ensemble members, the corresponding intervals have length zero and observations with the same metric cannot be unambiguously assigned to one bin. When considering metrics like precipitation amounts, this is a common occurrence due to the atom in the distribution at $0$\,mm. To mend this, we use the approach presented in Equation~\eqref{sklar} in \cite{czado2009predictive}, where such observations are randomly assigned to one of the bins in question.

For both raw input ensemble forecasts, many observed precipitation amounts are in the first and last bin, respectively, revealing an underdispersive behaviour, i.e., the ensemble does not cover the whole range of observations. The overestimation at the lower percentile might be induced by non-precipitation cases but also by an overforecast of the precipitation amount, the latter, especially for ICON forecasts. However, it should be noted that this result may be sensitive to the QPE that we have used as observation. On the other hand, STEPS shows deficiencies with higher precipitation amounts. This may be induced by the cascade of autoregressive processes that reduces the maxima of intensity and cannot cover the range of growth/decay processes of precipitation.

With precipitation sums based on synthetic rainfall fields generated by COPULA, we can at least cover the observed  range of values at the upper bound. However, many cases with less precipitation are not covered. For this, also non-precipitation cases may play a role. Since at least a few values drawn from the upper tail of the probability distributions predicted by the C$^3$-model are always positive, the total precipitation amount within synthetic ensemble members is often positive and not zero. This could be underpinned by the sorted arrangement depicted in the right column of Figure~\ref{fig.pit_precipitation}. In these diagrams, no outliers are visible at both bounds due to a larger range of predicted area precipitation sums compared to COPULA. However, the frequency of larger precipitation amounts is higher in the sorted arrangement, since most of the observed rainfall is below the median. Thus, the sorted arrangement achieves the best PIT diagrams compared to the  other three models.

\subsection{Object-based metrics and estimated variogram models}

\begin{figure}[H]
	\center
	\scriptsize
    \begin{tabular}{c|c||cccc||cccc}
    & & \makecell{ ICON} & \makecell{STEPS} & \makecell{COPULA} & \makecell{Sorted} & \makecell{ICON\\ $\geq 0.1$\,mm} & \makecell{STEPS\\ $\geq 0.1$\,mm} & \makecell{COPULA\\ $\geq 0.1$\,mm} & \makecell{Sorted\\ $\geq 0.1$\,mm}\\ \hline\hline
    & \makecell{Area} &
    \makecell{\includegraphics[width=0.05\textwidth]{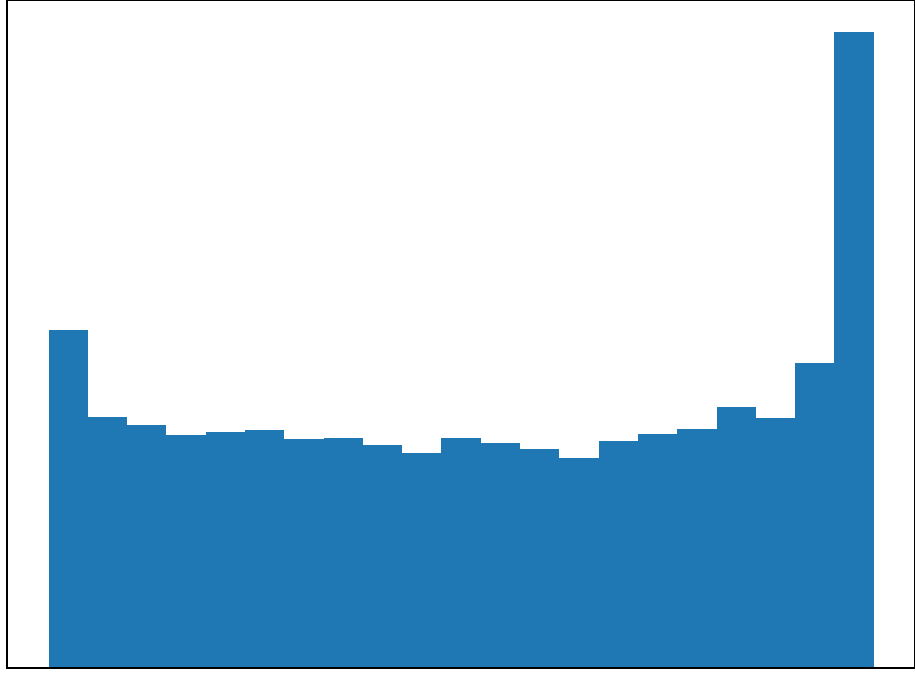}} &
    \makecell{\includegraphics[width=0.05\textwidth]{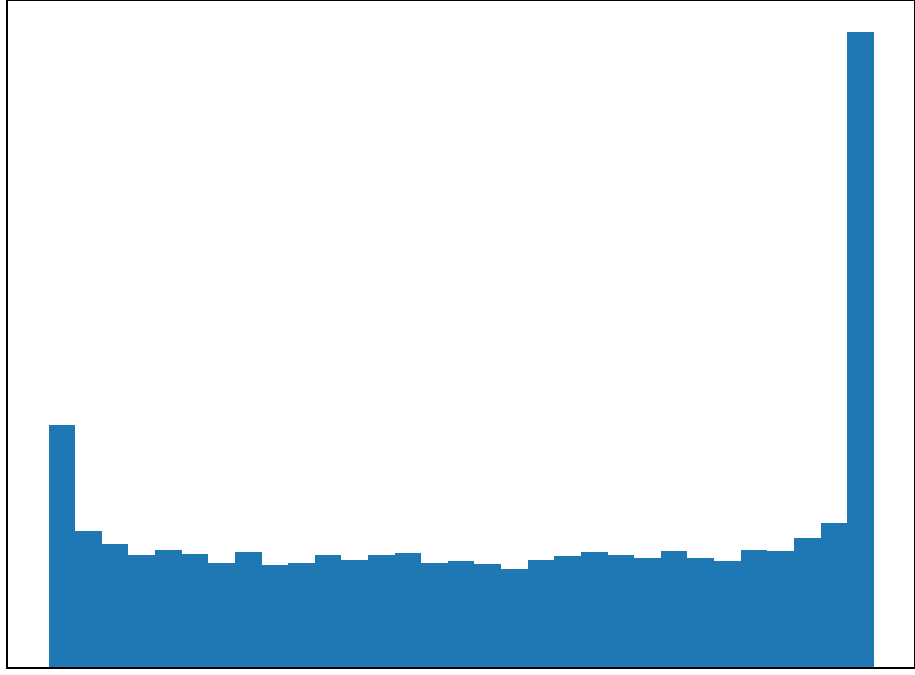}} &
    \makecell{\includegraphics[width=0.05\textwidth]{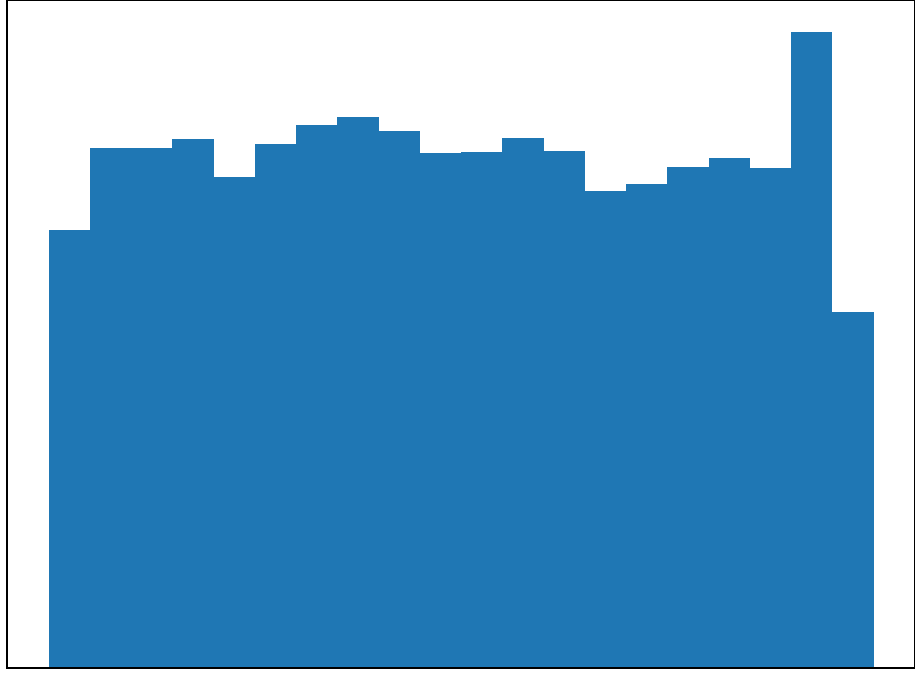}} &
    \makecell{\includegraphics[width=0.05\textwidth]{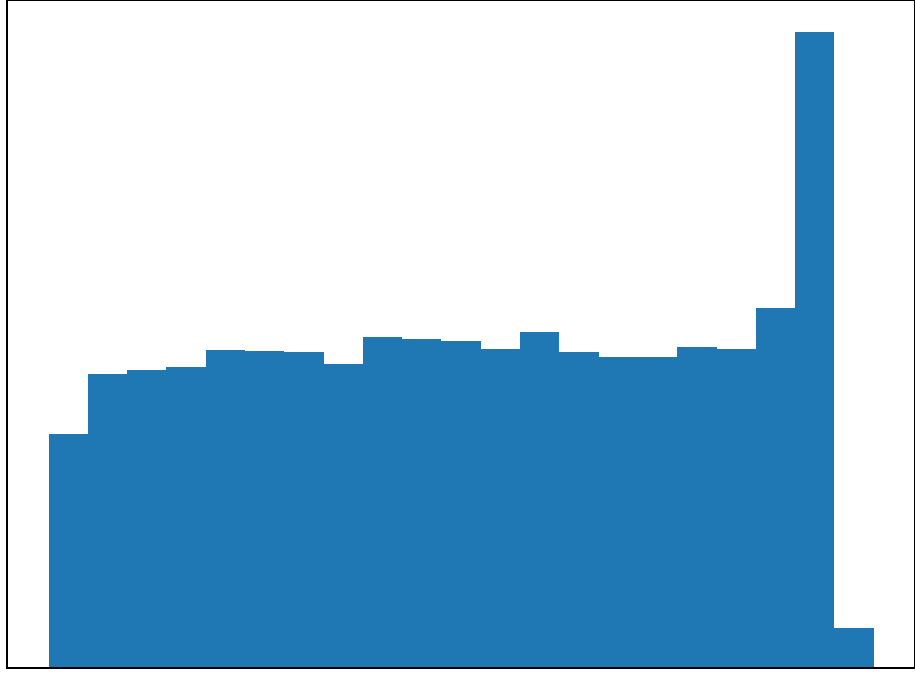}} &
    \makecell{\includegraphics[width=0.05\textwidth]{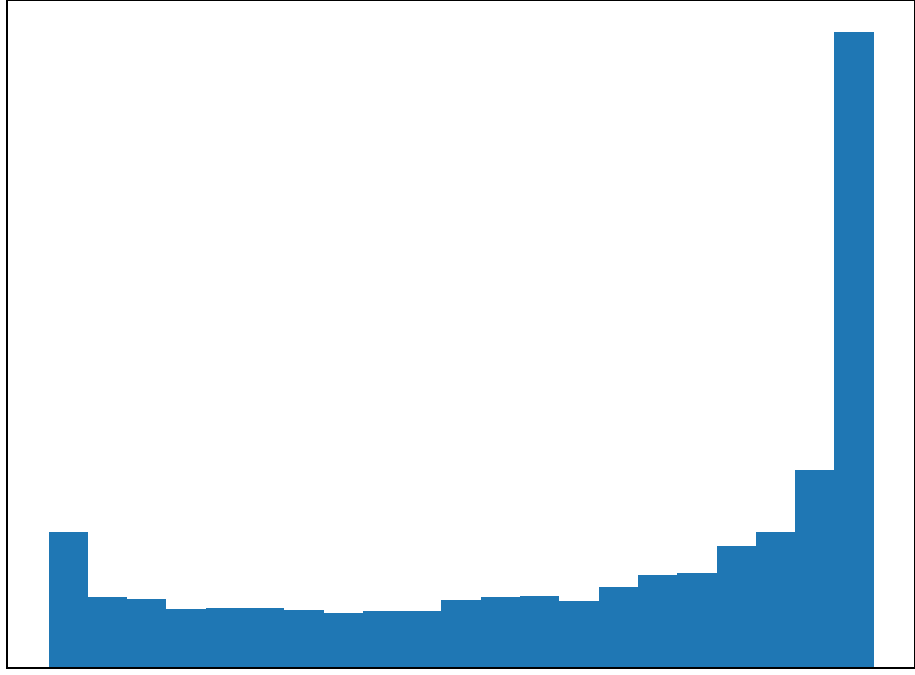}} &
    \makecell{\includegraphics[width=0.05\textwidth]{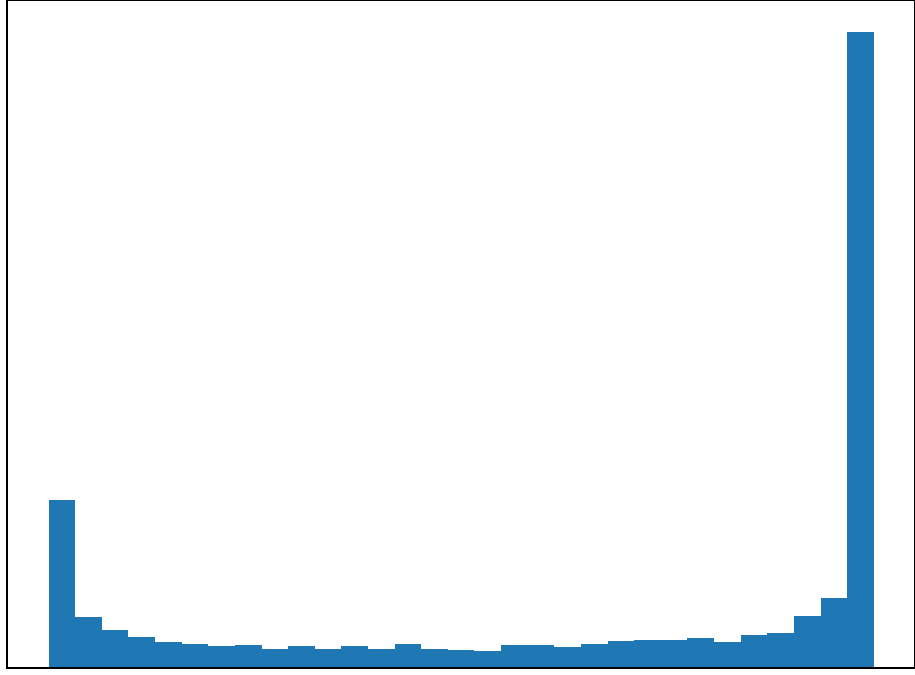}} &
    \makecell{\includegraphics[width=0.05\textwidth]{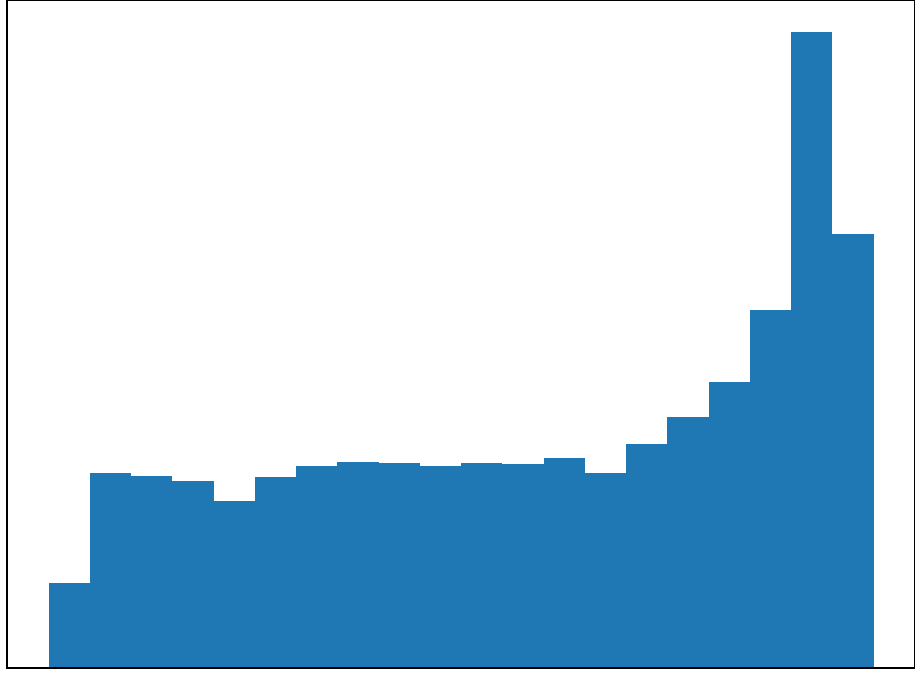}} &
    \makecell{\includegraphics[width=0.05\textwidth]{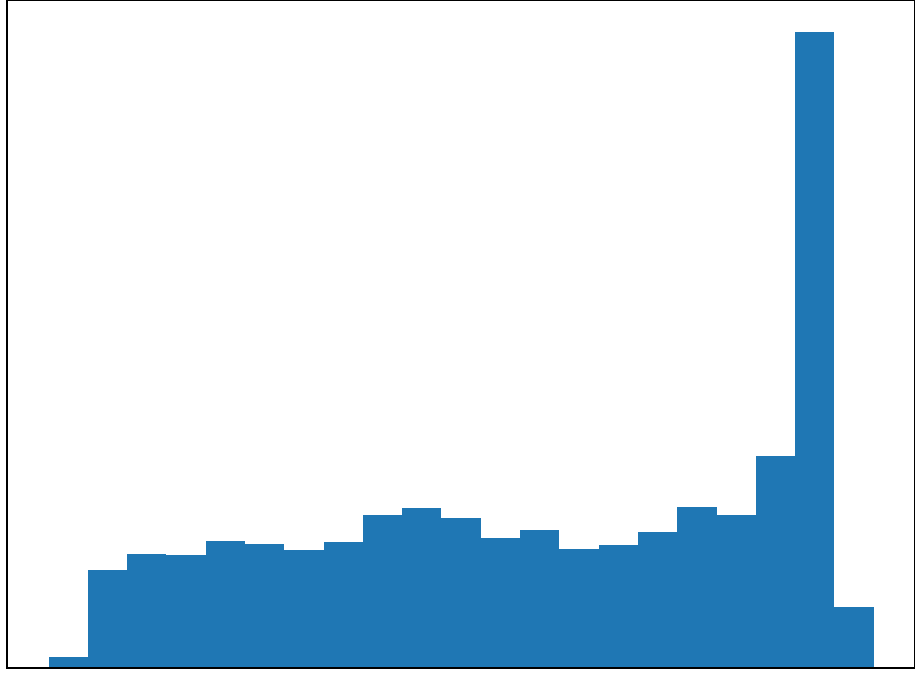}}\\
    & \makecell{Number of\\ objects} &
    \makecell{\includegraphics[width=0.05\textwidth]{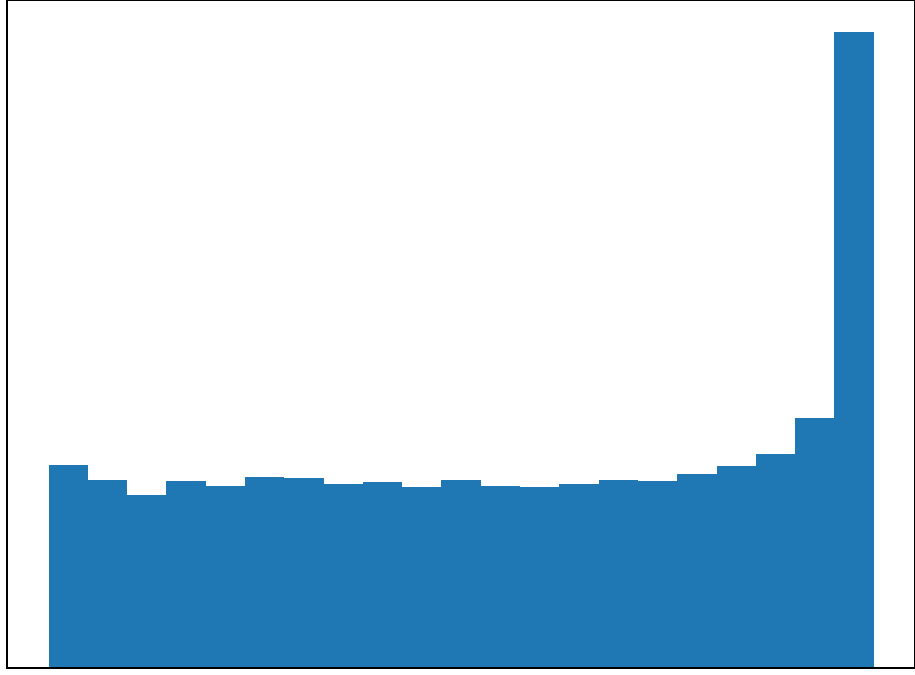}} &
    \makecell{\includegraphics[width=0.05\textwidth]{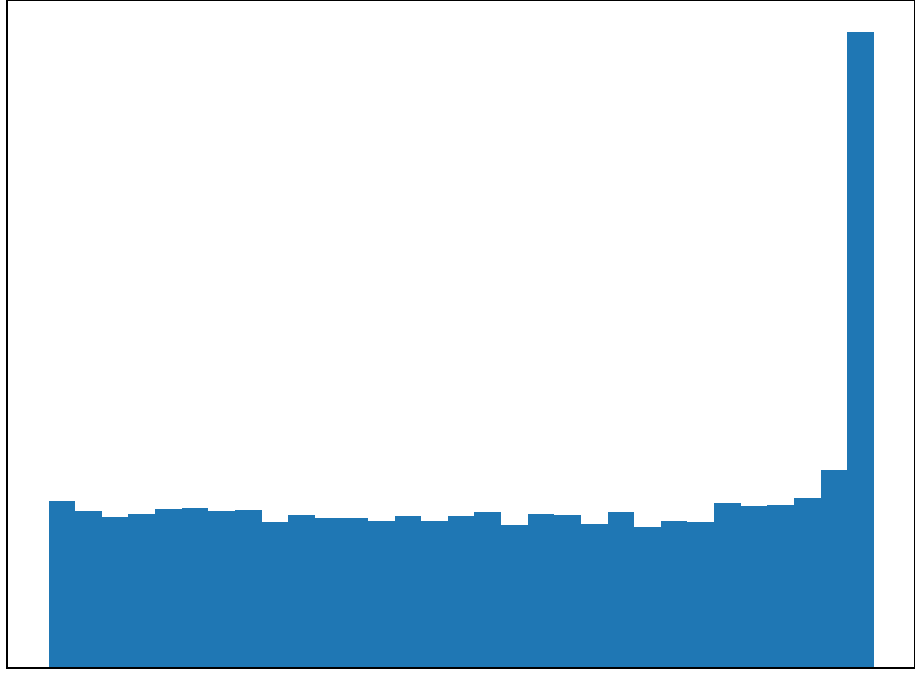}} &
    \makecell{\includegraphics[width=0.05\textwidth]{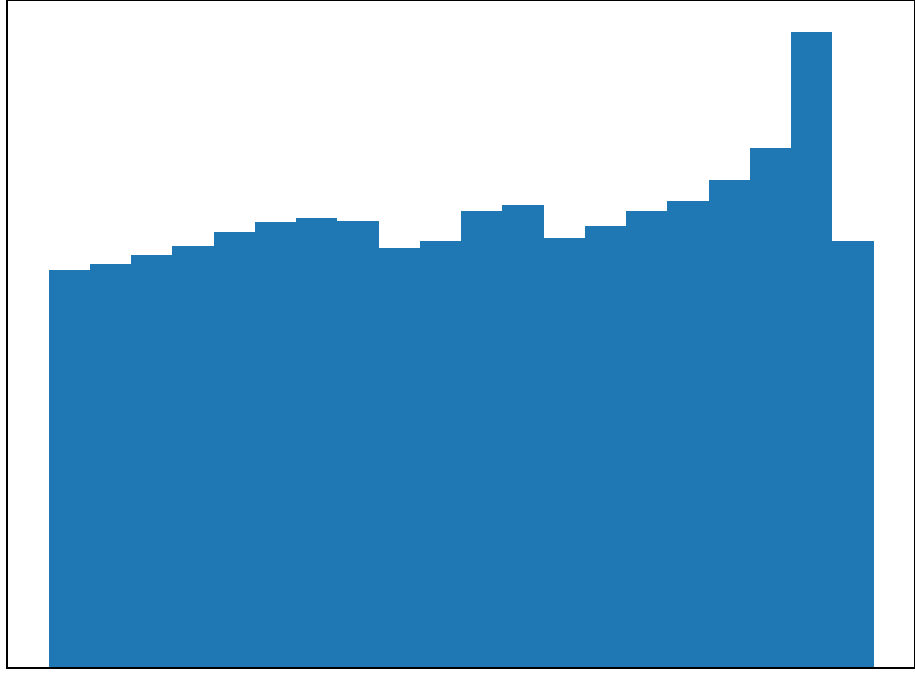}} &
    \makecell{\includegraphics[width=0.05\textwidth]{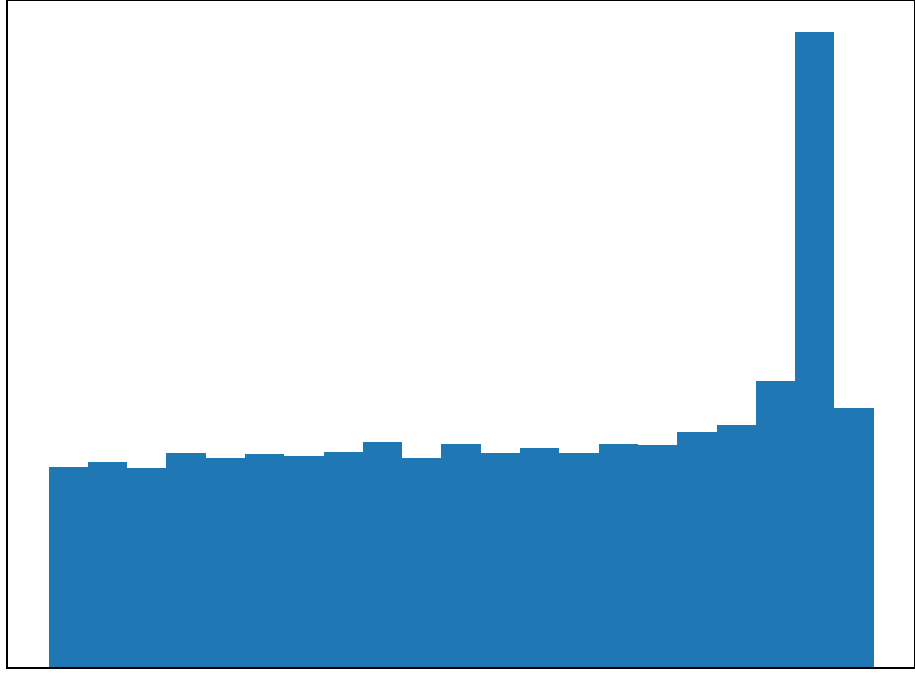}} &
    \makecell{\includegraphics[width=0.05\textwidth]{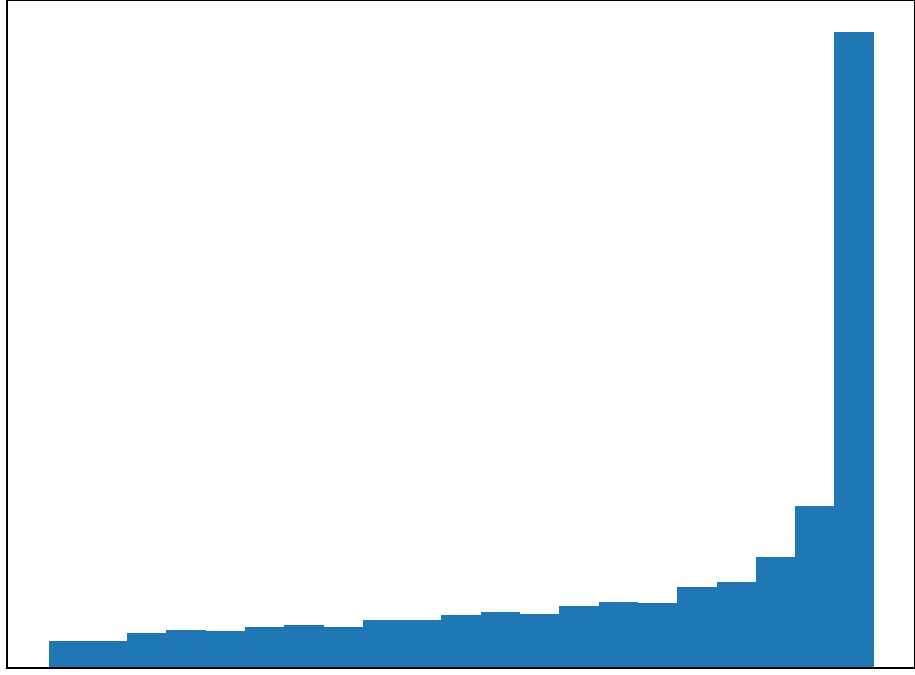}} &
    \makecell{\includegraphics[width=0.05\textwidth]{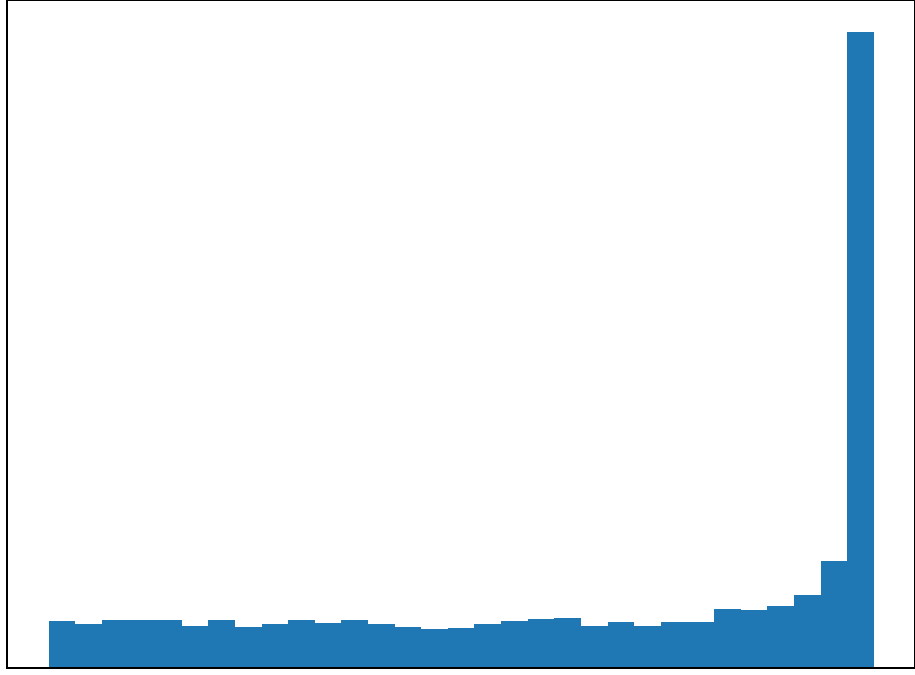}} &
    \makecell{\includegraphics[width=0.05\textwidth]{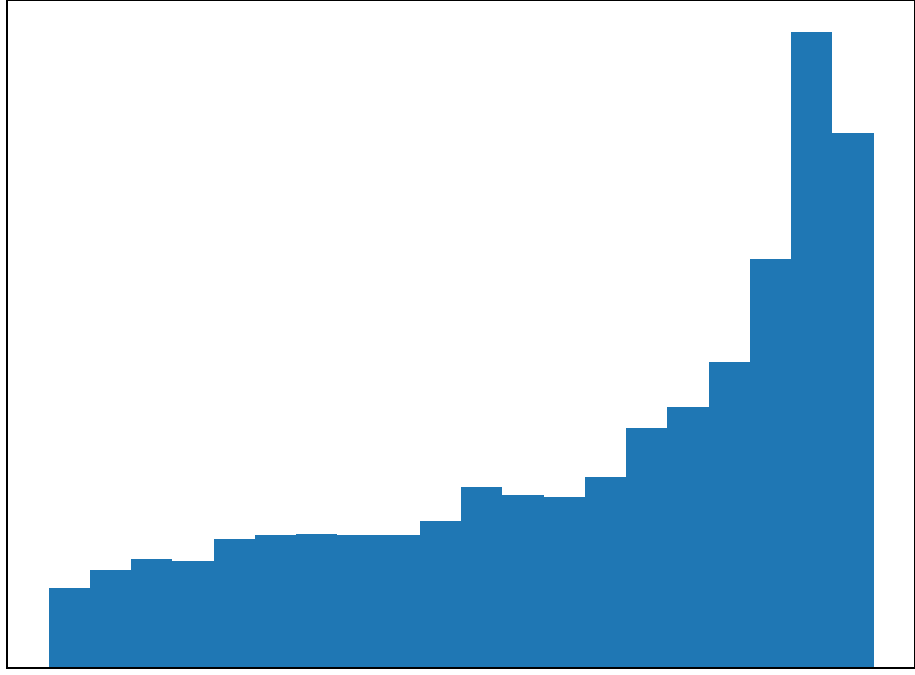}} &
    \makecell{\includegraphics[width=0.05\textwidth]{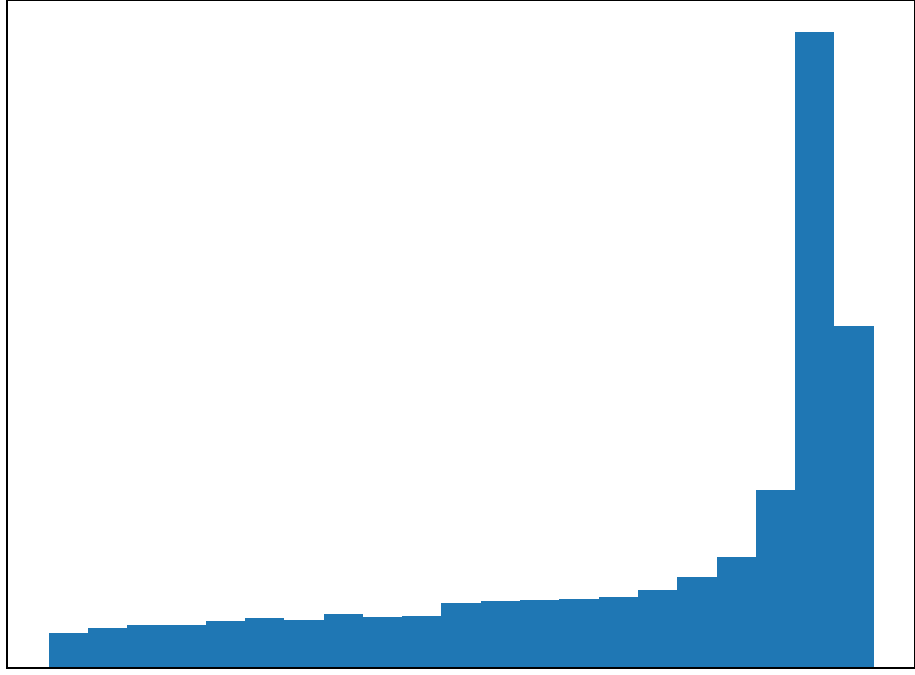}}\\
    \hline
    \multirow{3}{*}{\rotatebox[origin=c]{90}{SAL}} & \makecell{Scaled\\ volume} &
    \makecell{\includegraphics[width=0.05\textwidth]{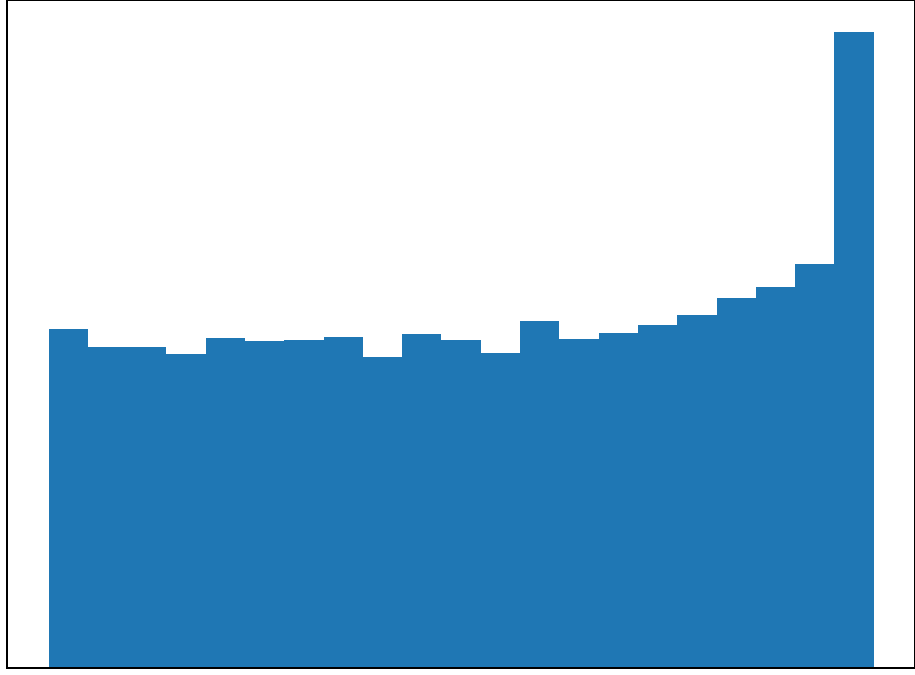}} &
    \makecell{\includegraphics[width=0.05\textwidth]{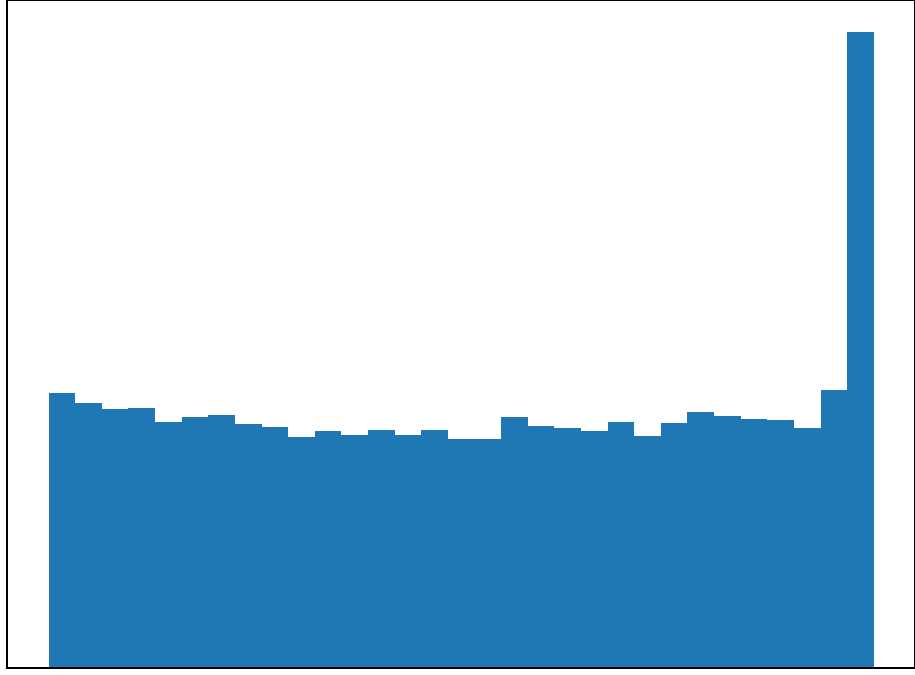}} &
    \makecell{\includegraphics[width=0.05\textwidth]{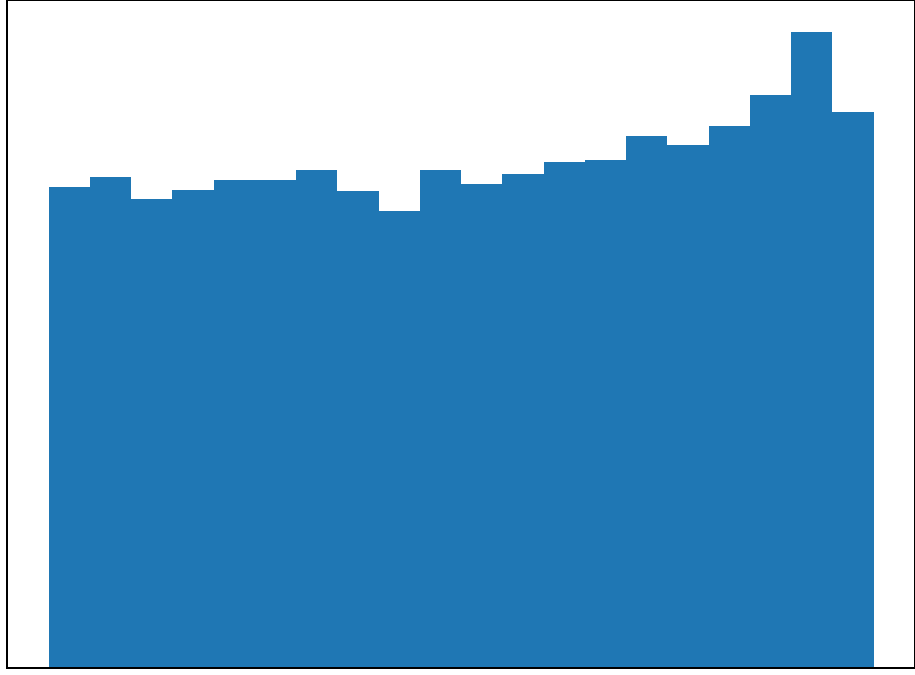}} &
    \makecell{\includegraphics[width=0.05\textwidth]{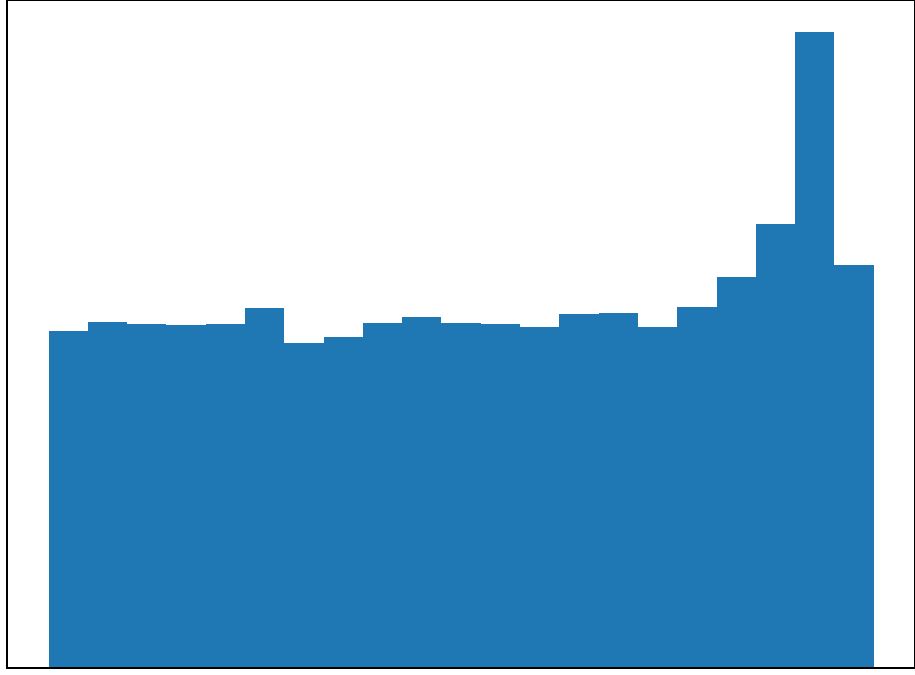}} &
    \makecell{\includegraphics[width=0.05\textwidth]{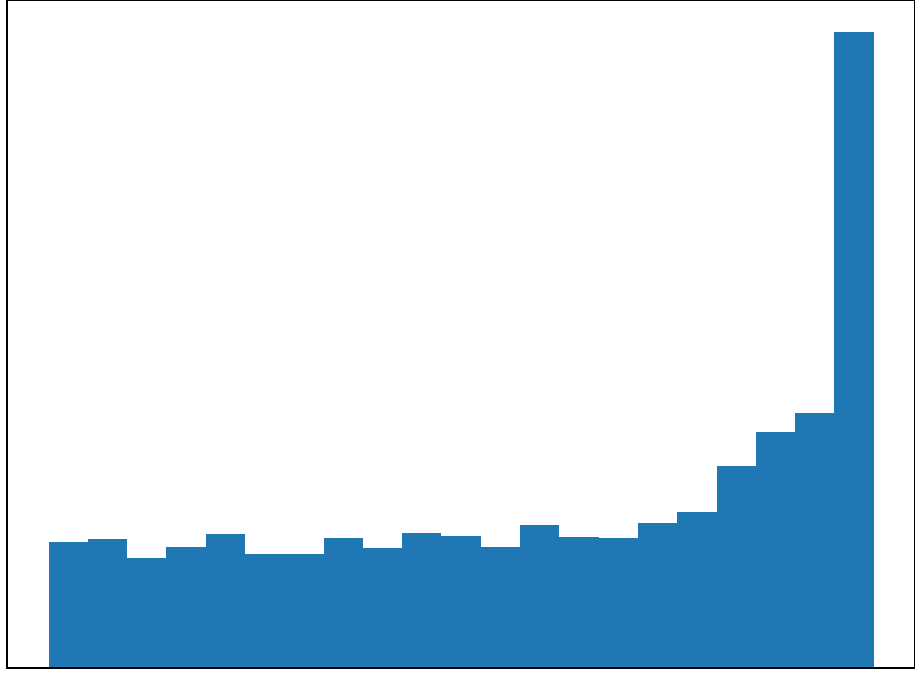}} &
    \makecell{\includegraphics[width=0.05\textwidth]{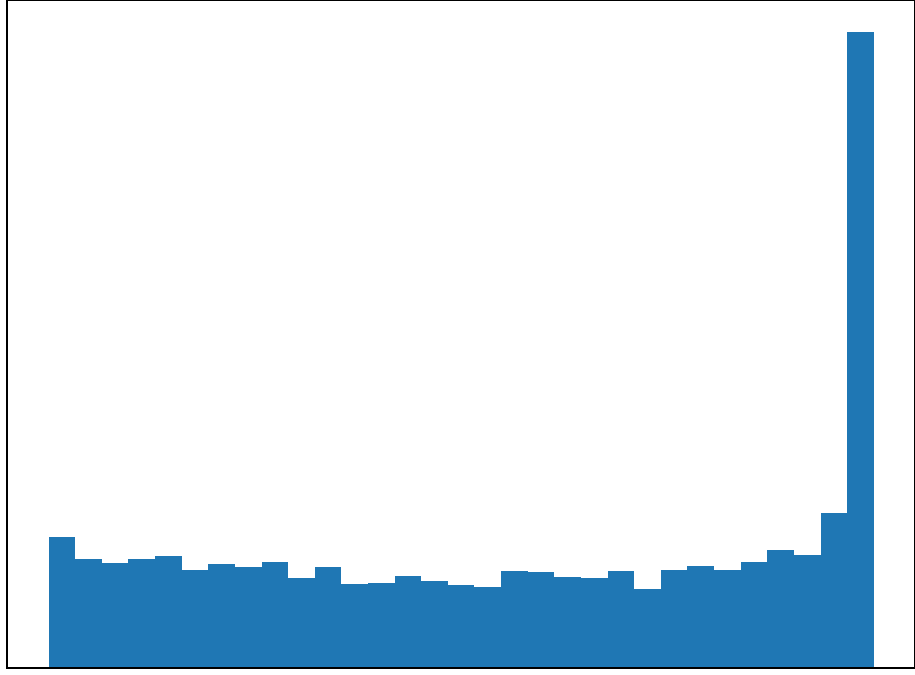}} &
    \makecell{\includegraphics[width=0.05\textwidth]{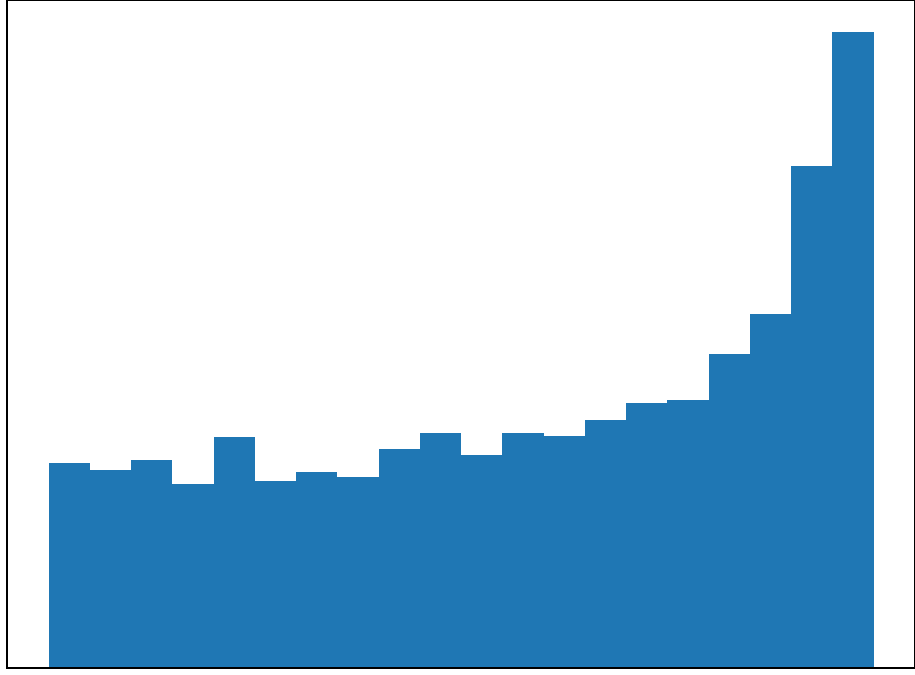}} &
    \makecell{\includegraphics[width=0.05\textwidth]{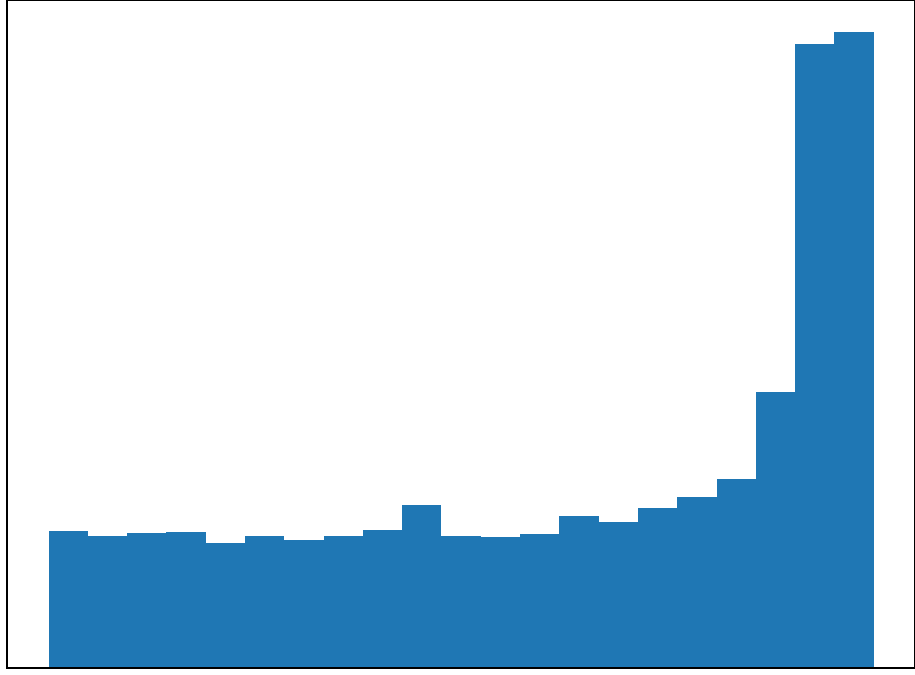}}\\
    & \makecell{Weighted\\ center\\ distance} &
    \makecell{\includegraphics[width=0.05\textwidth]{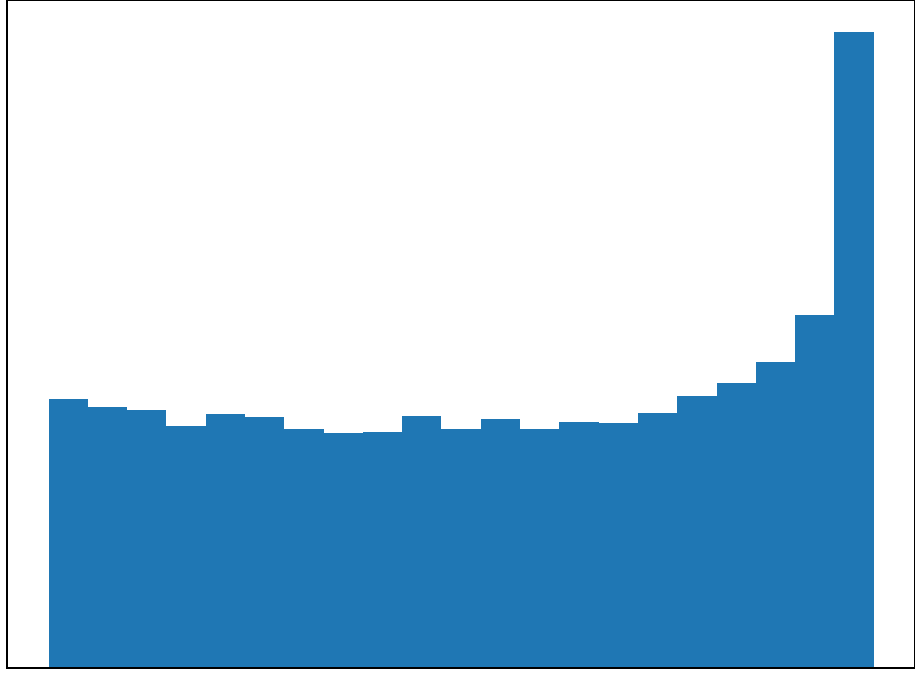}} &
    \makecell{\includegraphics[width=0.05\textwidth]{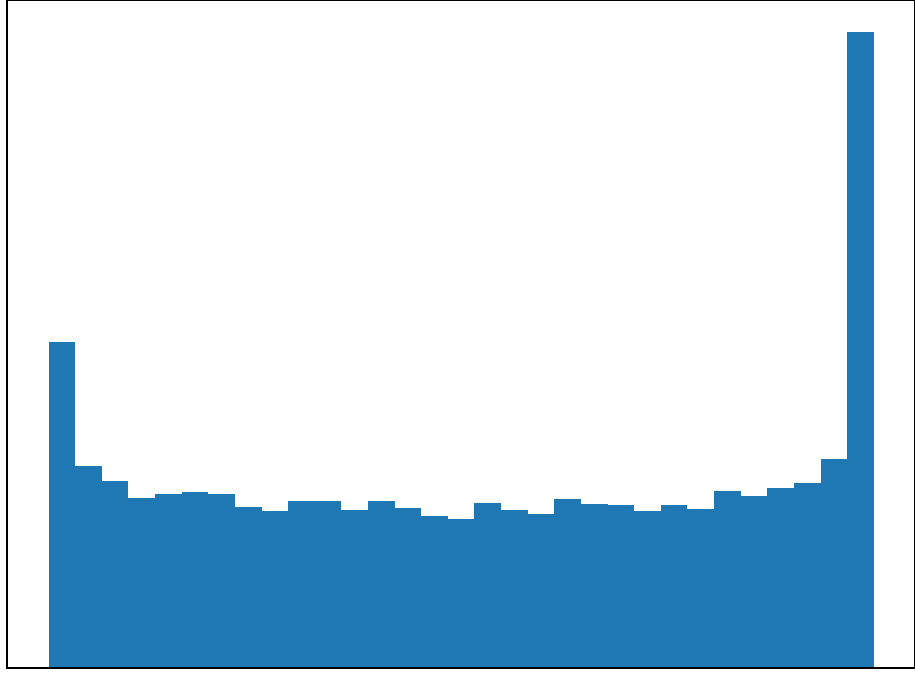}} &
    \makecell{\includegraphics[width=0.05\textwidth]{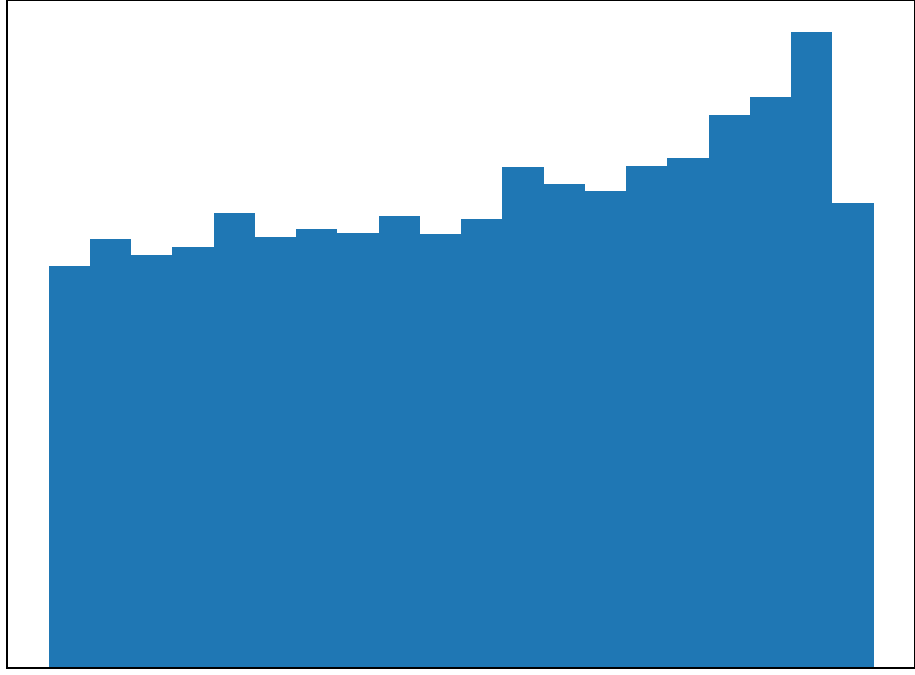}} &
    \makecell{\includegraphics[width=0.05\textwidth]{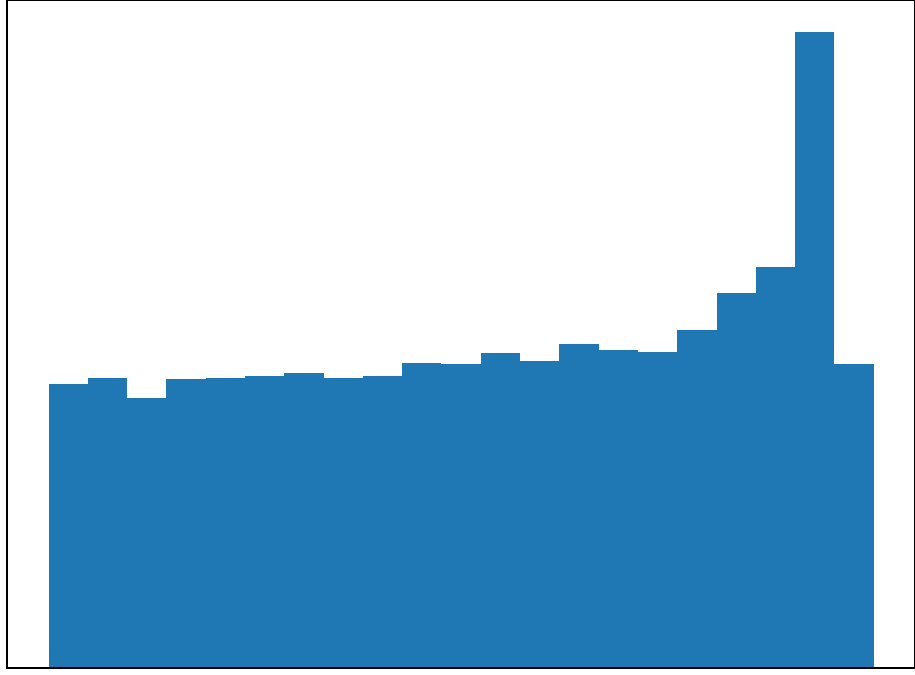}} &
    \makecell{\includegraphics[width=0.05\textwidth]{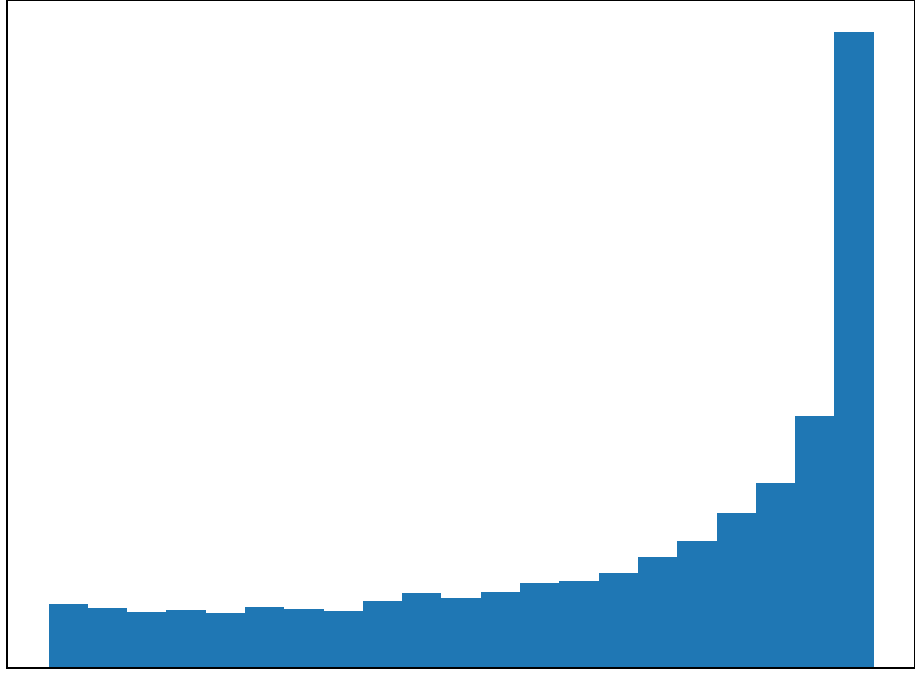}} &
    \makecell{\includegraphics[width=0.05\textwidth]{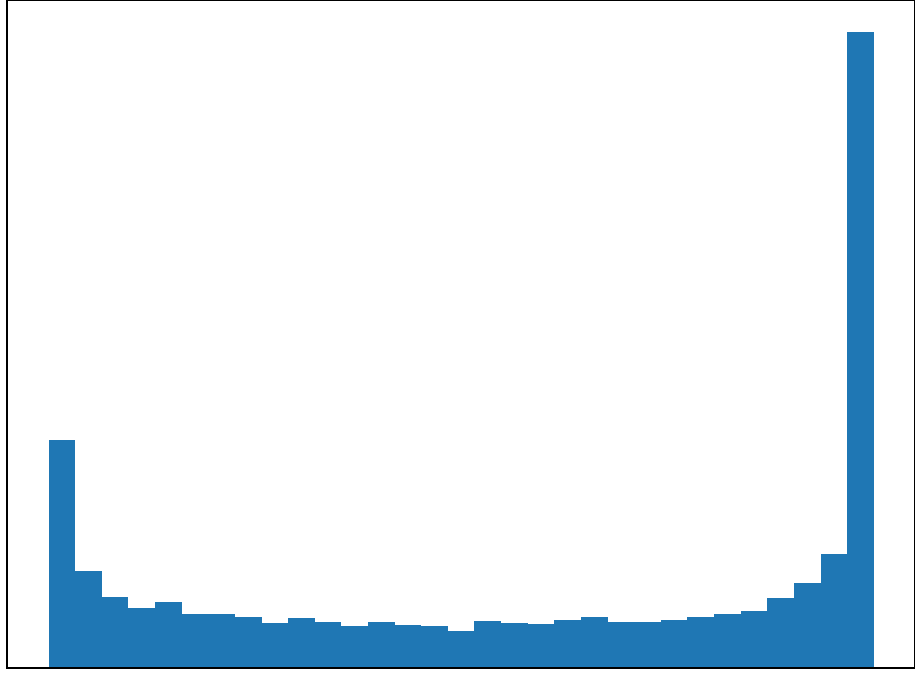}} &
    \makecell{\includegraphics[width=0.05\textwidth]{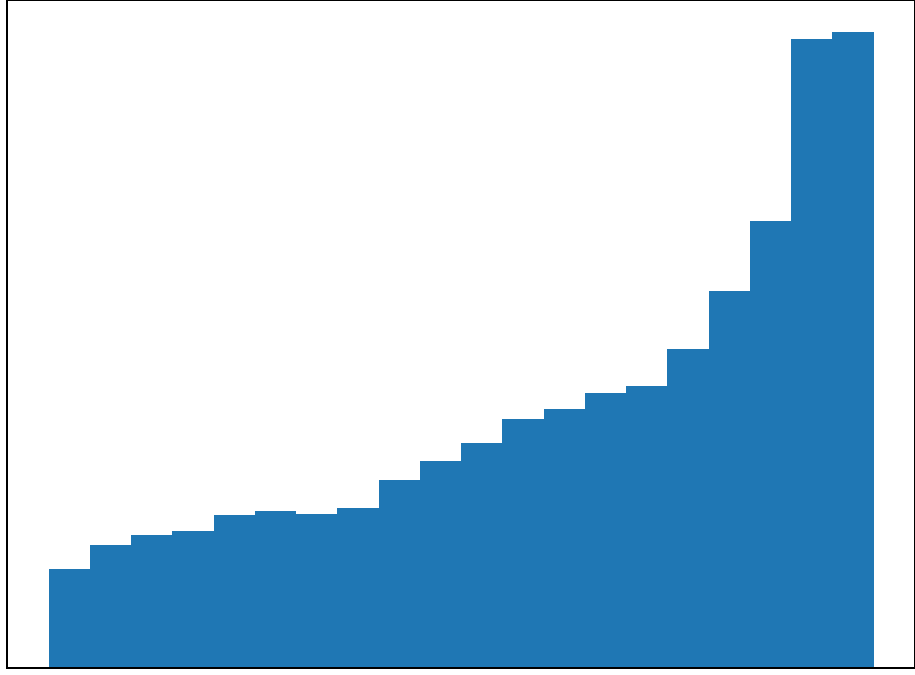}} &
    \makecell{\includegraphics[width=0.05\textwidth]{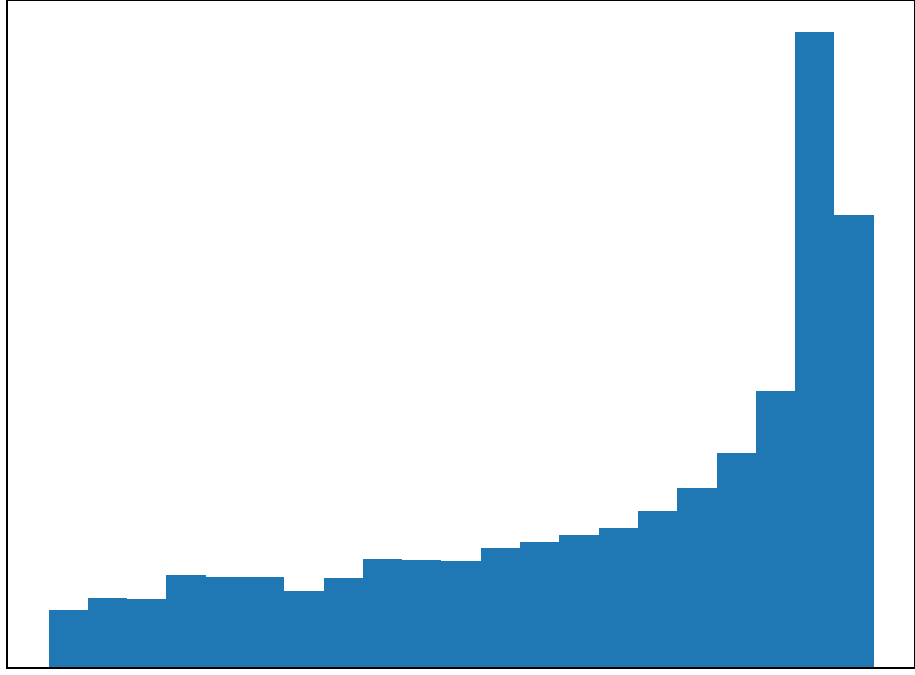}}\\ \hline
    \multirow{3}{*}{\rotatebox[origin=c]{90}{SCAI}} & \makecell{$D_0$} &
    \makecell{\includegraphics[width=0.05\textwidth]{pit_1h_icon__SCAI_D0.pdf}} &
    \makecell{\includegraphics[width=0.05\textwidth]{pit_1h_steps__SCAI_D0.pdf}} &
    \makecell{\includegraphics[width=0.05\textwidth]{pit_1h_copula__SCAI_D0.pdf}} &
    \makecell{\includegraphics[width=0.05\textwidth]{pit_1h_copula_sorted_SCAI_D0.pdf}} &
    \makecell{\includegraphics[width=0.05\textwidth]{pit_cond_1h_icon__SCAI_D0.pdf}} &
    \makecell{\includegraphics[width=0.05\textwidth]{pit_cond_1h_steps__SCAI_D0.pdf}} &
    \makecell{\includegraphics[width=0.05\textwidth]{pit_cond_1h_copula__SCAI_D0.pdf}} &
    \makecell{\includegraphics[width=0.05\textwidth]{pit_cond_1h_copula_sorted_SCAI_D0.pdf}}\\
    & \makecell{$D_1$} &
    \makecell{\includegraphics[width=0.05\textwidth]{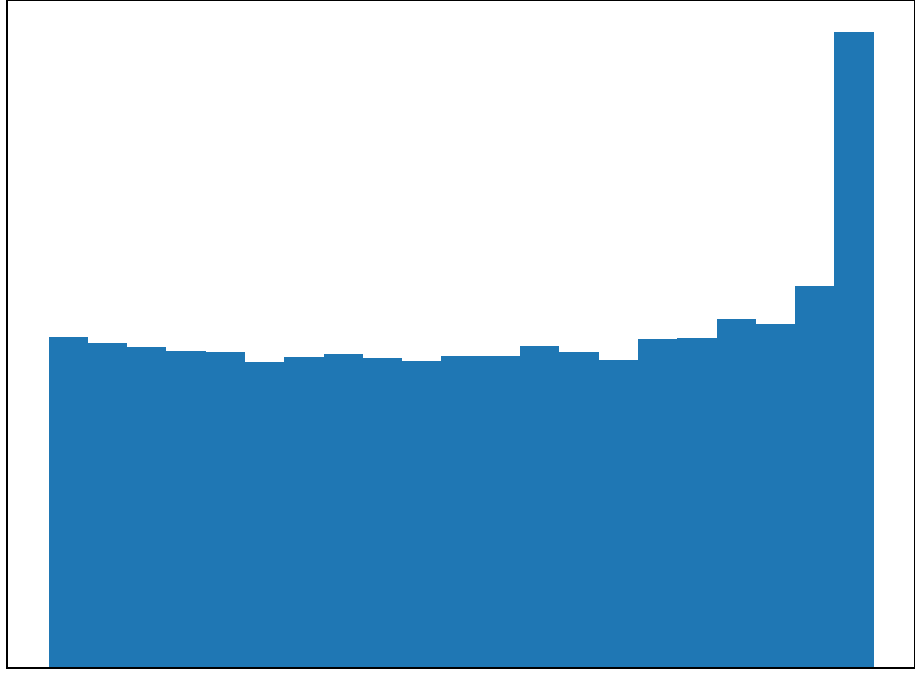}} &
    \makecell{\includegraphics[width=0.05\textwidth]{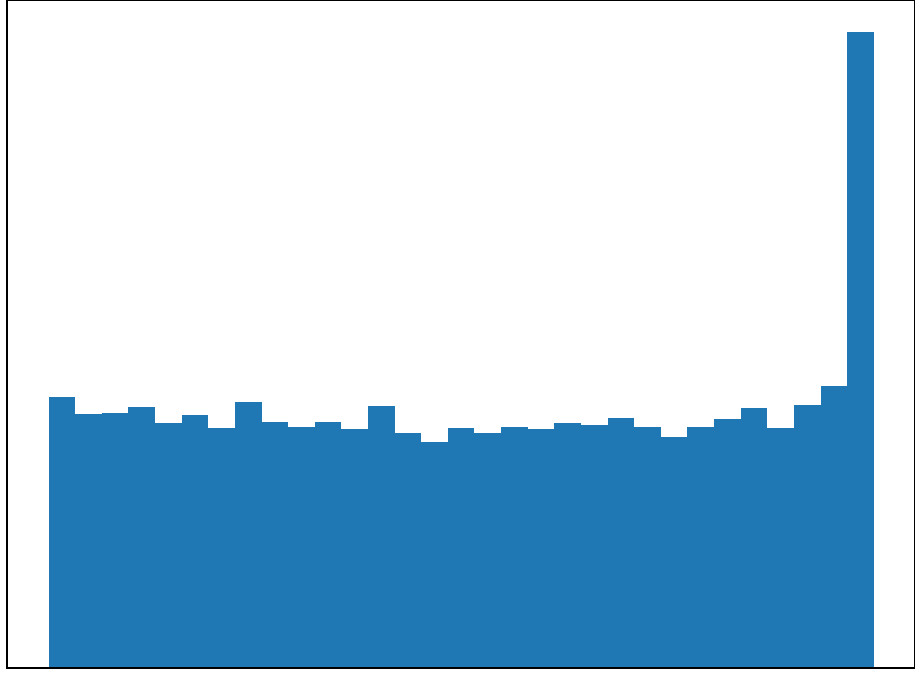}} &
    \makecell{\includegraphics[width=0.05\textwidth]{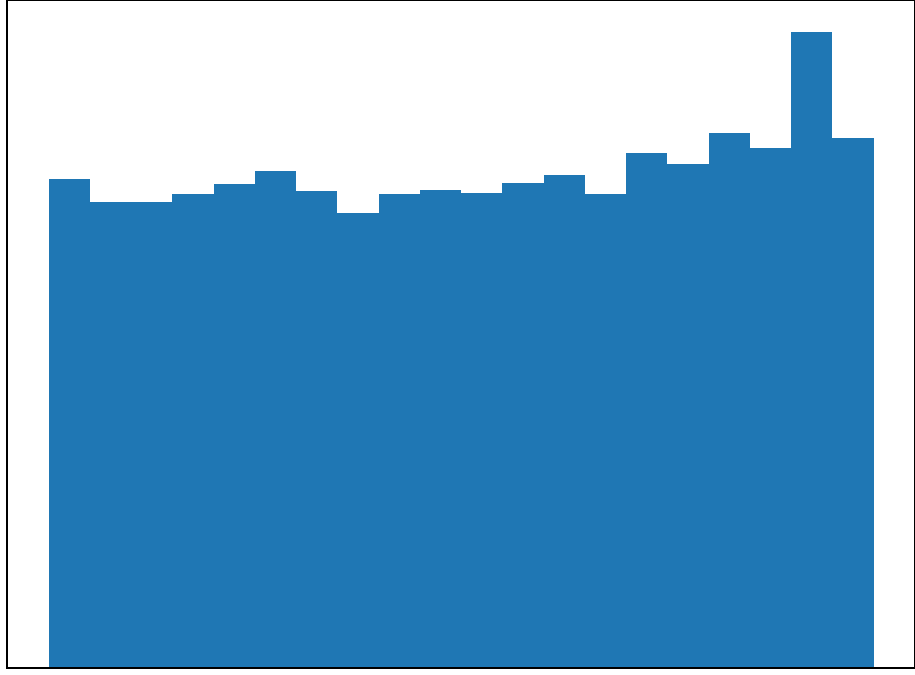}} &
    \makecell{\includegraphics[width=0.05\textwidth]{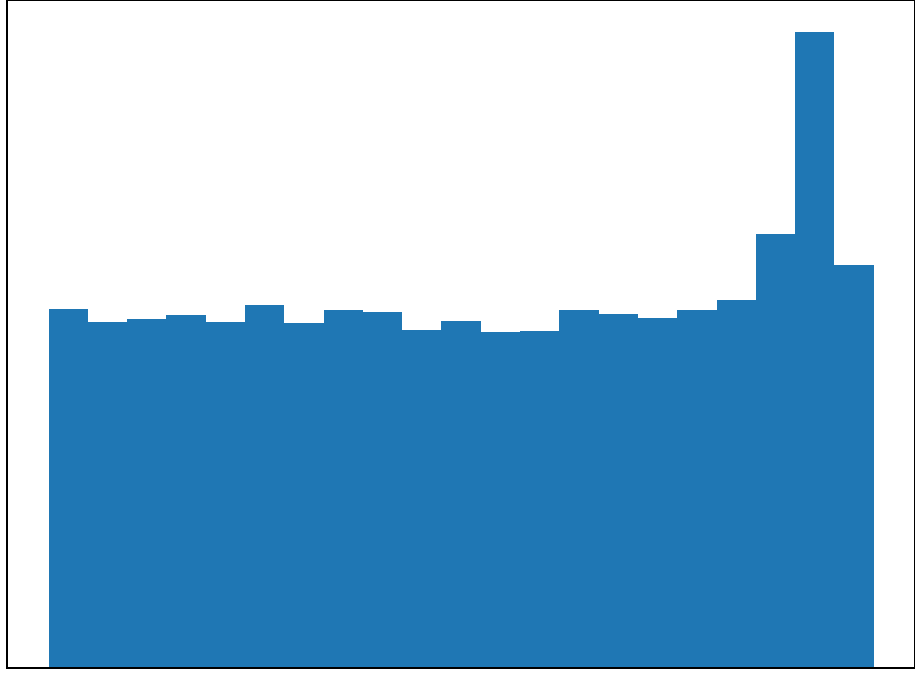}} &
    \makecell{\includegraphics[width=0.05\textwidth]{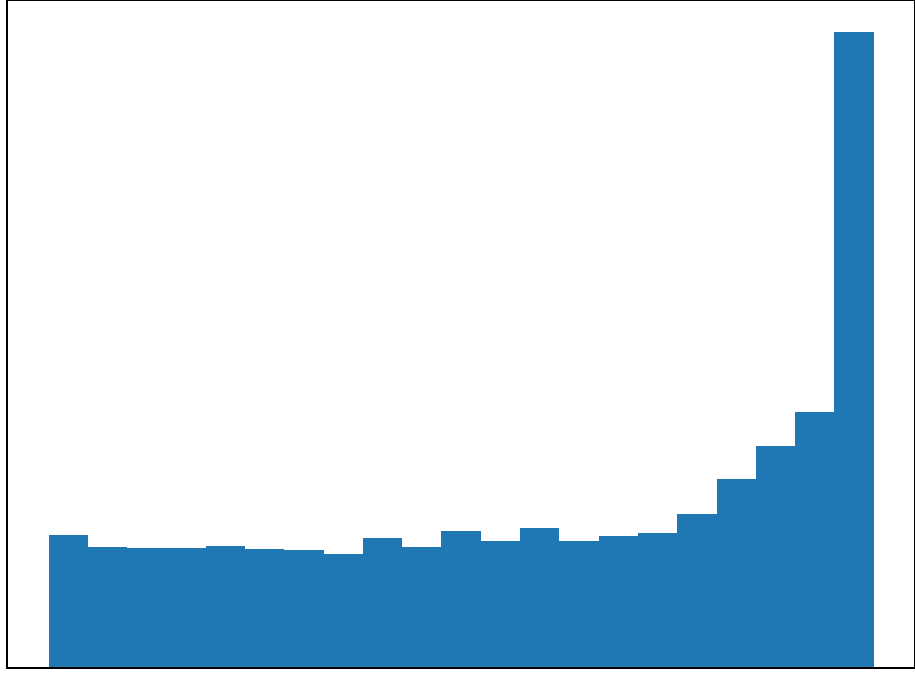}} &
    \makecell{\includegraphics[width=0.05\textwidth]{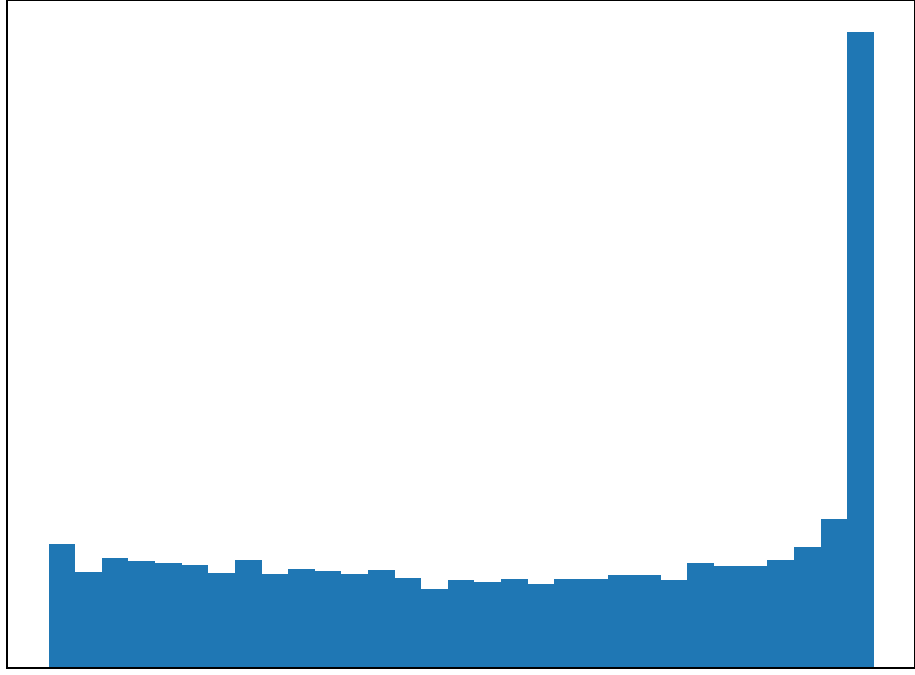}} &
    \makecell{\includegraphics[width=0.05\textwidth]{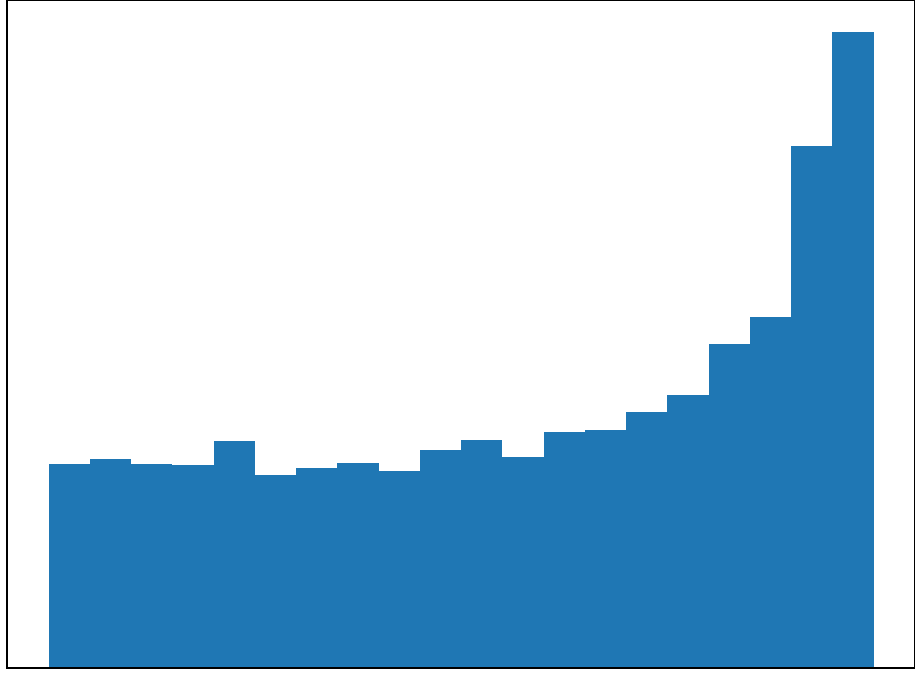}} &
    \makecell{\includegraphics[width=0.05\textwidth]{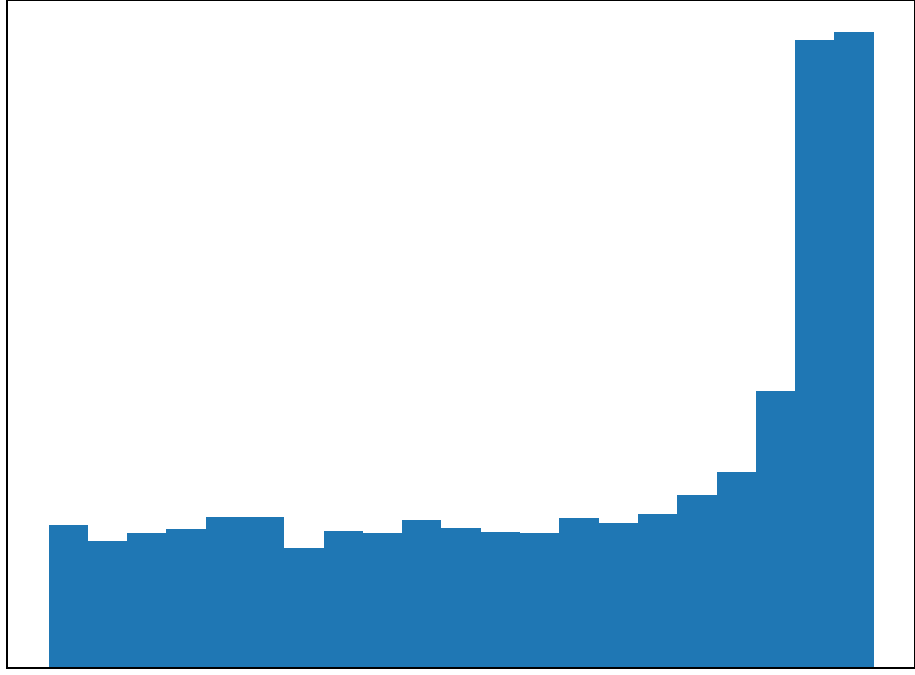}}
    \end{tabular}
    \caption{PIT diagrams as in Figure~\ref{fig.pit_precipitation} but for various object-based metrics. Top-down are the area and the number of objects, scaled volume and weighted centre distance of SAL, D$_0$ and D$_1$ of SCAI. Results are shown for STEPS, ICON, COPULA and COPULA sorted for $+1$\,h lead time. Results for other lead times are omitted since they are essentially identical. The four columns on the left-hand side show the results for the whole dataset, while the four columns on the right-hand side are restricted to instances where precipitation was observed within the considered $9 \times 9$ sub-region $V\subset W$.}
    \label{fig.pit_obj}
\end{figure}

To highlight the benefits of the restoration of spatial dependencies and to compare the resulting precipitation structures in each ensemble member with that of the radar observation, we consider PIT diagrams of various object-based metrics in Figure~\ref{fig.pit_obj}. As an object, we define a contiguous area in the sense of a 4-connectivity where grid points exceed an hourly precipitation sum of $0.1\,$mm. The metrics consist of the total area of all objects and their number. Further, we consider two metrics of the SAL-index (Structure, Amplitude, Location; \citep{wernli2008sal}). First, the scaled volume provides information about the average shape of identified objects. It is given by the precipitation mass of an object normalized by its maximum and additionally weighted with this mass. Here, the precipitation mass of an object is defined as the sum of the rainfall amount at each associated grid point. Second, the weighted centre distance describes the average distance between single objects and the total centre of mass and provides information about the aggregation of precipitation. The averaging is also based on the precipitation mass to favour larger objects. In addition, we assess $D_0$ and $D_1$, which are components of SCAI (Simple Convective Aggregation Index; \citep{tobin2012observational}). The metrics $D_0$ and $D_1$ represent the geometric and arithmetic mean of the distances for all possible pairs of objects.

The evaluation results for the four models, ICON, STEPS, COPULA and COPULA sorted, are divided into two groups. First, we consider results based on the whole dataset, depicted in the four left columns of Figure~\ref{fig.pit_obj}. Second, instances without any observed precipitation are removed for the four right columns. This implies that at least at one grid point within the $9 \times 9$ sub-region $V\subset W$, the observed hourly rainfall sum must be $\geq 0.1$\,mm. ICON and STEPS reveal a tendency to underestimate the respective values in all metrics since there is a peak of observations in the last bin. This peak is even more distinct for the precipitation subset, indicating many cases in which no precipitation is forecast or observed. The peak may be induced by cases where no precipitation is forecast at all. One should be aware that we statistically evaluated sub-regions with an edge length of $\approx$ 20\,km so that errors in location (e.g. spatial shifts in forecast precipitation) strongly influence the depicted results.
In some cases, ICON and STEPS overestimate the area, whereby STEPS further overestimates the weighted centre distance. This may be attributable to situations where only one smaller object is identified in the observation. Since precipitation fields in ICON are smoother, one can assume that, in general, the number of objects is smaller. Therefore, if only one object is detected, the centre of masses is ``equal to itself'', and the weighted centre distance is zero.

The forecasts of  COPULA sorted also show an underestimation of the metrics. However, at least a few ensemble members are above the observed values since the frequency in bins of the upper percentiles is also increased. Furthermore, at least the maximum value of the COPULA sorted ensemble forecast seems to reproduce the observed values since the maximum peak is in the second last bin. Compared to the raw drawn values, the forecasts of COPULA reveal a nearly uniform distribution for the whole dataset. For the precipitation subset, the distribution of observed values is still skewed. However, the frequency begins to increase at lower percentiles. This indicates that the forecasts of Copula exhibit a more realistic representation of spatial structures in terms of such a set of object-based metrics.

\begin{figure}[H]
	\center
	\scriptsize
    \begin{tabular}{c|c||cccc||cccc}
    & & \makecell{ICON} & \makecell{STEPS} & \makecell{COPULA} & \makecell{Sorted} & \makecell{ICON\\ $\geq 0.1$\,mm} & \makecell{STEPS\\ $\geq 0.1$\,mm} & \makecell{COPULA\\ $\geq 0.1$\,mm} & \makecell{Sorted\\ $\geq 0.1$\,mm}\\ \hline\hline
    \multirow{5}{*}{\rotatebox[origin=c]{90}{$+1$\,h}} & \makecell{Nugget} &
    \makecell{\includegraphics[width=0.05\textwidth]{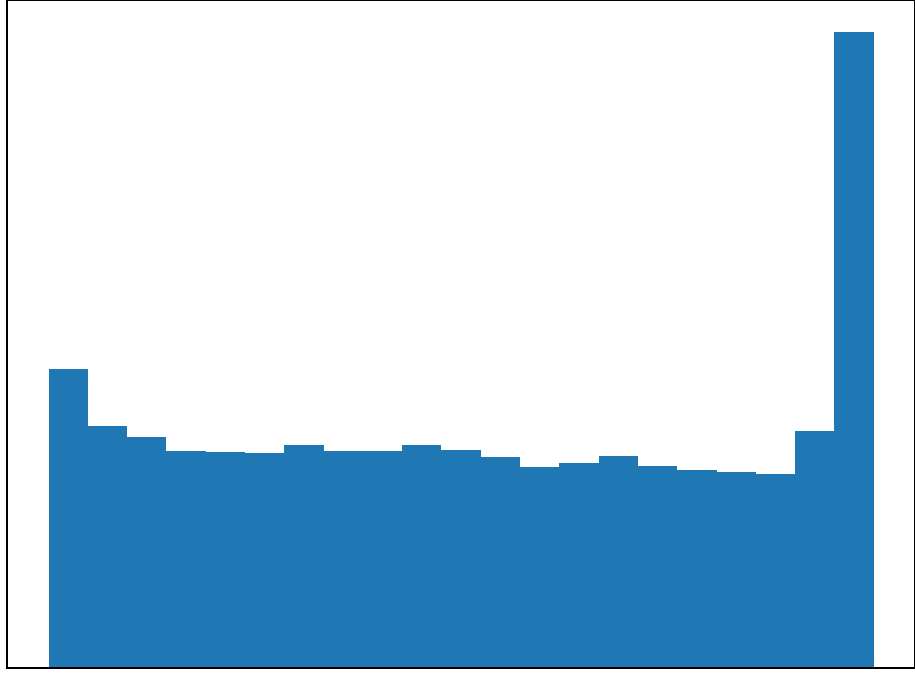}} &
    \makecell{\includegraphics[width=0.05\textwidth]{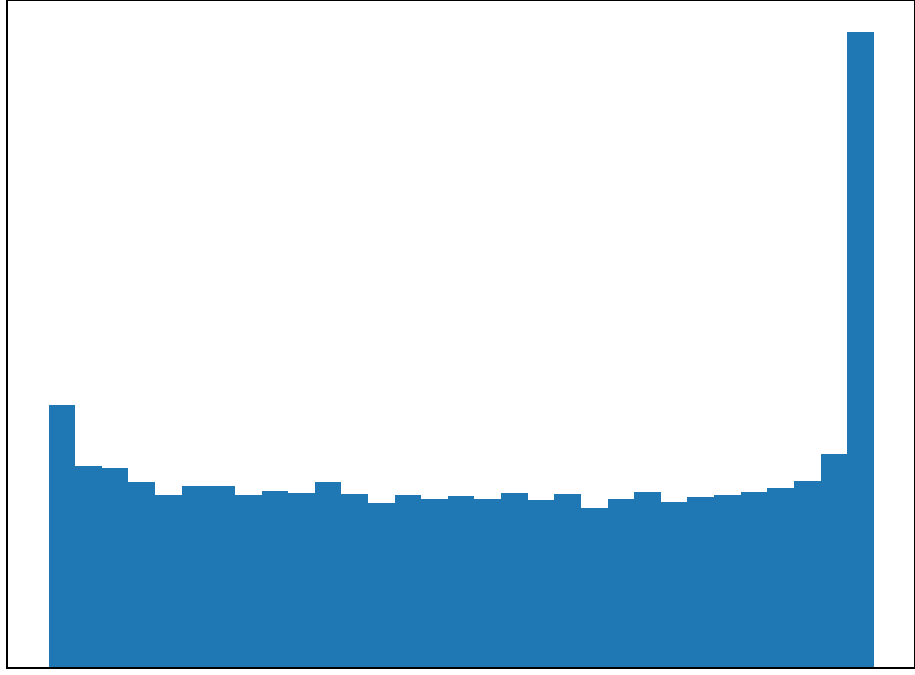}} &
    \makecell{\includegraphics[width=0.05\textwidth]{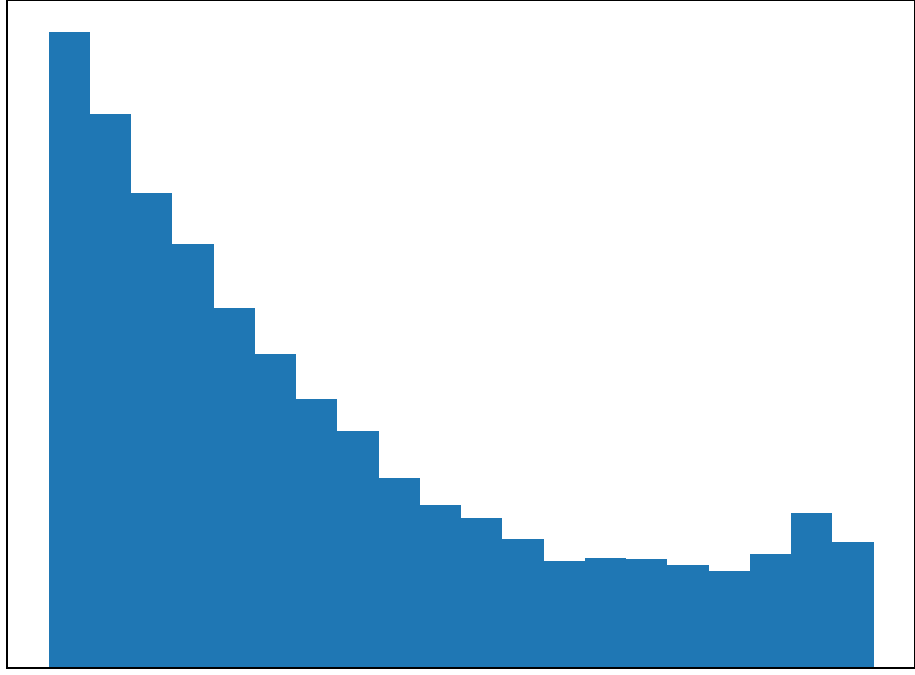}} &
    \makecell{\includegraphics[width=0.05\textwidth]{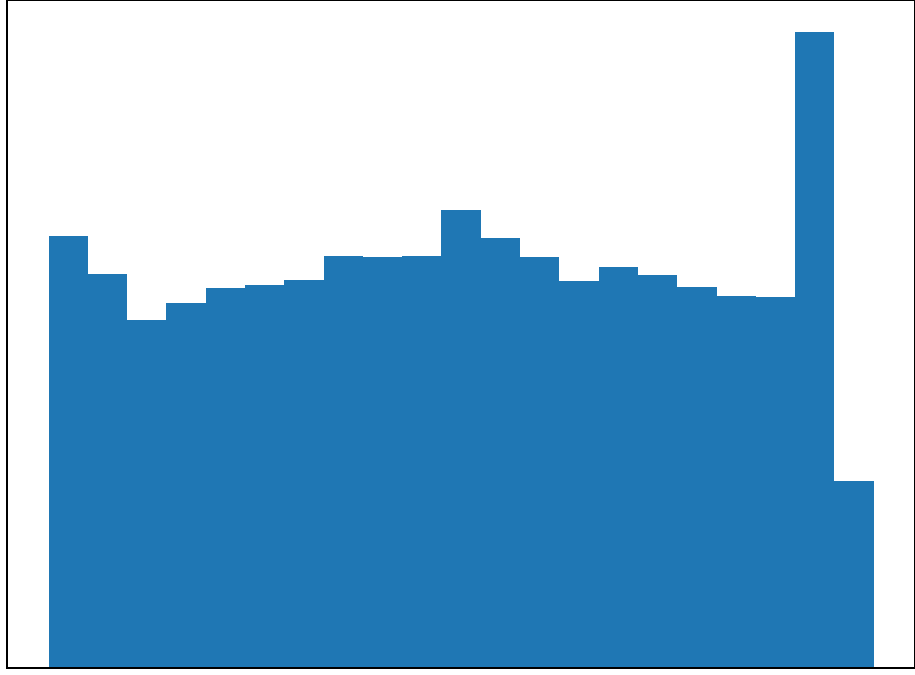}} &
    \makecell{\includegraphics[width=0.05\textwidth]{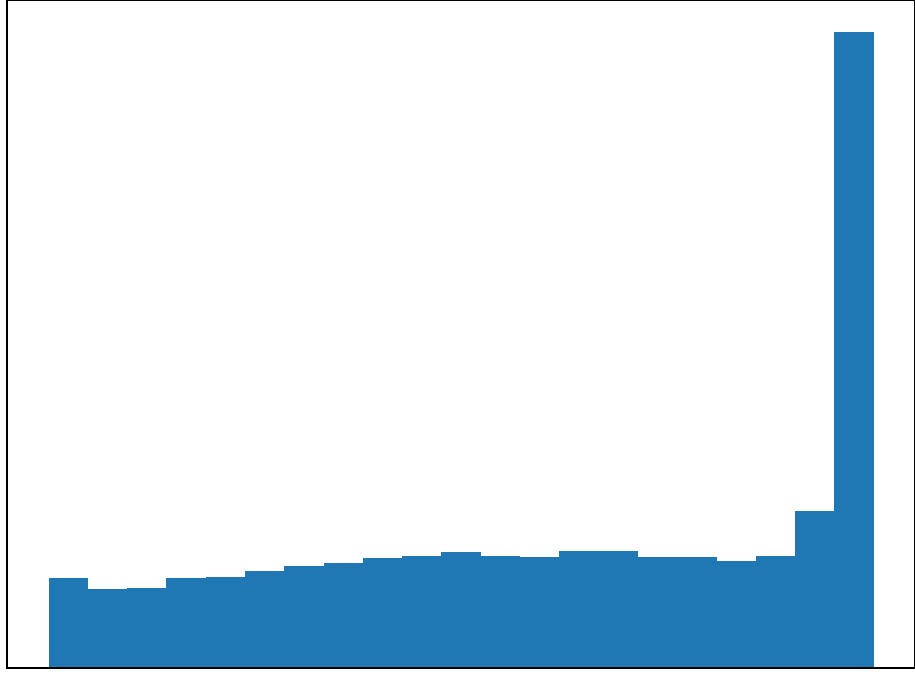}} &
    \makecell{\includegraphics[width=0.05\textwidth]{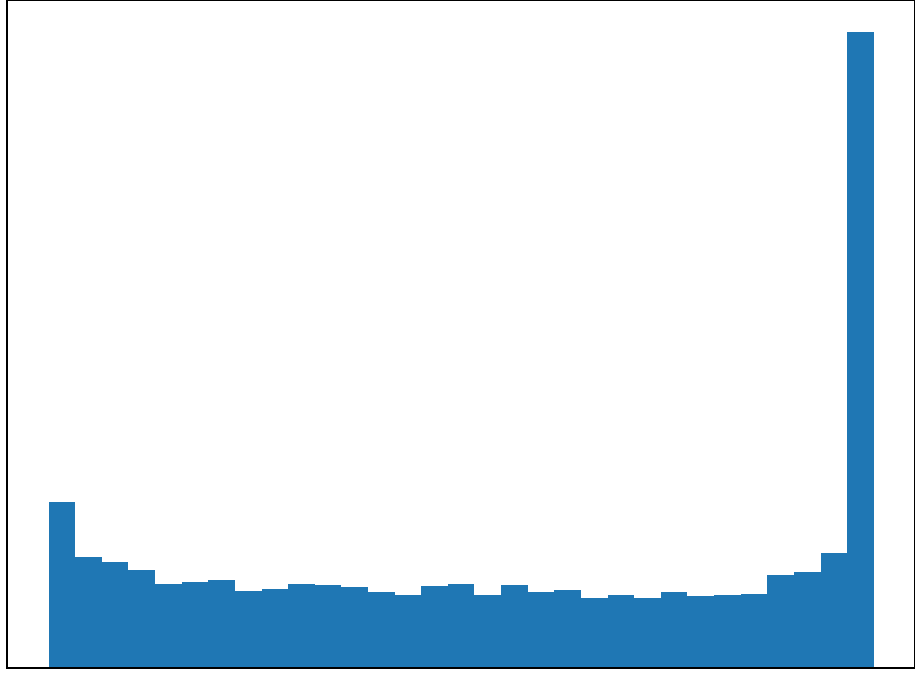}} &
    \makecell{\includegraphics[width=0.05\textwidth]{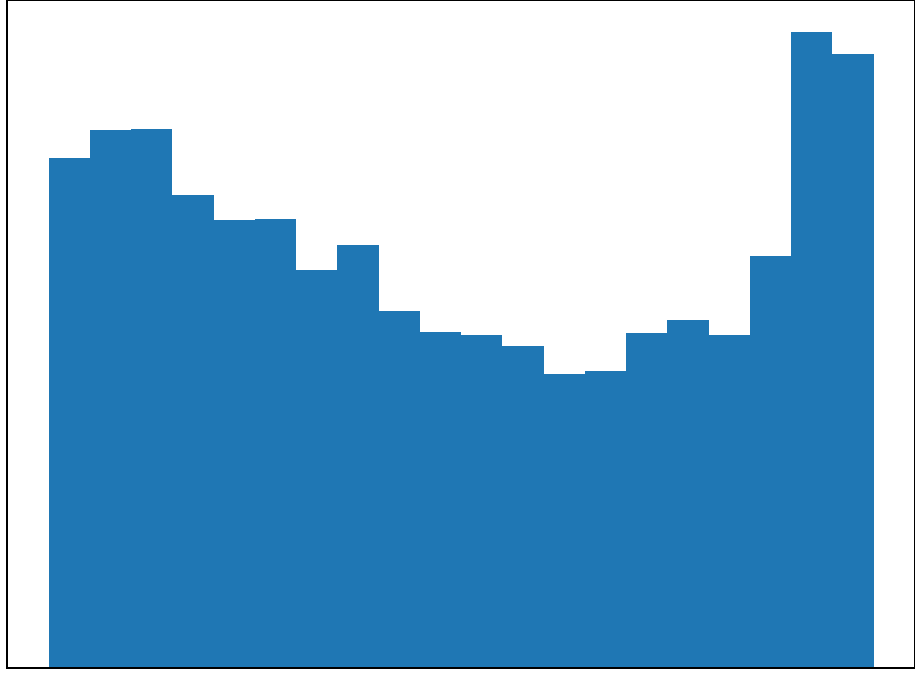}} &
    \makecell{\includegraphics[width=0.05\textwidth]{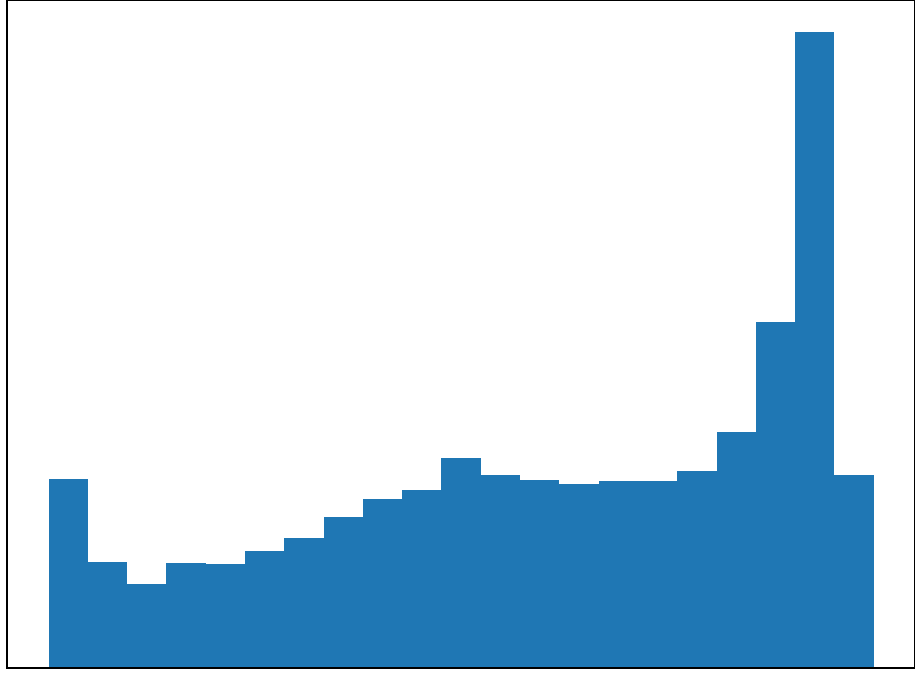}}\\
    & \makecell{Effective range} &
    \makecell{\includegraphics[width=0.05\textwidth]{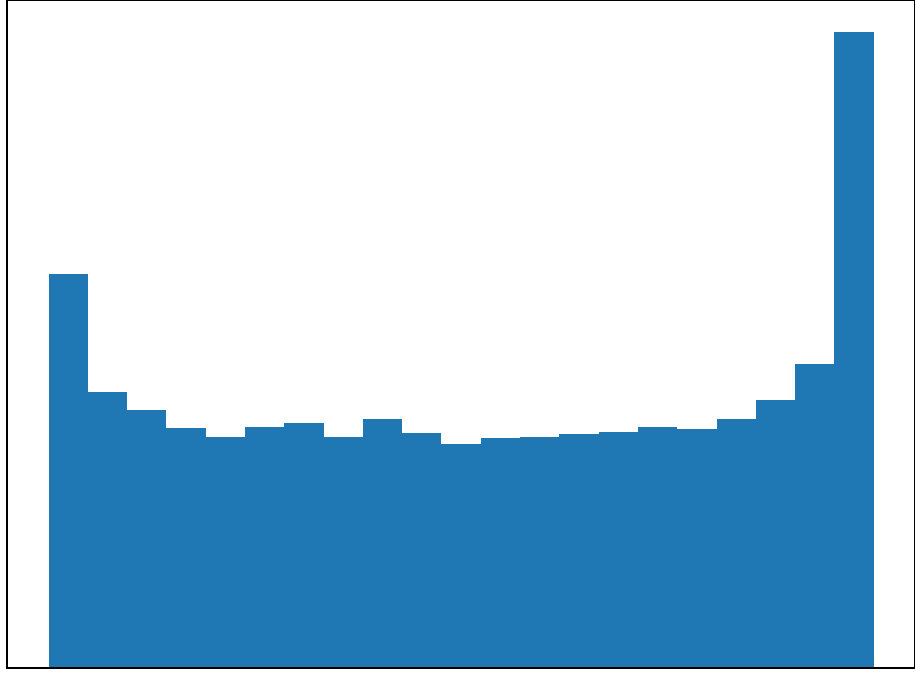}} &
    \makecell{\includegraphics[width=0.05\textwidth]{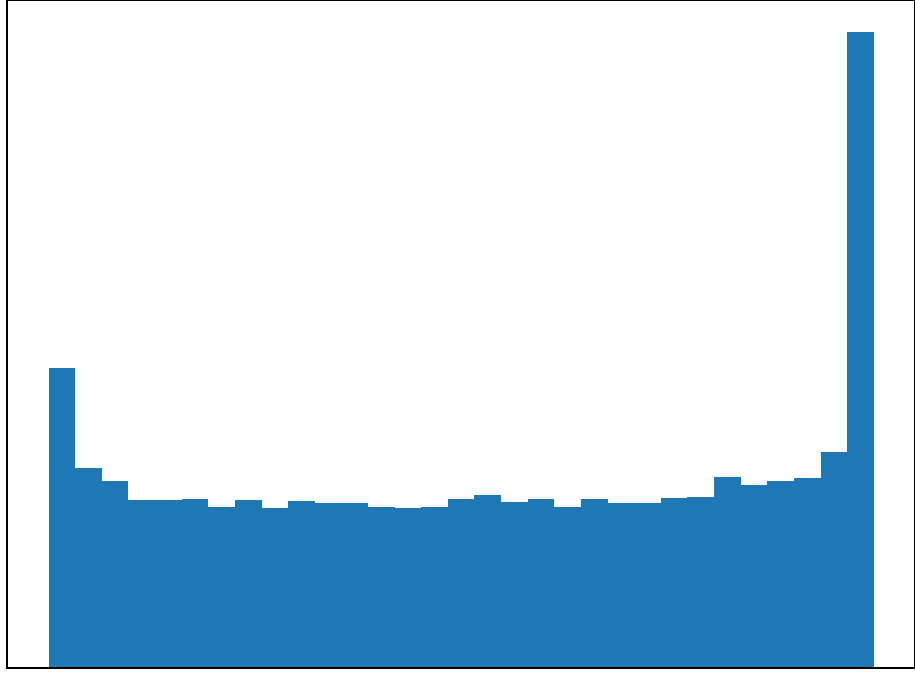}} &
    \makecell{\includegraphics[width=0.05\textwidth]{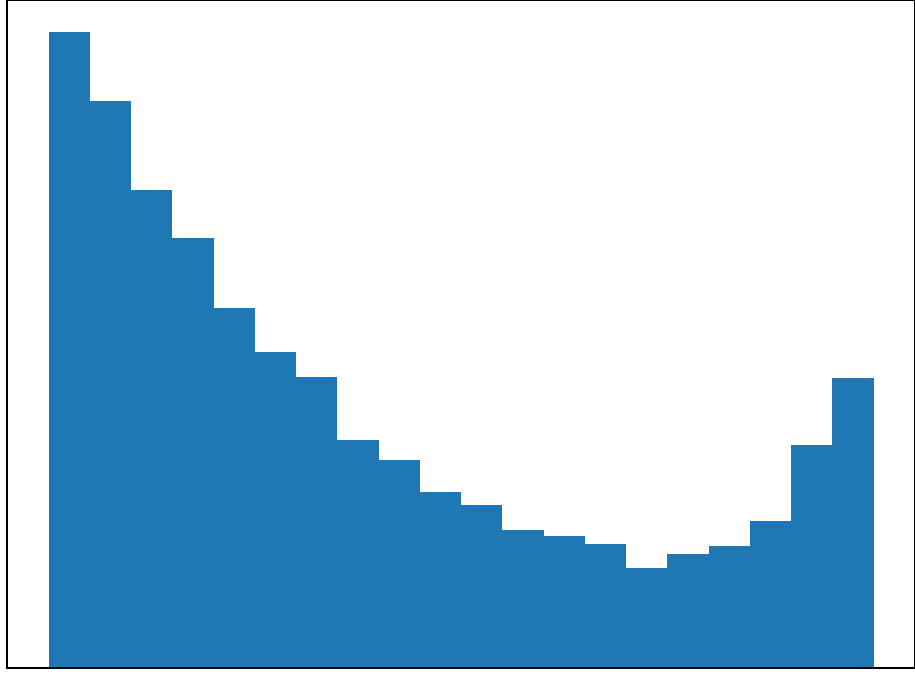}} &
    \makecell{\includegraphics[width=0.05\textwidth]{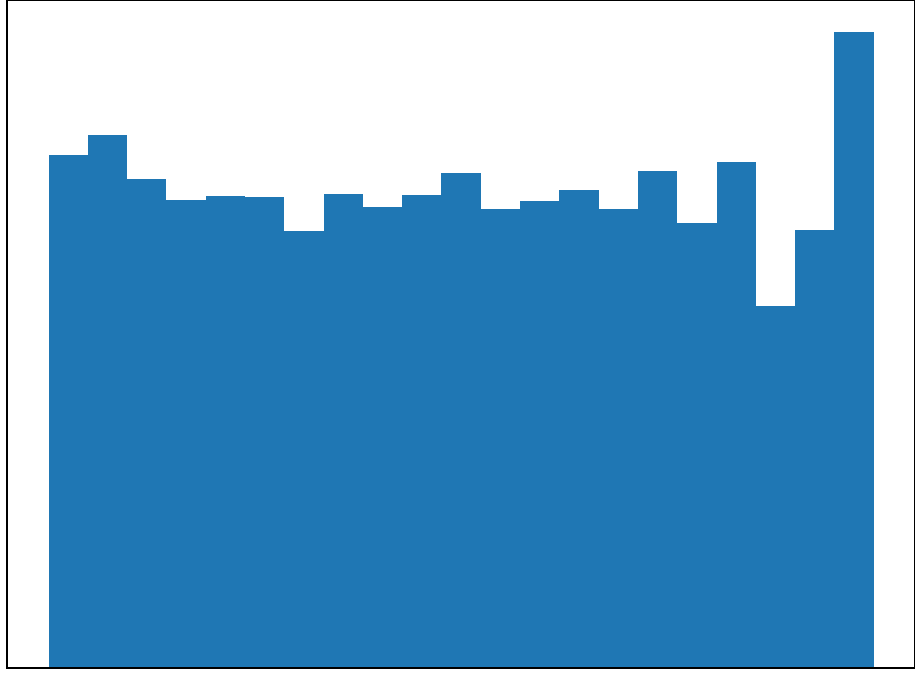}} &
    \makecell{\includegraphics[width=0.05\textwidth]{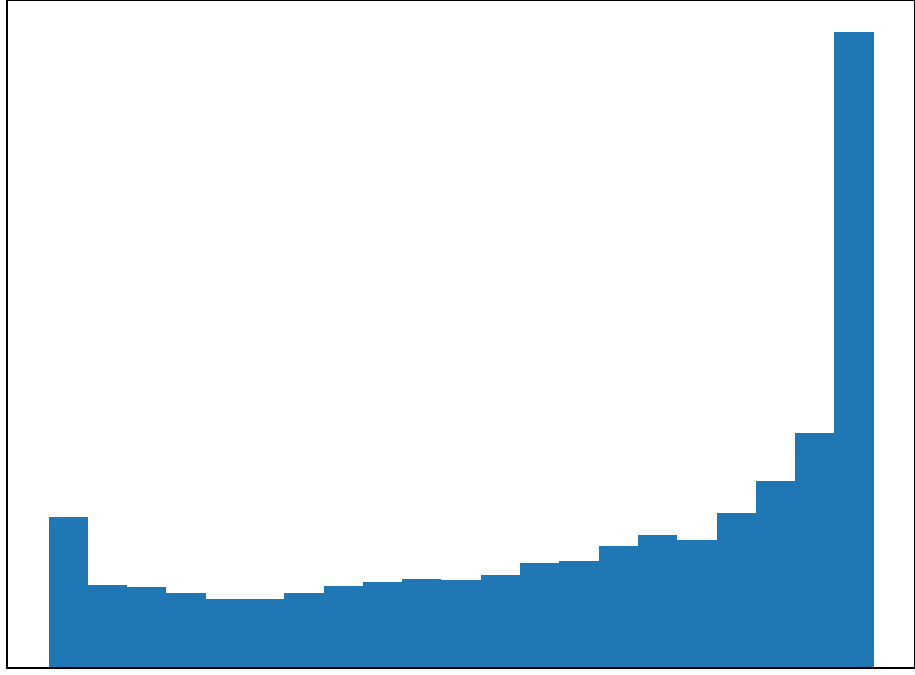}} &
    \makecell{\includegraphics[width=0.05\textwidth]{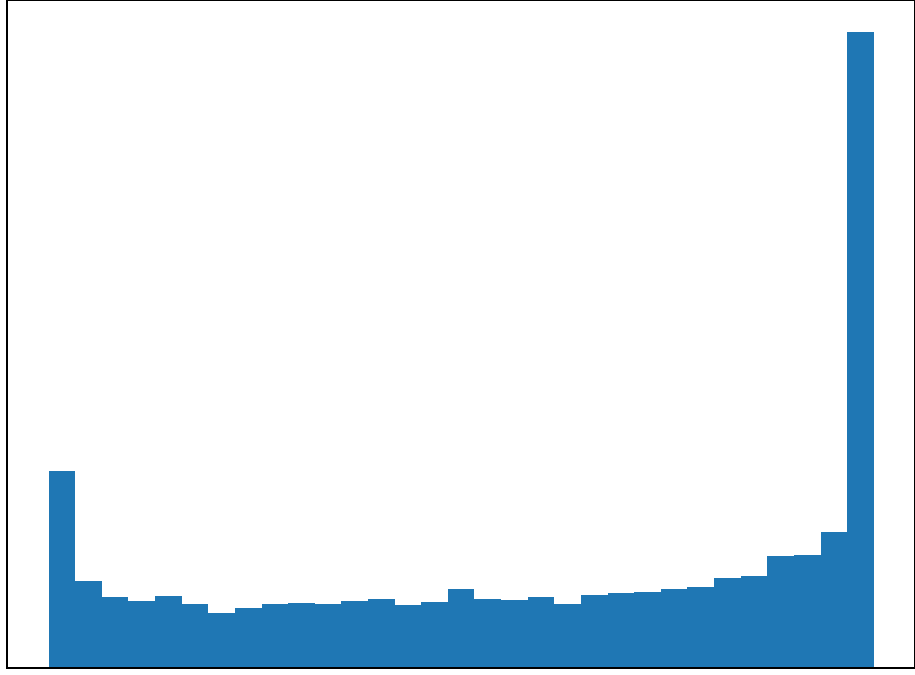}} &
    \makecell{\includegraphics[width=0.05\textwidth]{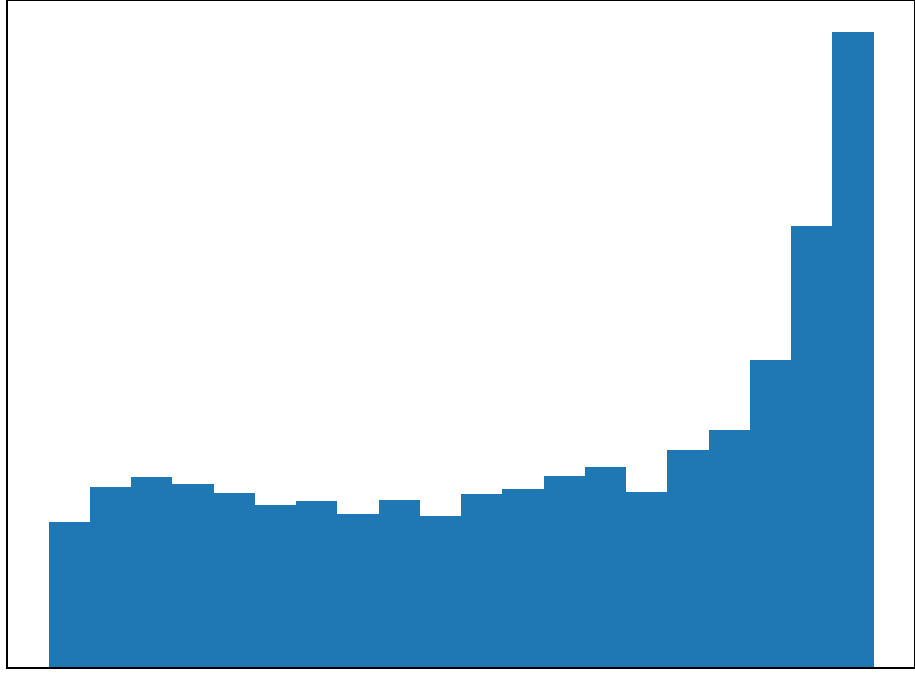}} &
    \makecell{\includegraphics[width=0.05\textwidth]{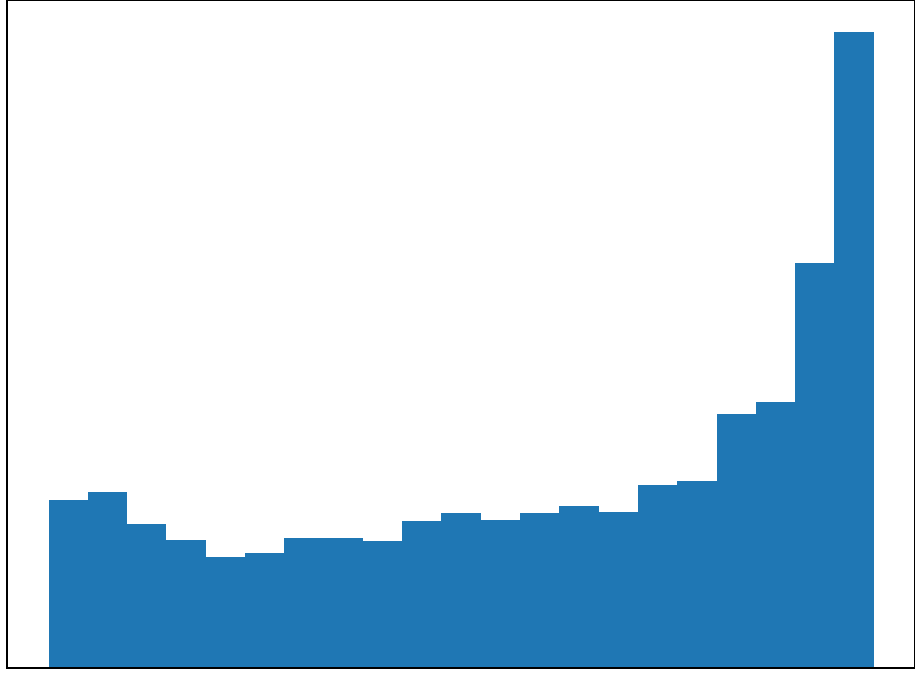}}\\
    & \makecell{Sill} &
    \makecell{\includegraphics[width=0.05\textwidth]{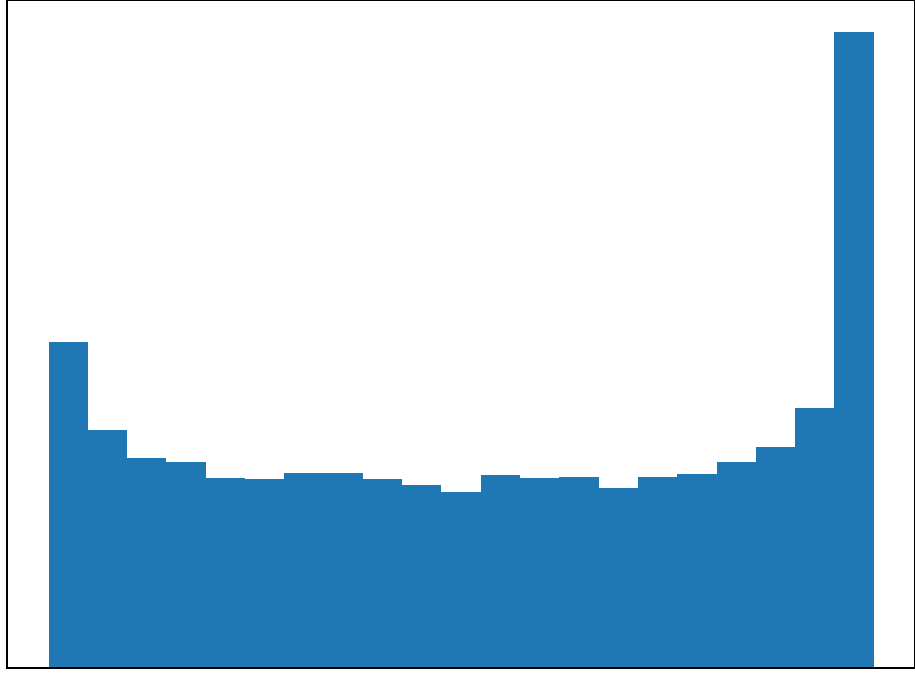}} &
    \makecell{\includegraphics[width=0.05\textwidth]{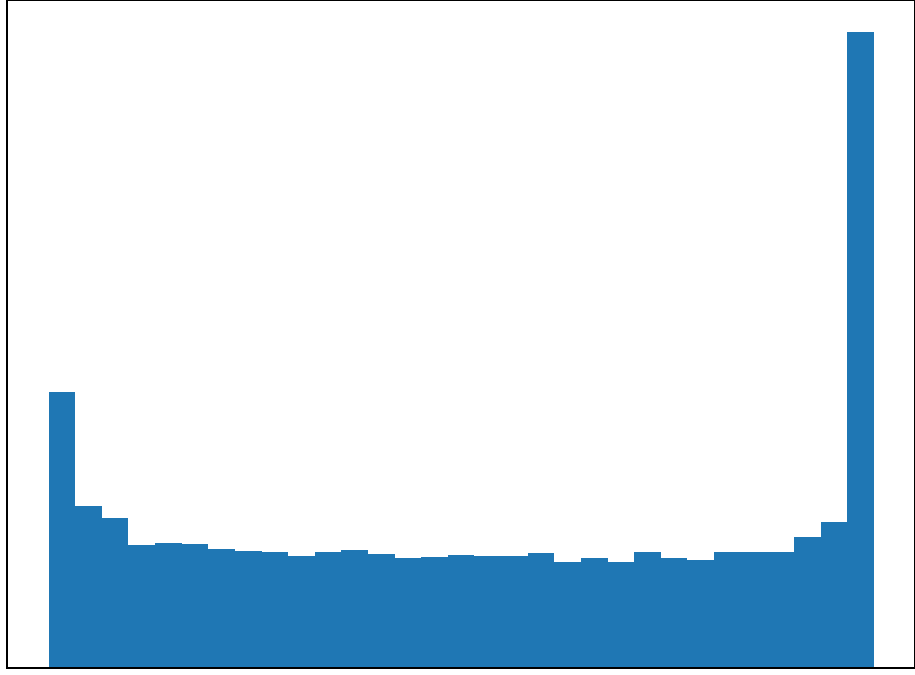}} &
    \makecell{\includegraphics[width=0.05\textwidth]{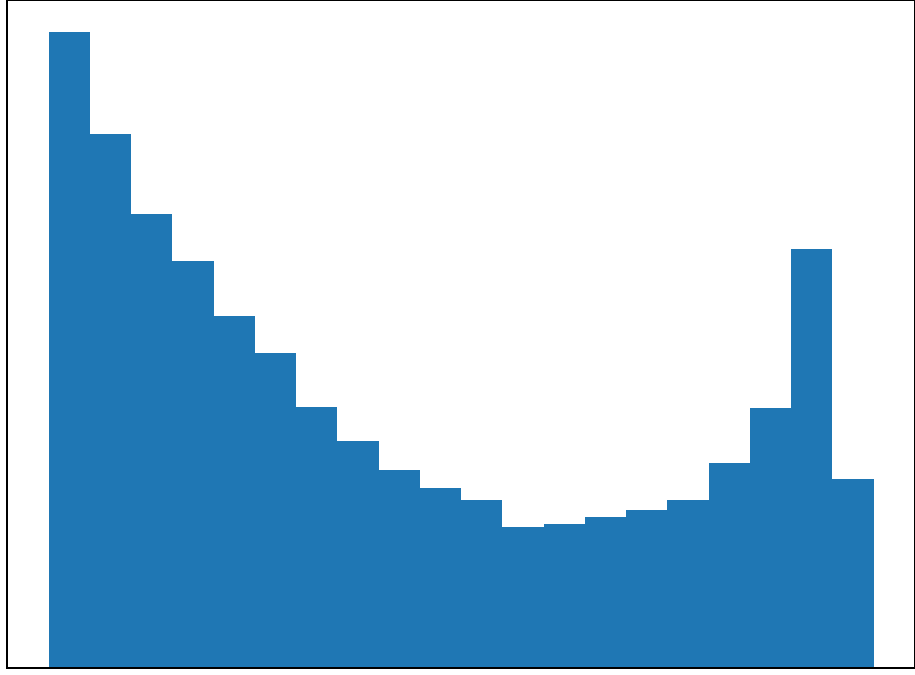}} &
    \makecell{\includegraphics[width=0.05\textwidth]{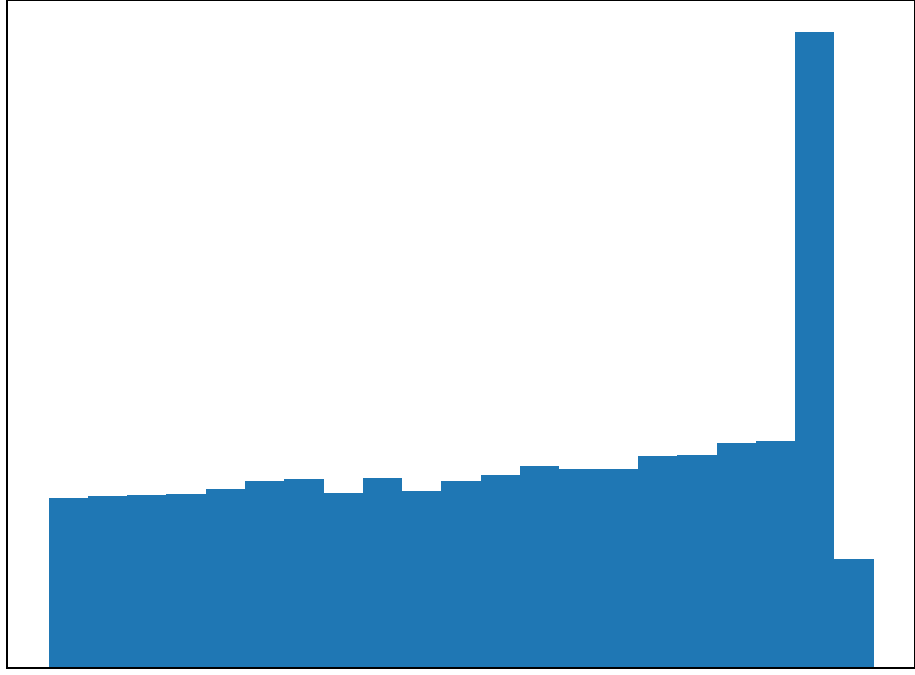}} &
    \makecell{\includegraphics[width=0.05\textwidth]{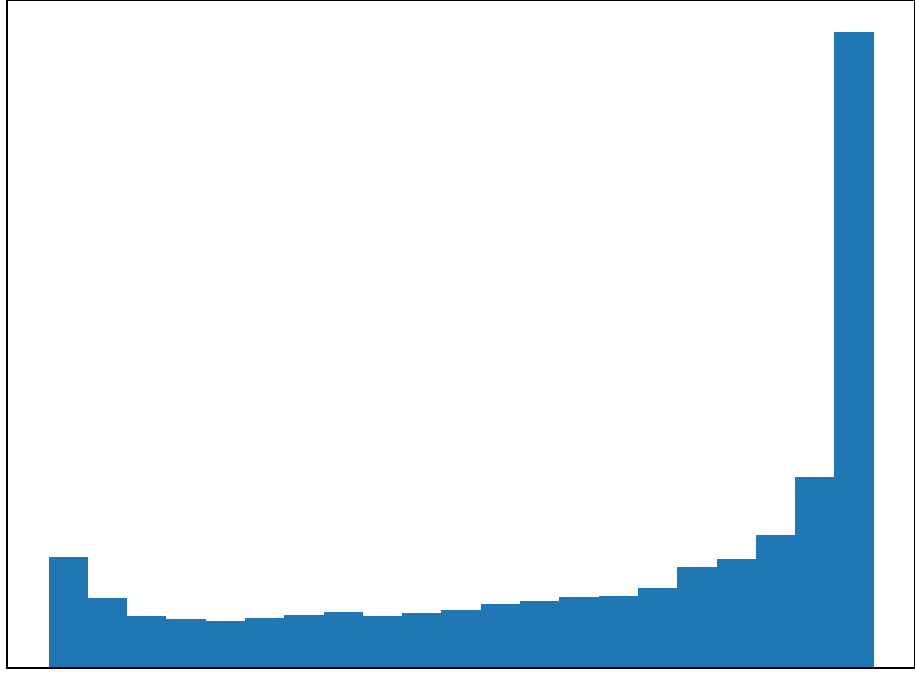}} &
    \makecell{\includegraphics[width=0.05\textwidth]{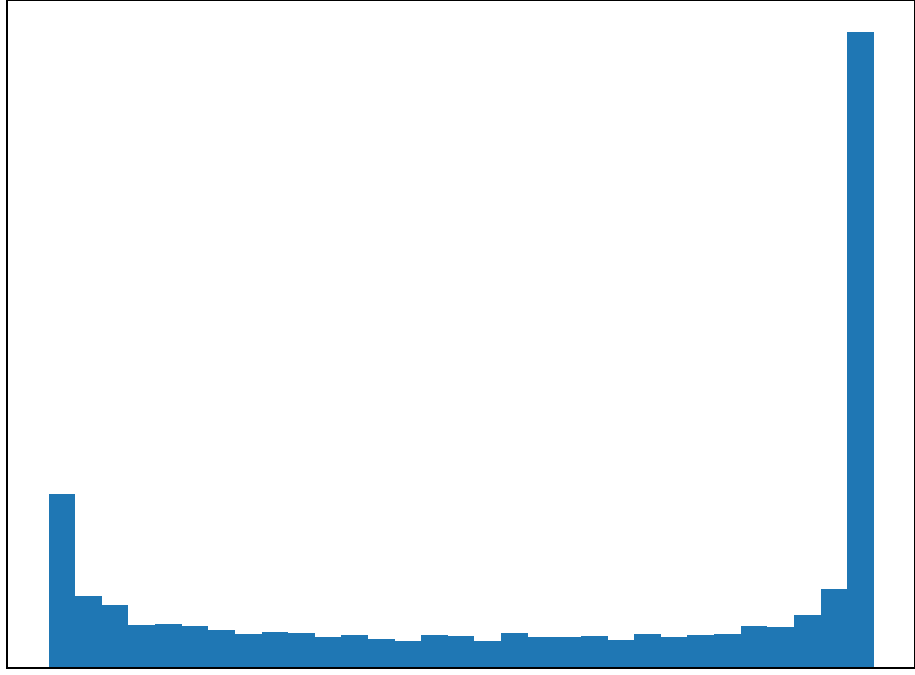}} &
    \makecell{\includegraphics[width=0.05\textwidth]{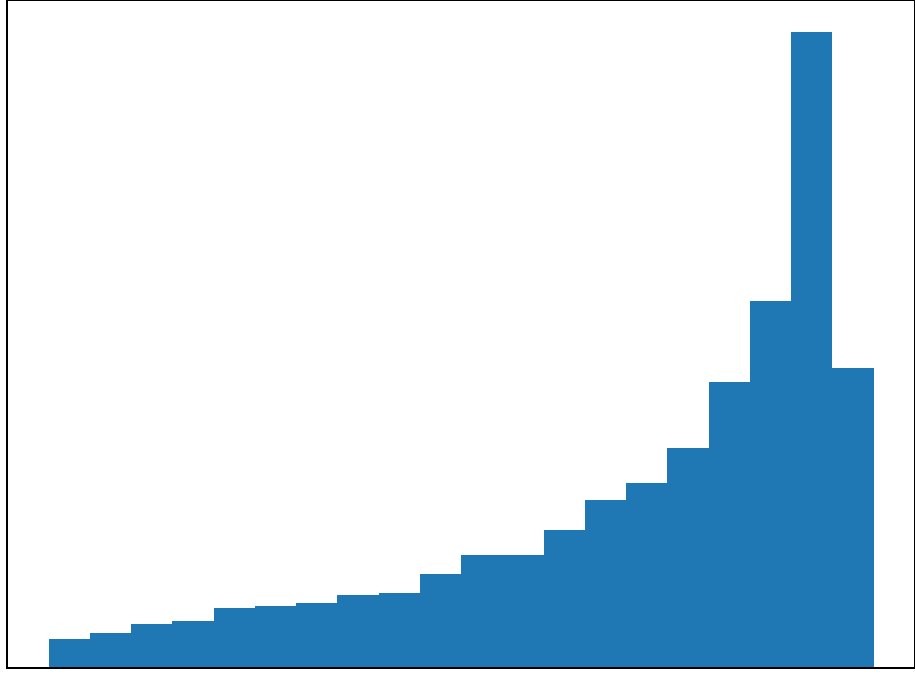}} &
    \makecell{\includegraphics[width=0.05\textwidth]{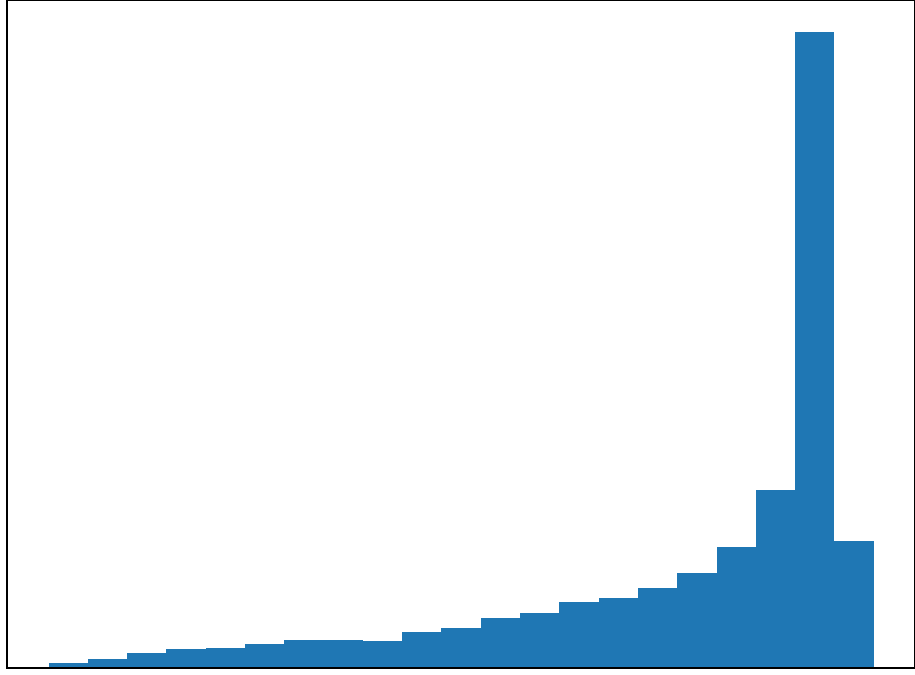}}\\ \hline
    \multirow{5}{*}{\rotatebox[origin=c]{90}{$+3$\,h}} & \makecell{Nugget} &
    \makecell{\includegraphics[width=0.05\textwidth]{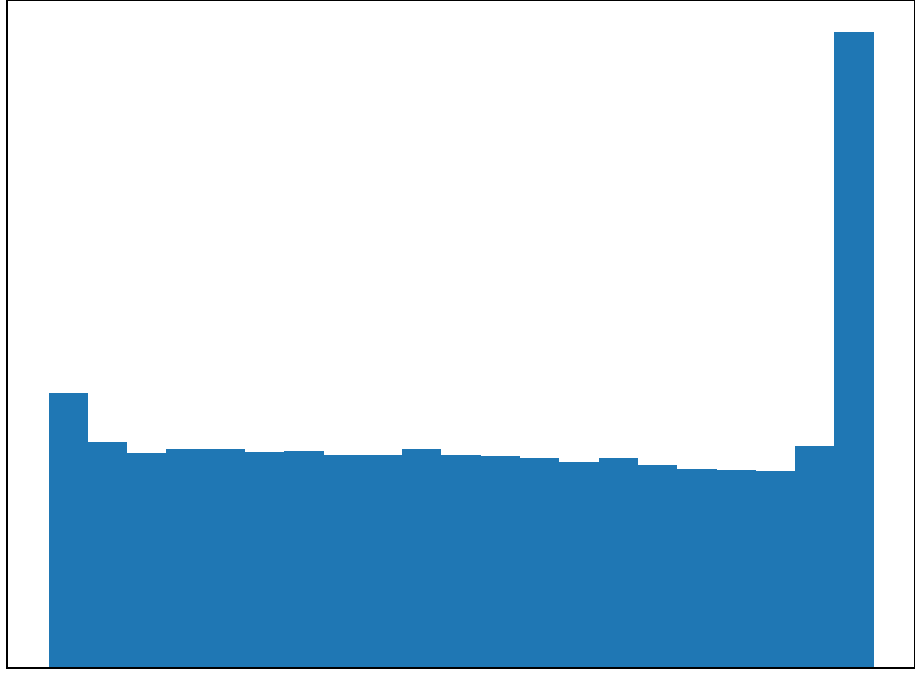}} &
    \makecell{\includegraphics[width=0.05\textwidth]{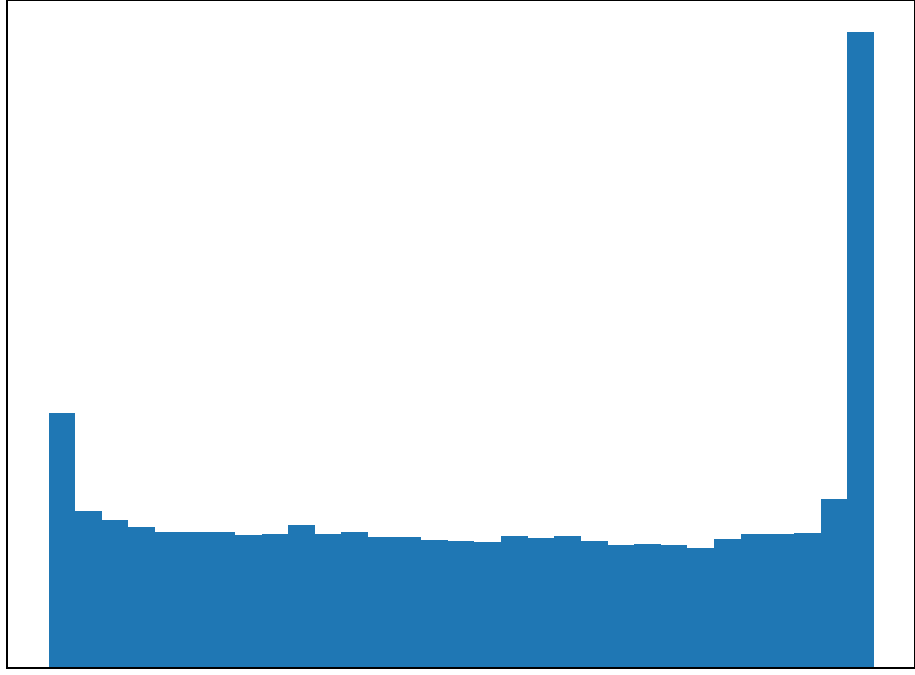}} &
    \makecell{\includegraphics[width=0.05\textwidth]{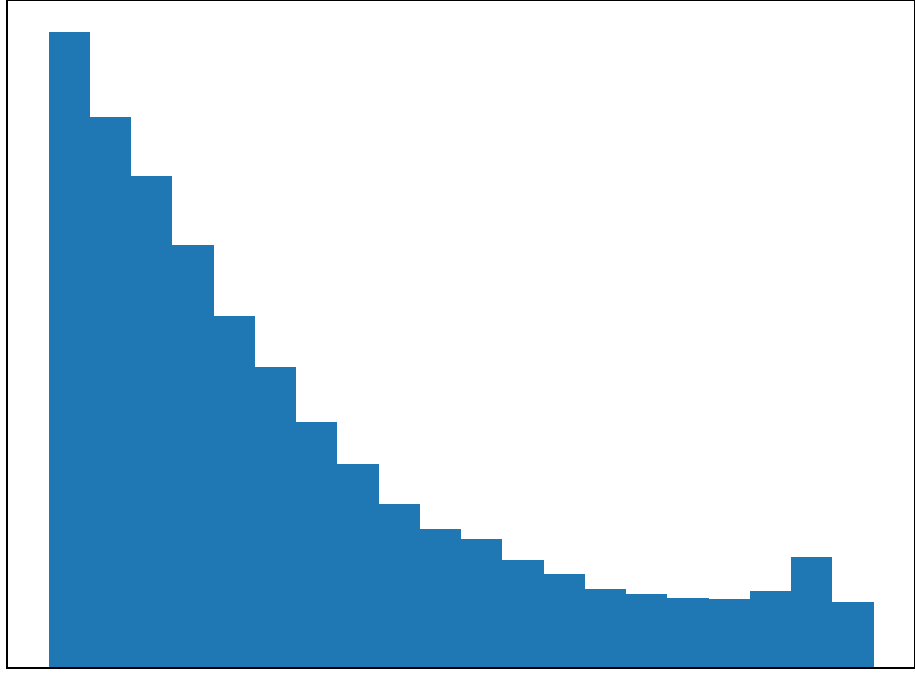}} &
    \makecell{\includegraphics[width=0.05\textwidth]{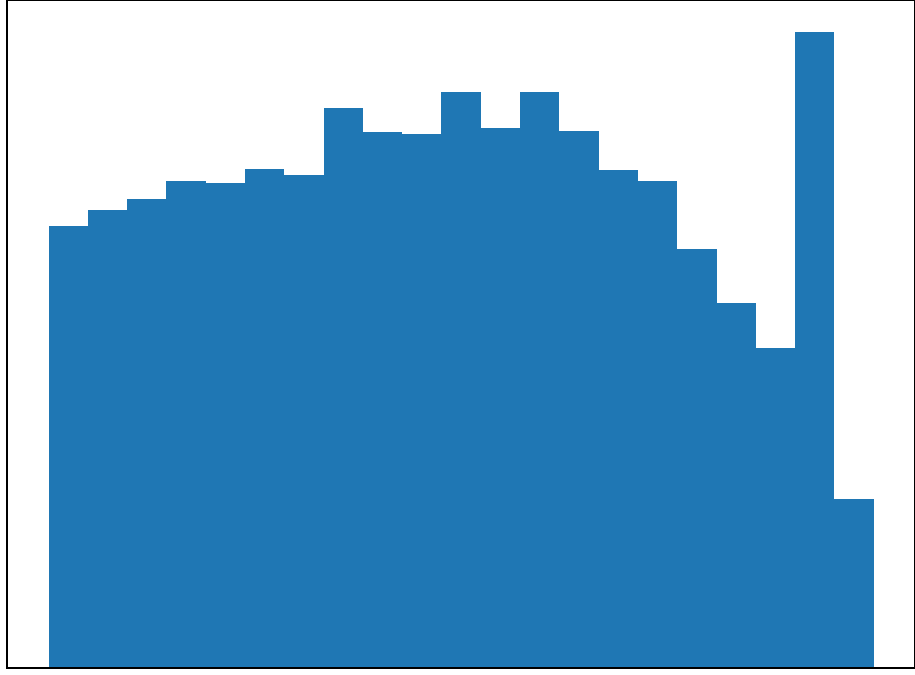}} &
    \makecell{\includegraphics[width=0.05\textwidth]{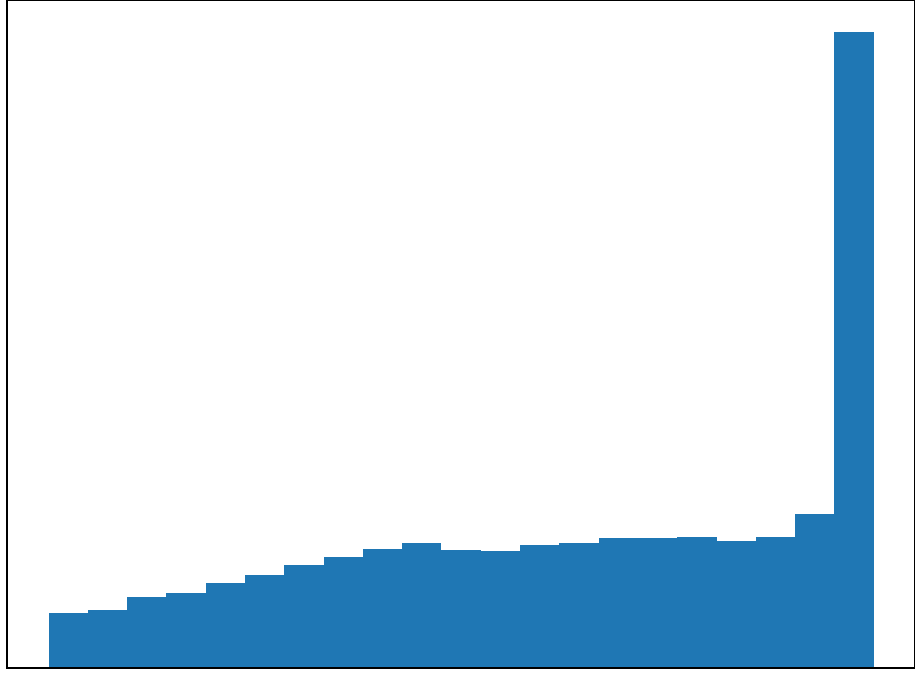}} &
    \makecell{\includegraphics[width=0.05\textwidth]{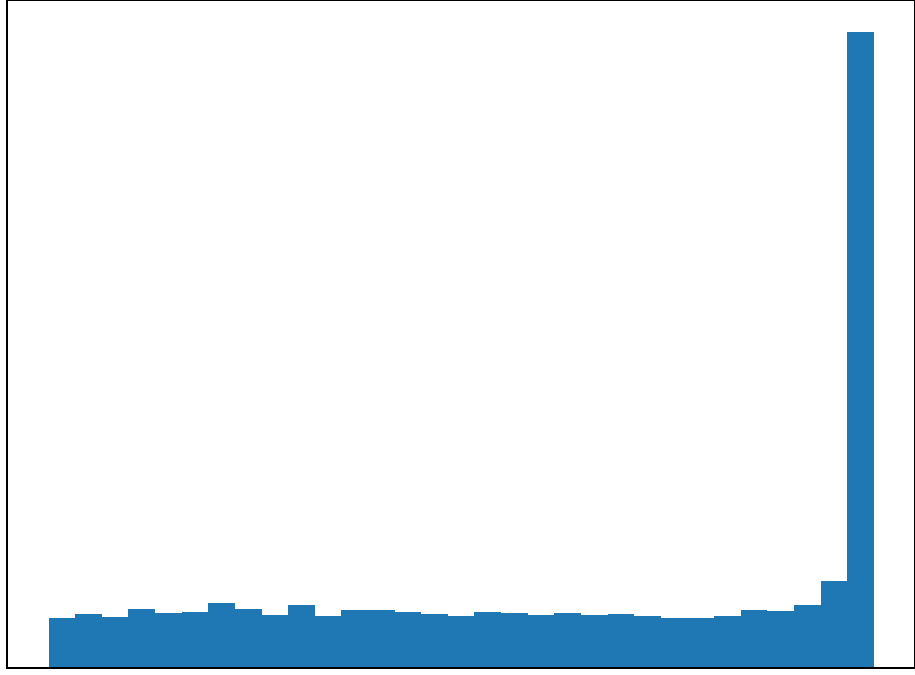}} &
    \makecell{\includegraphics[width=0.05\textwidth]{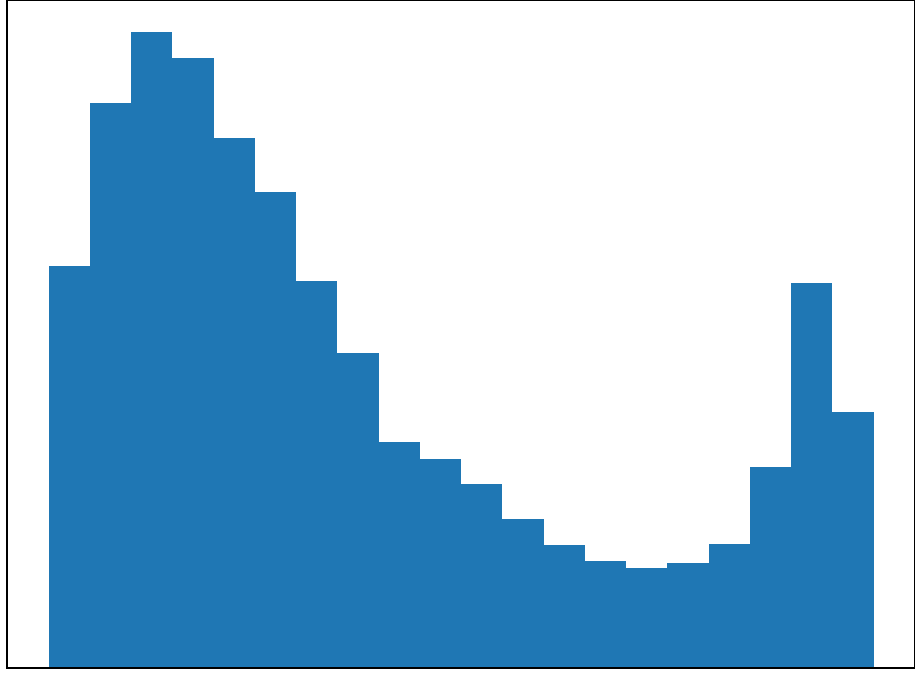}} &
    \makecell{\includegraphics[width=0.05\textwidth]{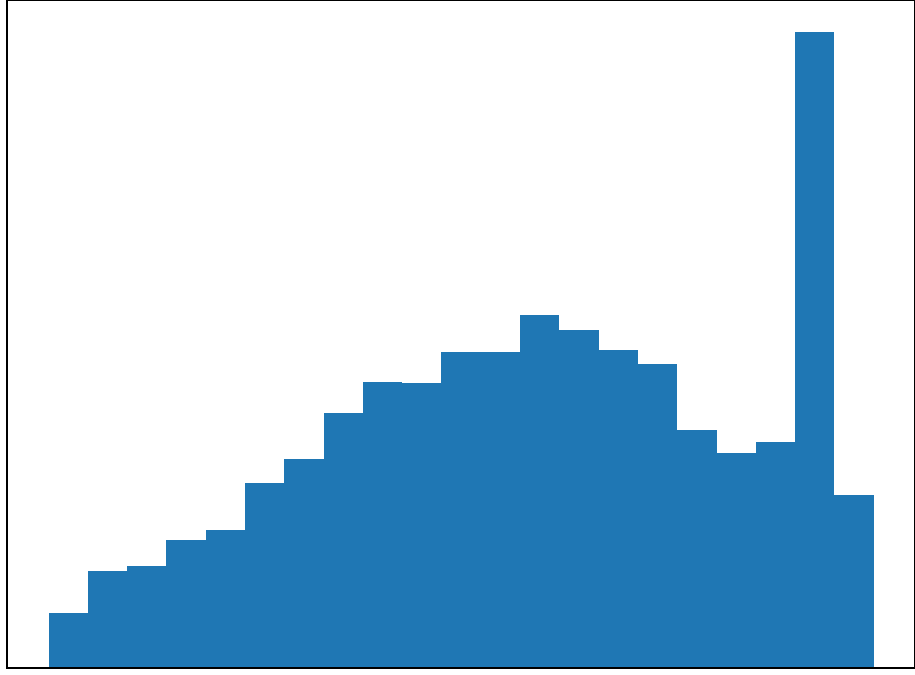}}\\
    & \makecell{Effective range} &
    \makecell{\includegraphics[width=0.05\textwidth]{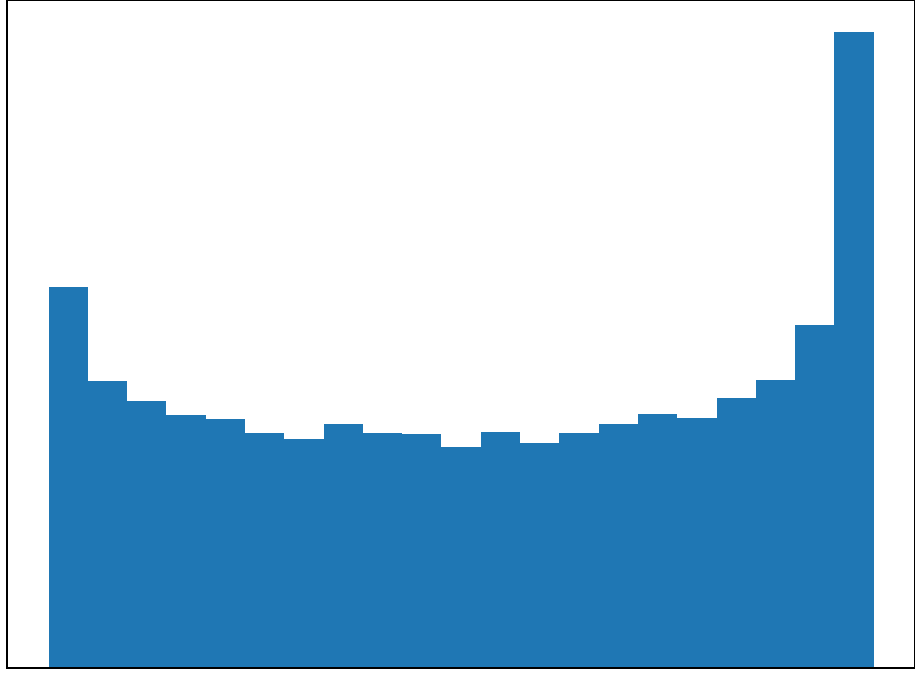}} &
    \makecell{\includegraphics[width=0.05\textwidth]{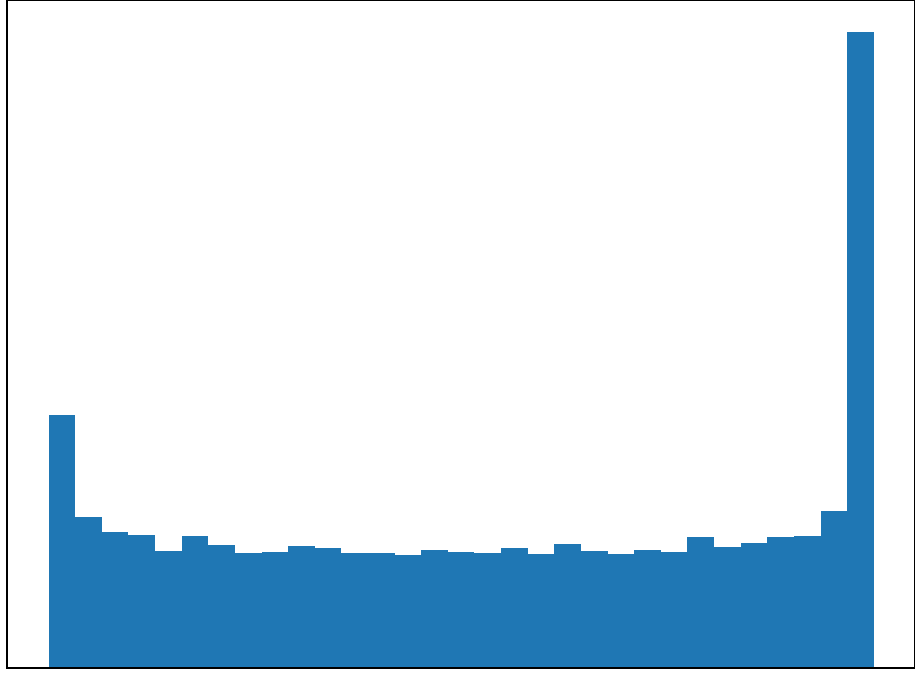}} &
    \makecell{\includegraphics[width=0.05\textwidth]{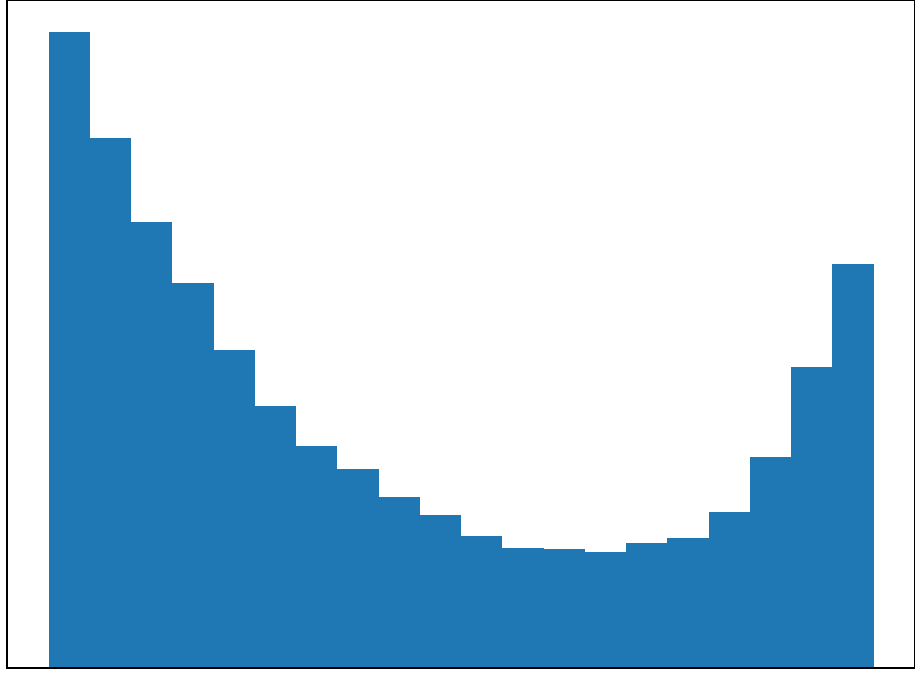}} &
    \makecell{\includegraphics[width=0.05\textwidth]{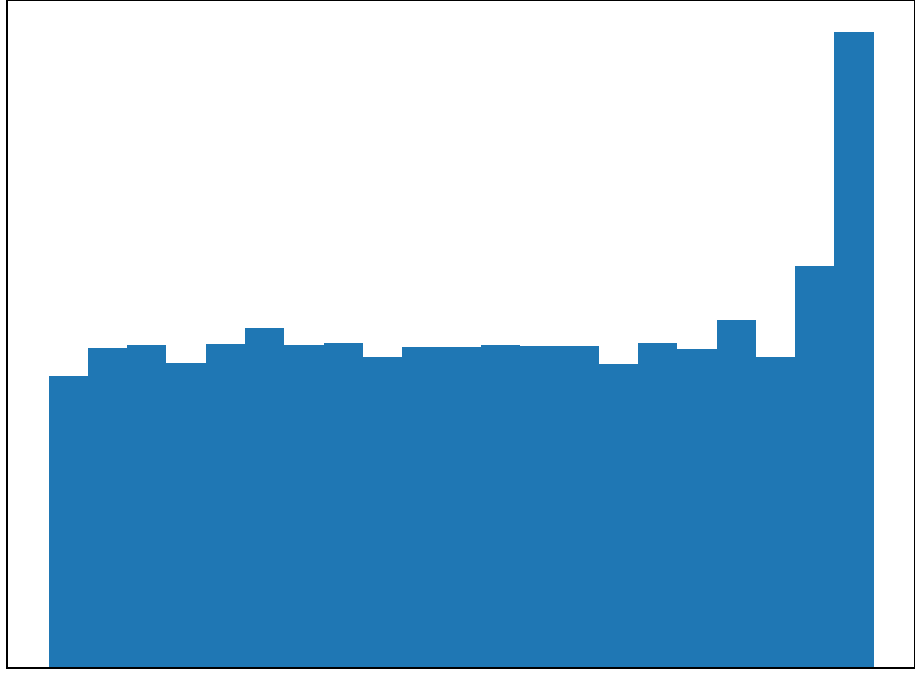}} &
    \makecell{\includegraphics[width=0.05\textwidth]{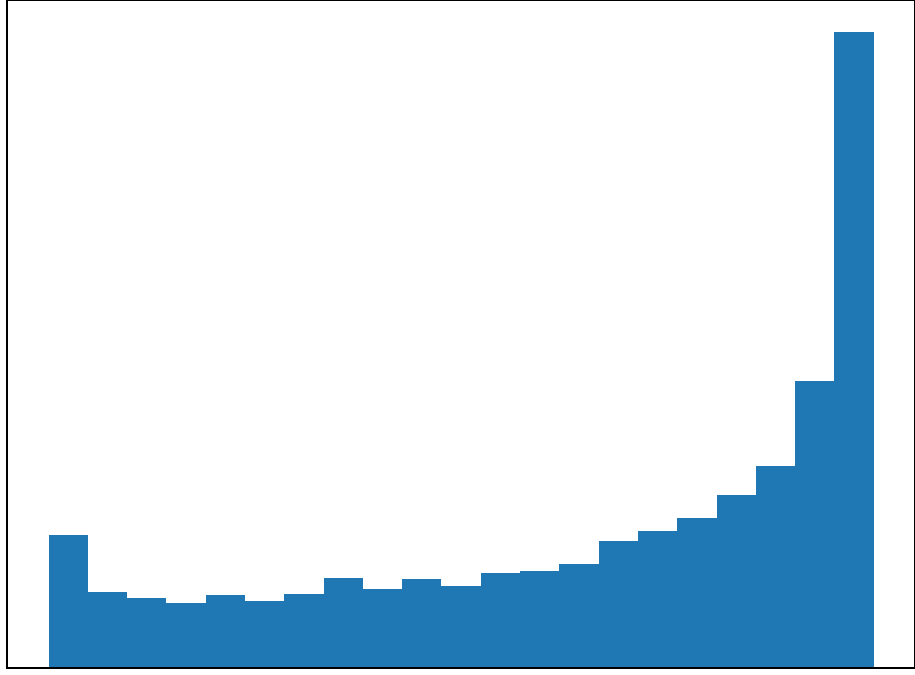}} &
    \makecell{\includegraphics[width=0.05\textwidth]{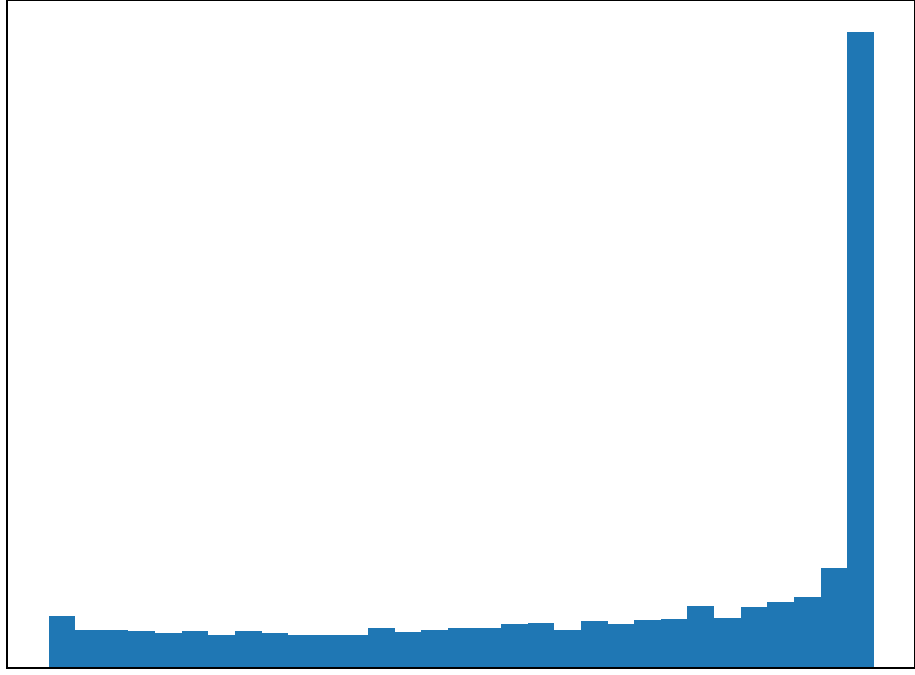}} &
    \makecell{\includegraphics[width=0.05\textwidth]{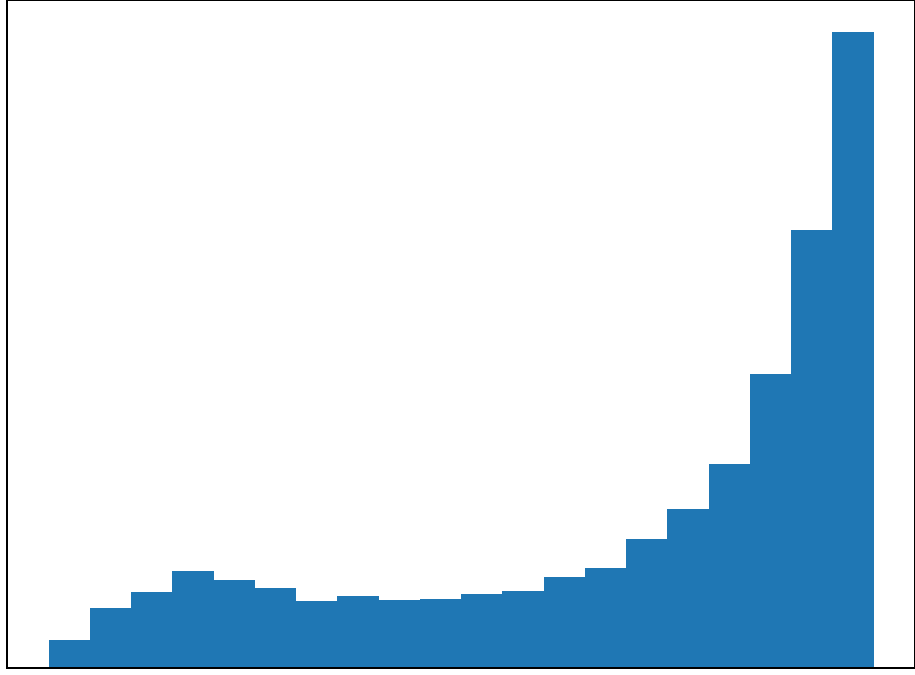}} &
    \makecell{\includegraphics[width=0.05\textwidth]{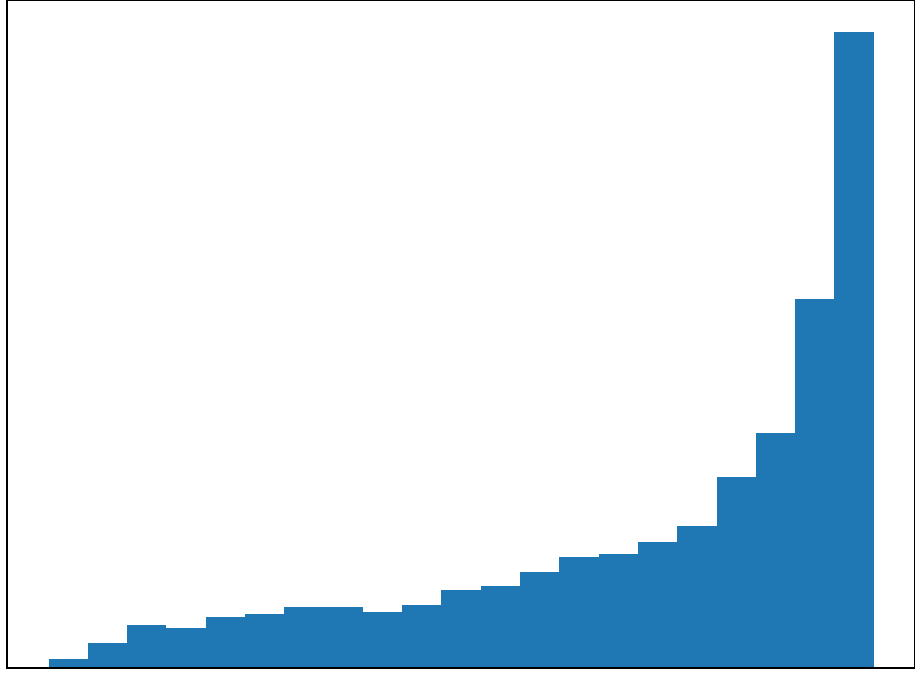}}\\
    & \makecell{Sill} &
    \makecell{\includegraphics[width=0.05\textwidth]{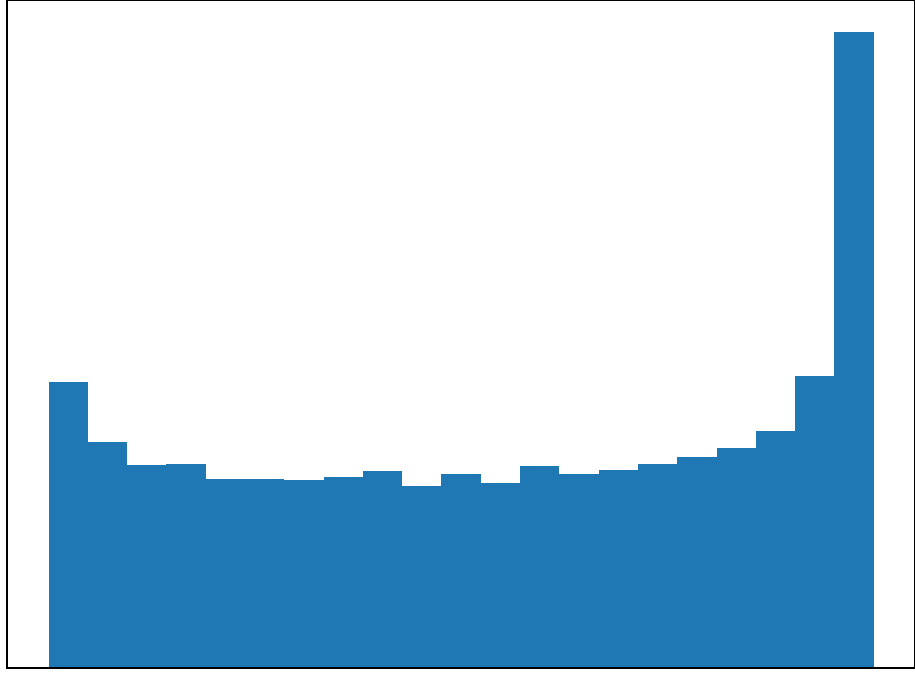}} &
    \makecell{\includegraphics[width=0.05\textwidth]{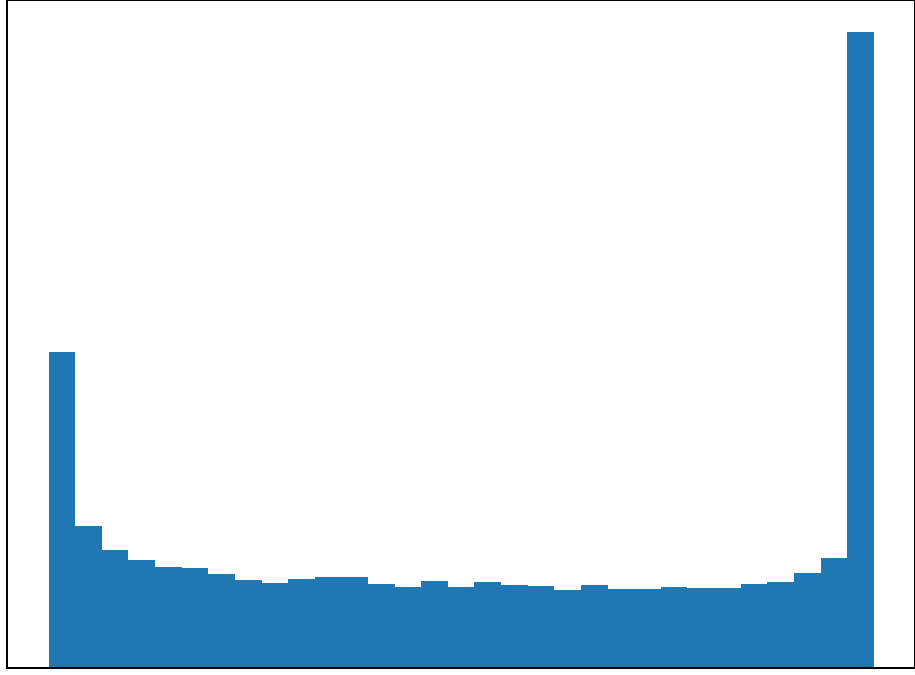}} &
    \makecell{\includegraphics[width=0.05\textwidth]{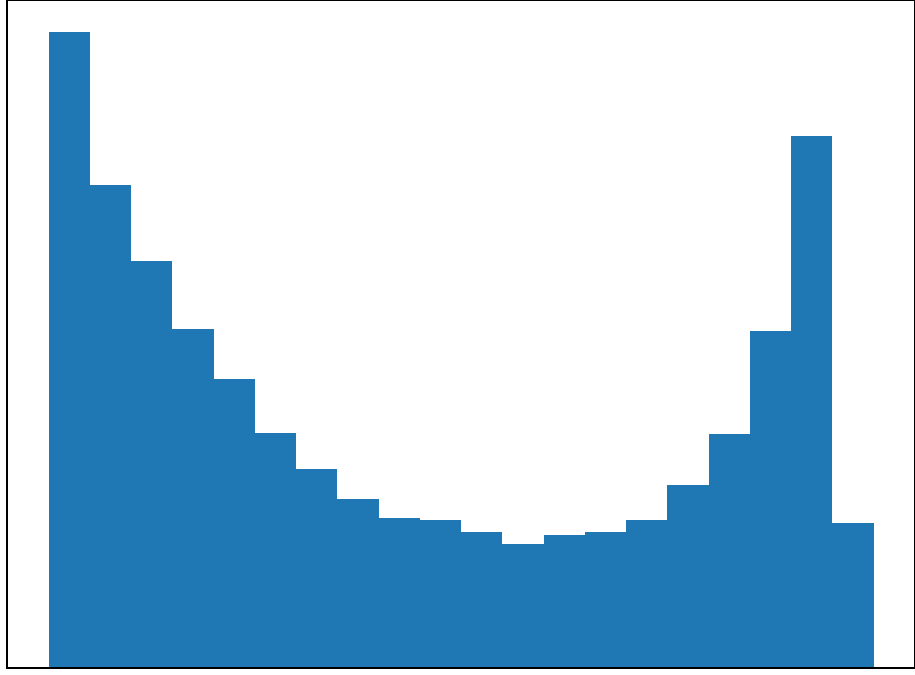}} &
    \makecell{\includegraphics[width=0.05\textwidth]{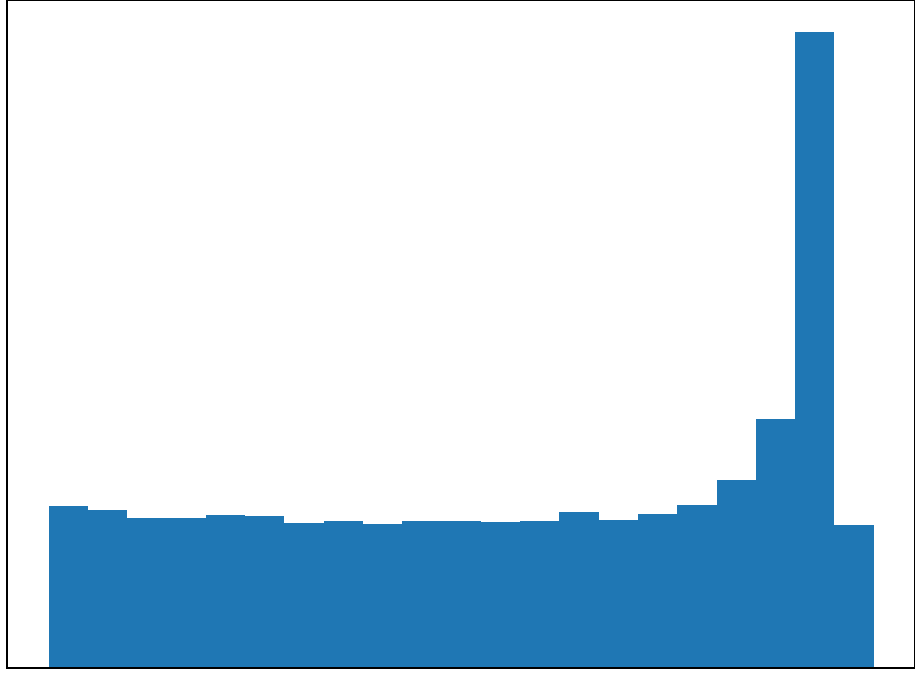}} &
    \makecell{\includegraphics[width=0.05\textwidth]{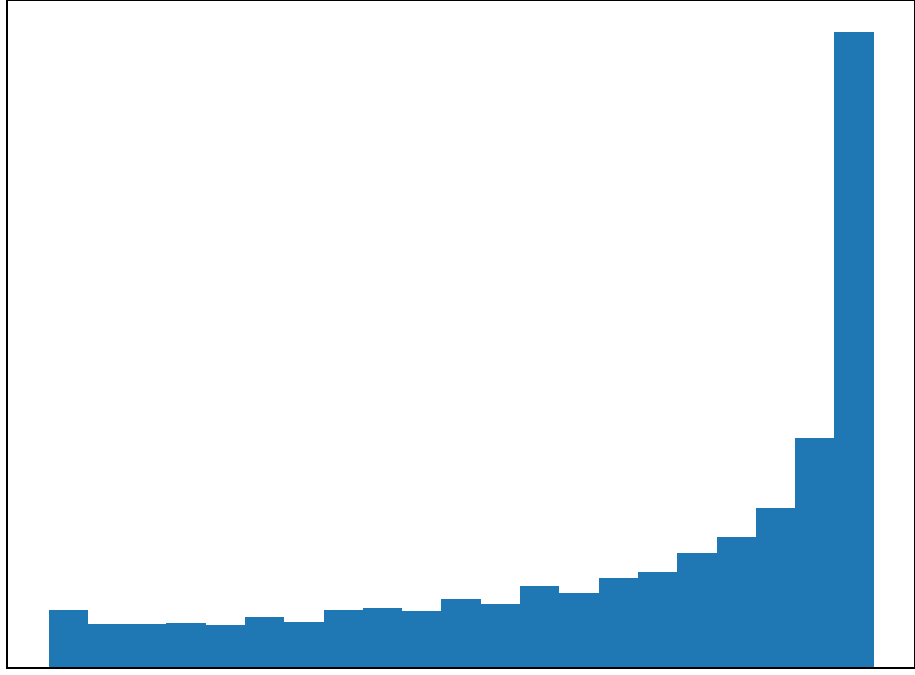}} &
    \makecell{\includegraphics[width=0.05\textwidth]{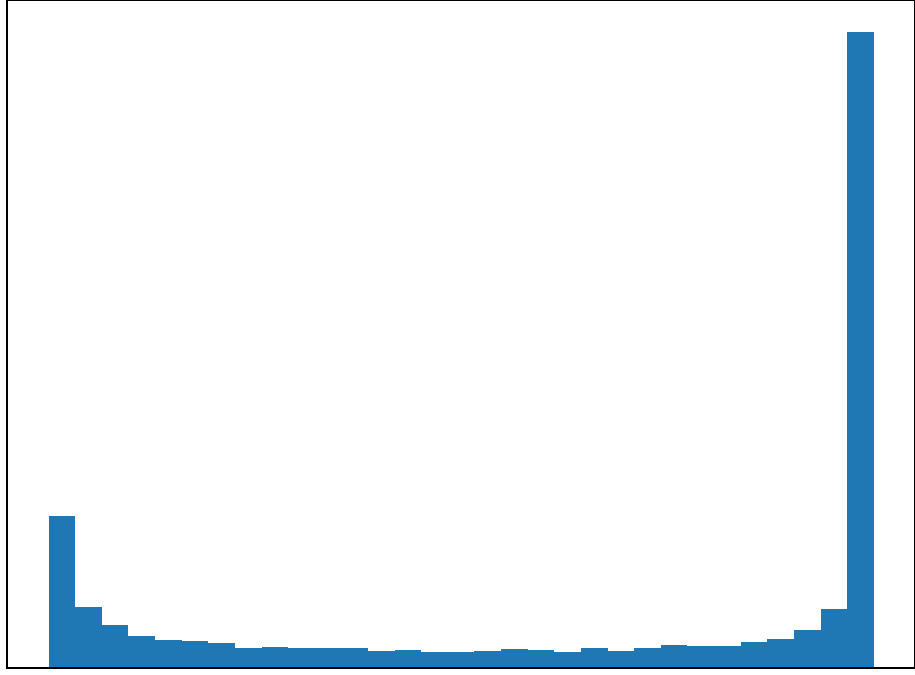}} &
    \makecell{\includegraphics[width=0.05\textwidth]{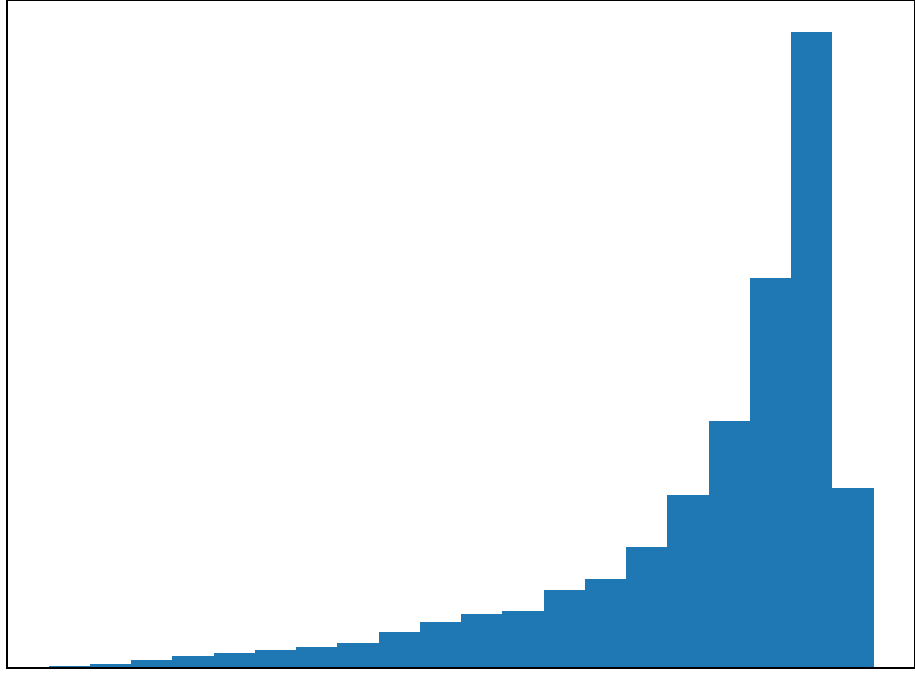}} &
    \makecell{\includegraphics[width=0.05\textwidth]{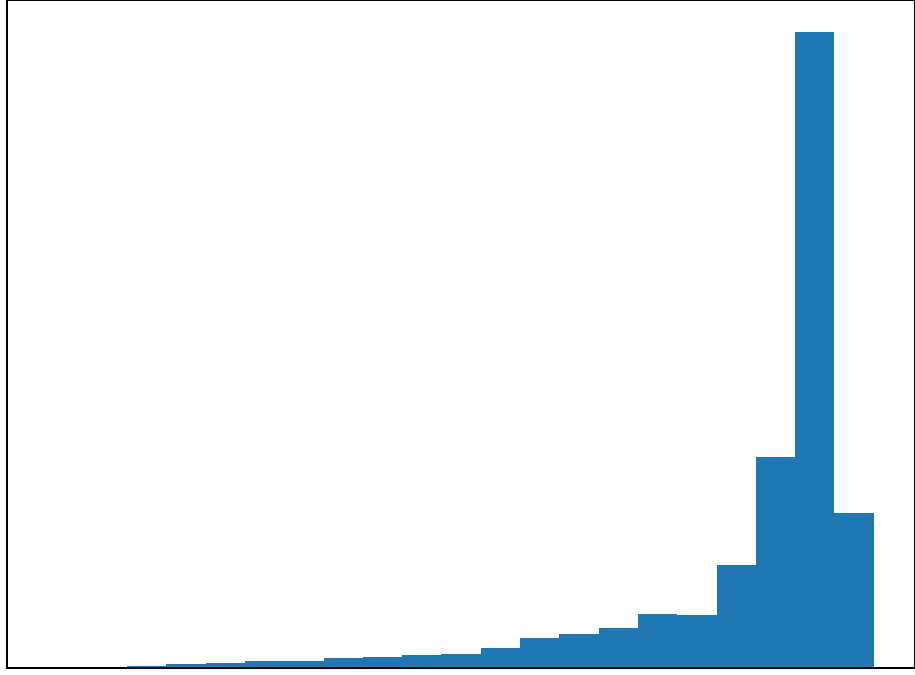}}\\ \hline
    \multirow{5}{*}{\rotatebox[origin=c]{90}{$+6$\,h}} & \makecell{Nugget} &
    \makecell{\includegraphics[width=0.05\textwidth]{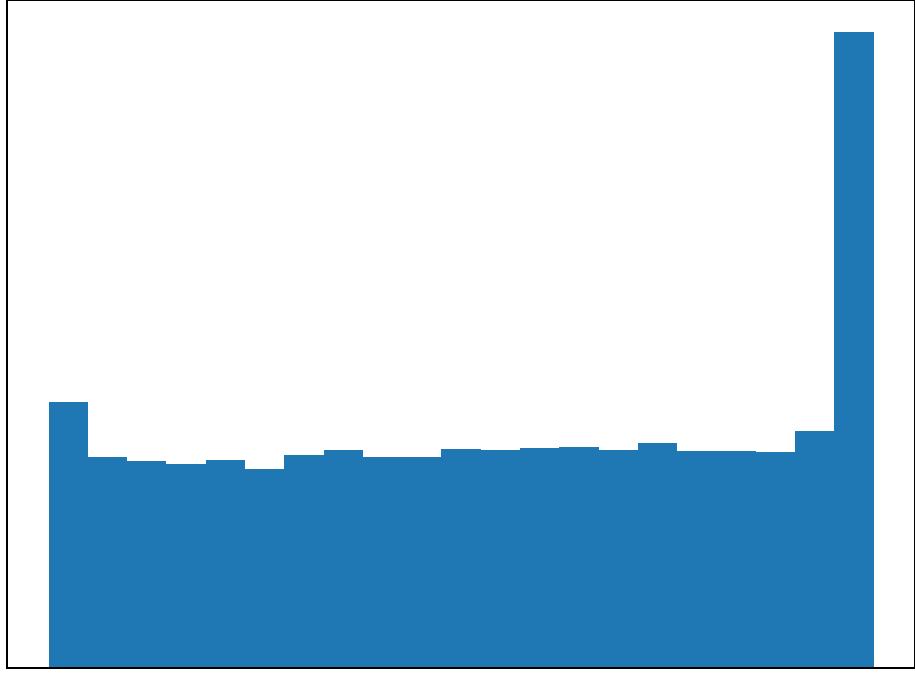}} &
    \makecell{\includegraphics[width=0.05\textwidth]{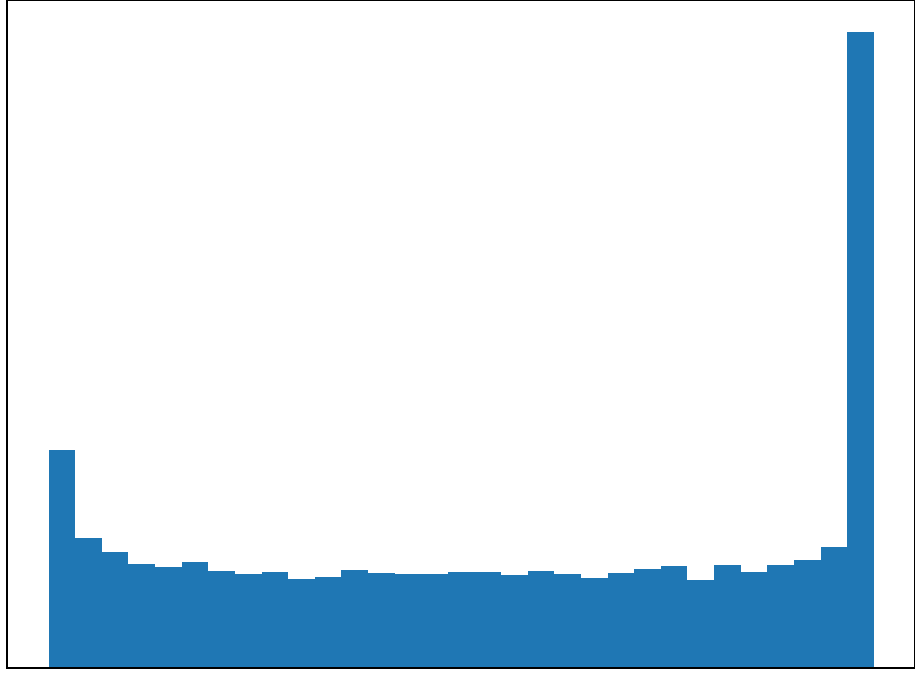}} &
    \makecell{\includegraphics[width=0.05\textwidth]{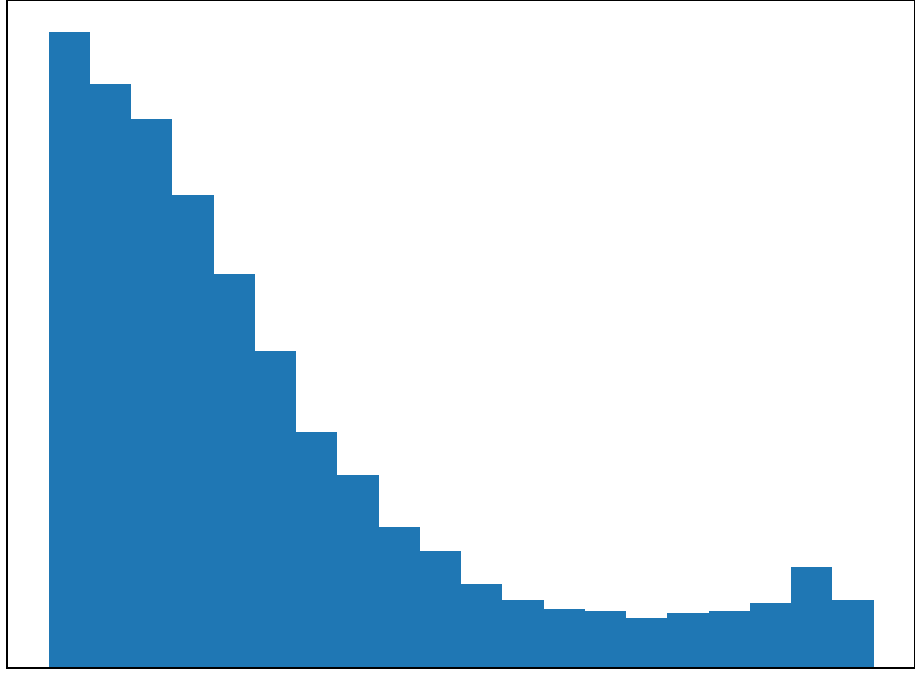}} &
    \makecell{\includegraphics[width=0.05\textwidth]{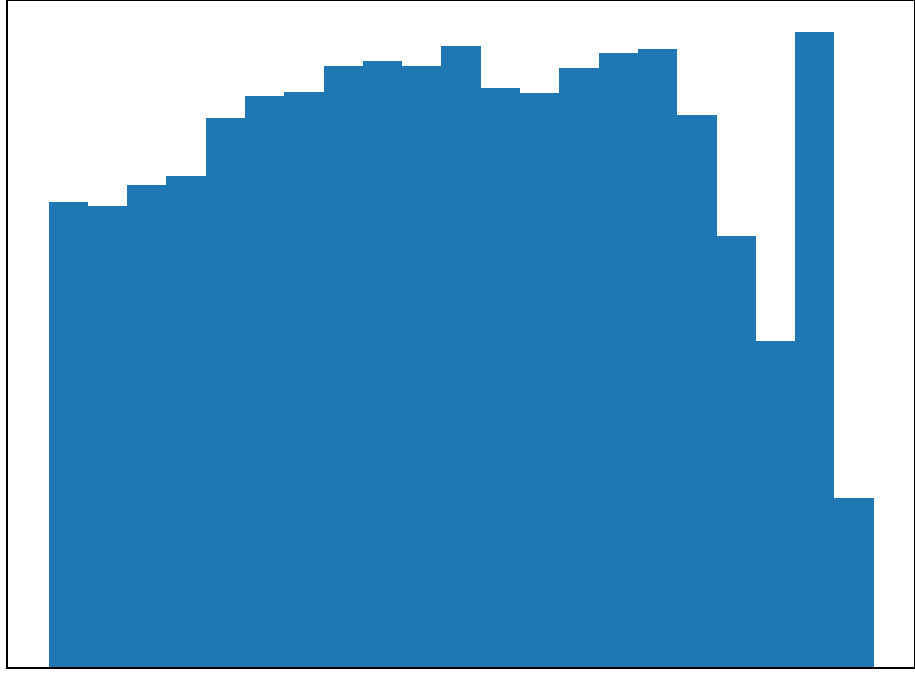}} &
    \makecell{\includegraphics[width=0.05\textwidth]{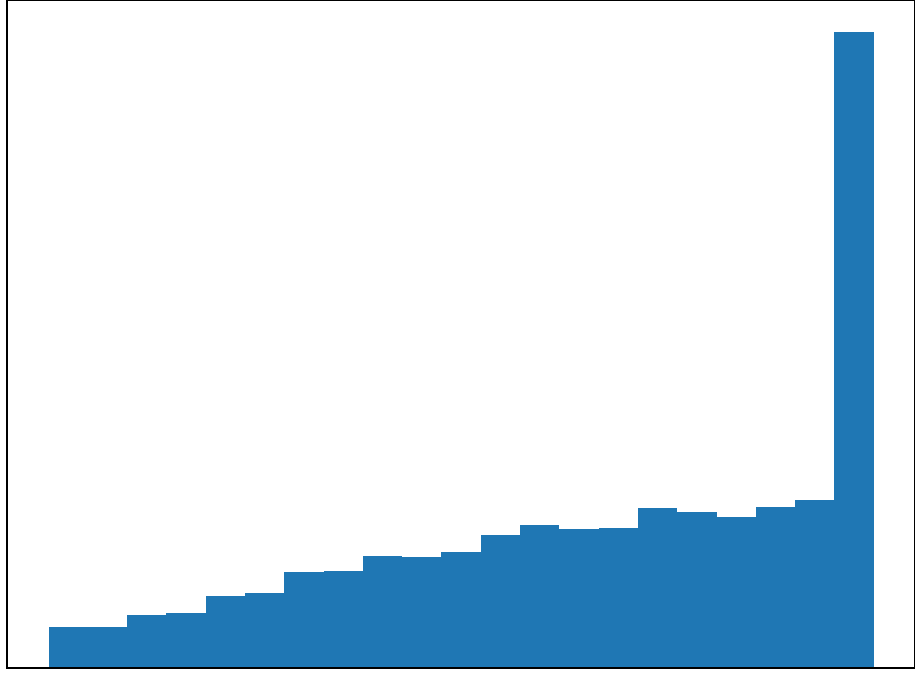}} &
    \makecell{\includegraphics[width=0.05\textwidth]{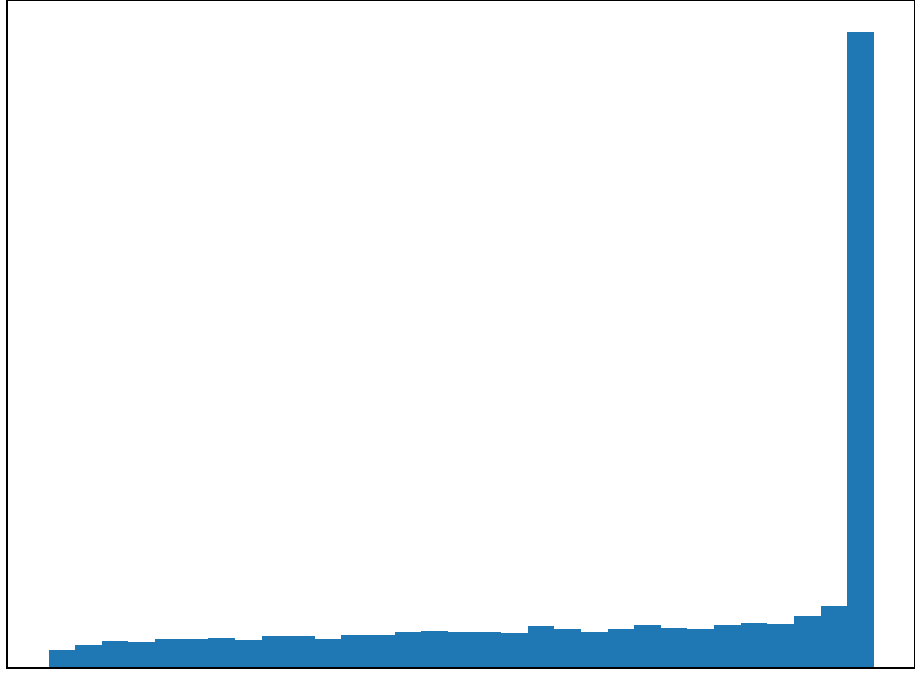}} &
    \makecell{\includegraphics[width=0.05\textwidth]{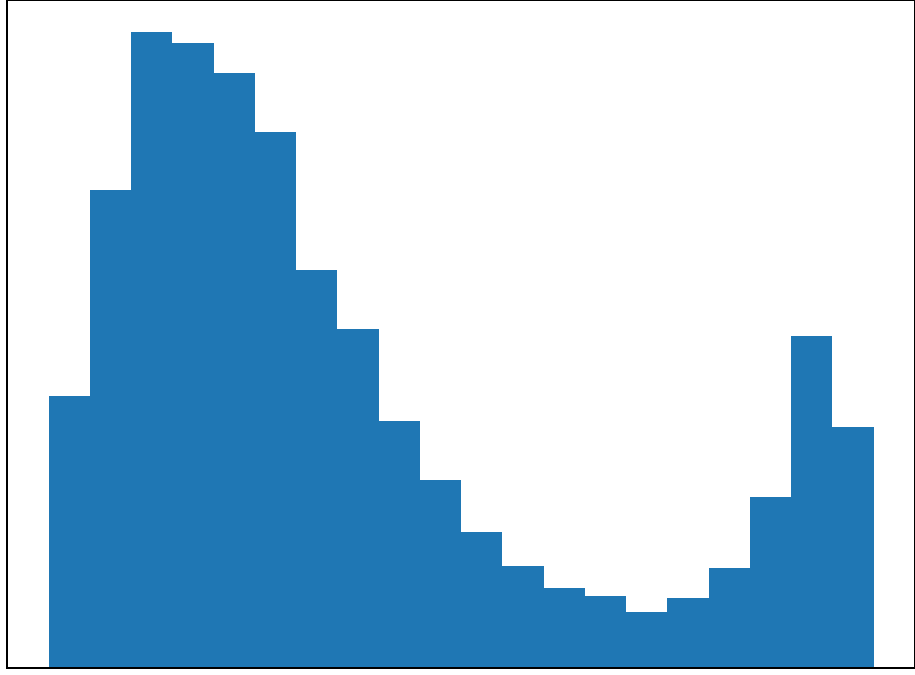}} &
    \makecell{\includegraphics[width=0.05\textwidth]{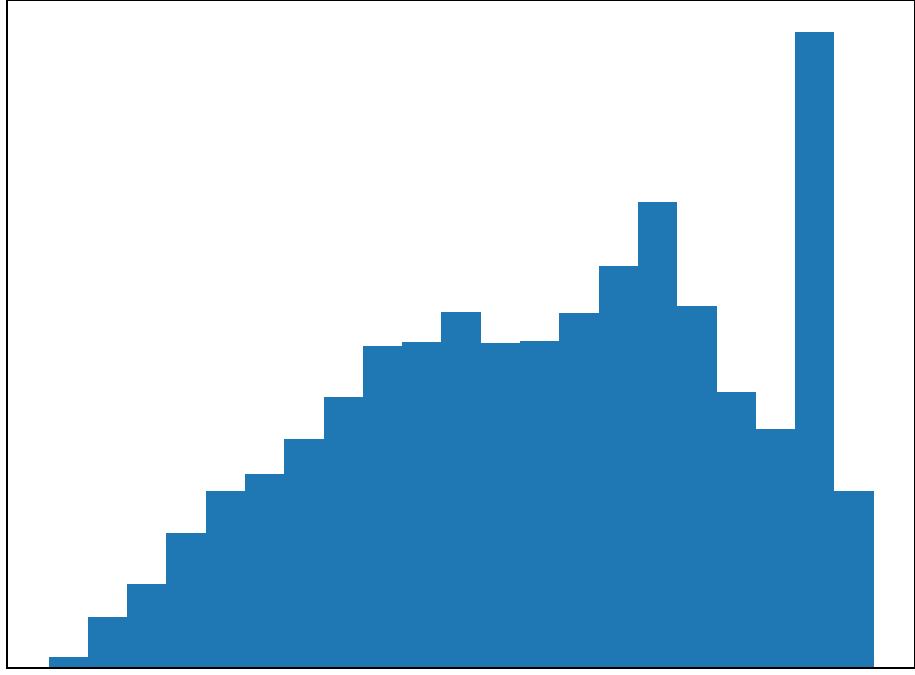}}\\
    & \makecell{Effective range} &
    \makecell{\includegraphics[width=0.05\textwidth]{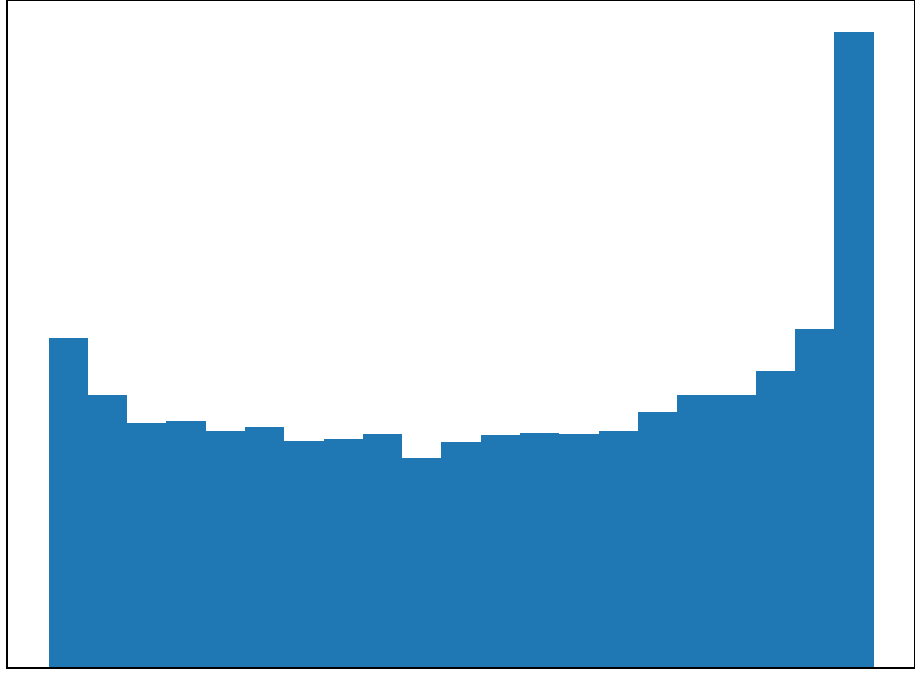}} &
    \makecell{\includegraphics[width=0.05\textwidth]{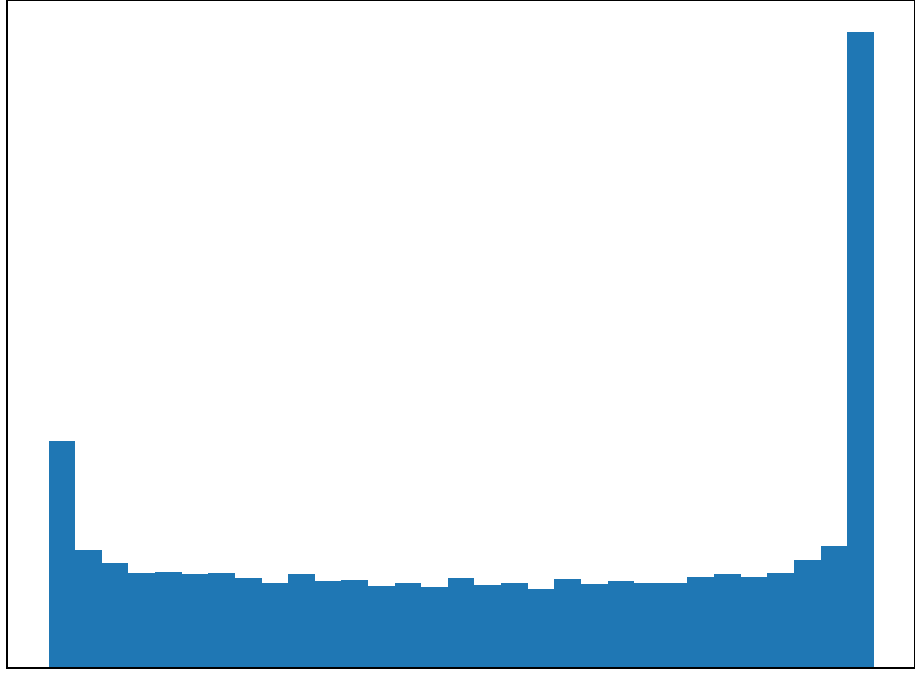}} &
    \makecell{\includegraphics[width=0.05\textwidth]{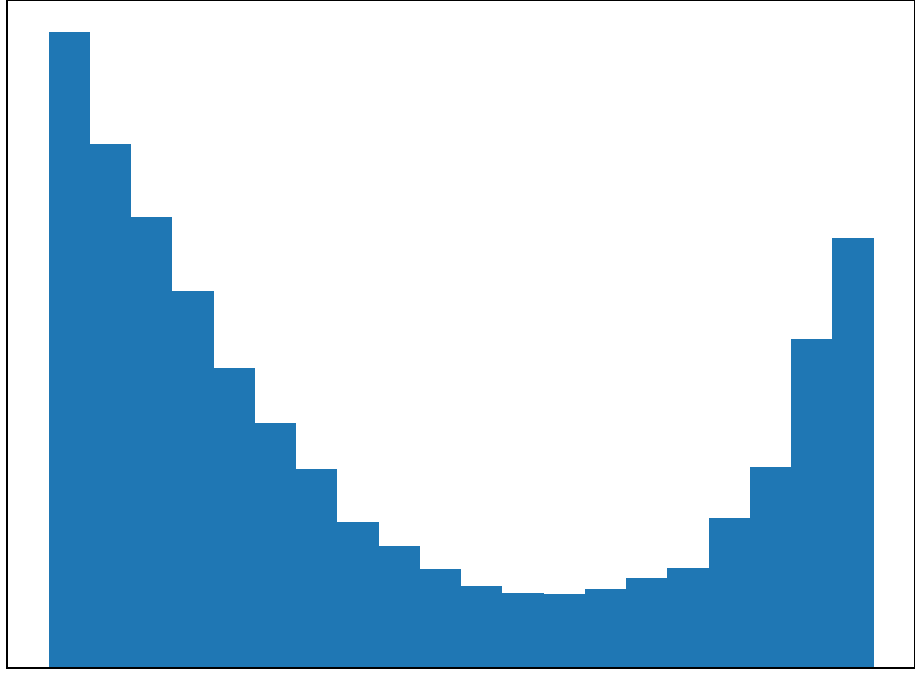}} &
    \makecell{\includegraphics[width=0.05\textwidth]{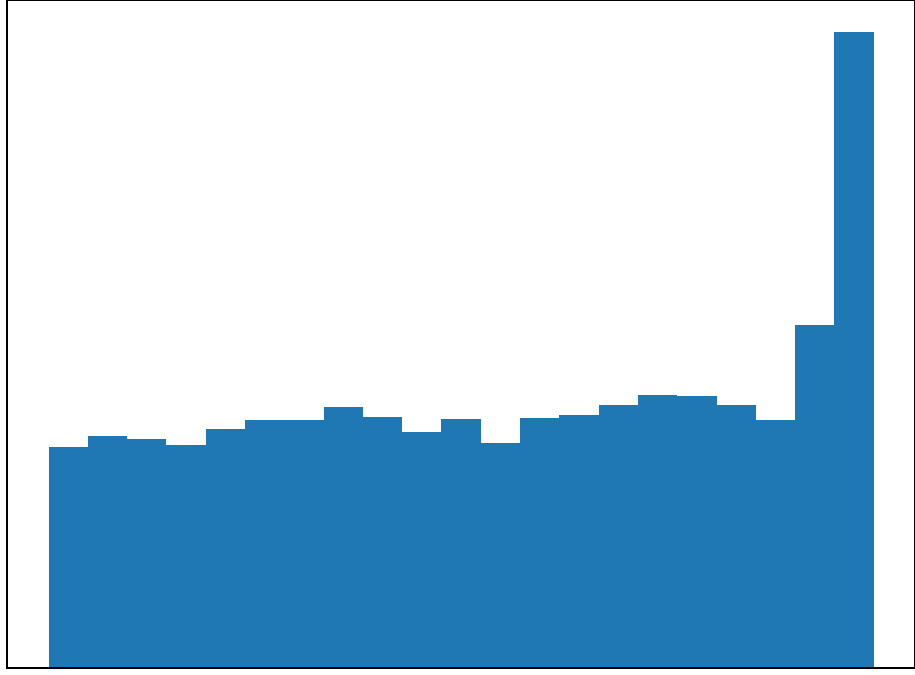}} &
    \makecell{\includegraphics[width=0.05\textwidth]{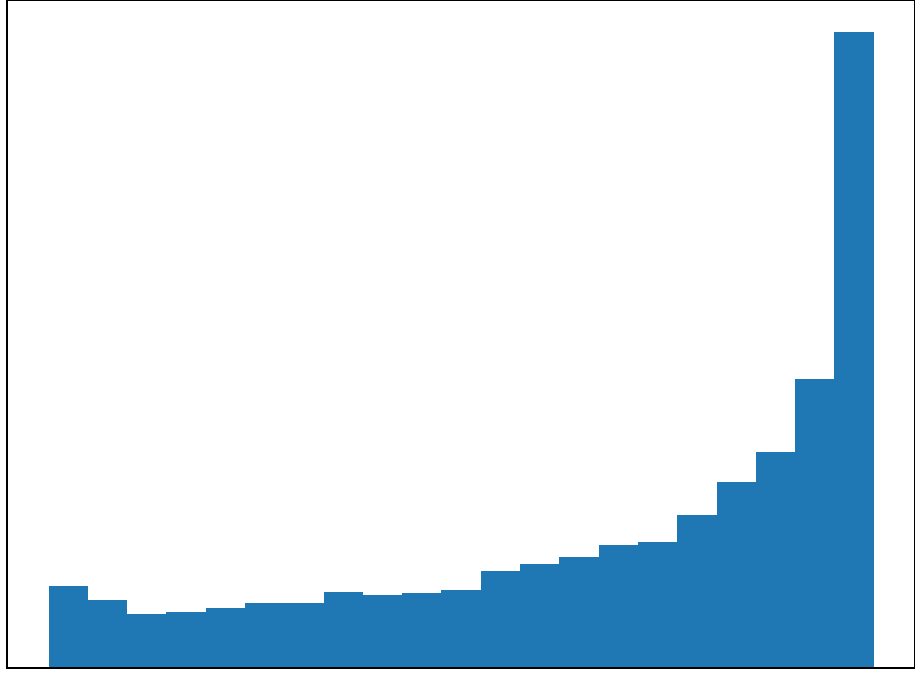}} &
    \makecell{\includegraphics[width=0.05\textwidth]{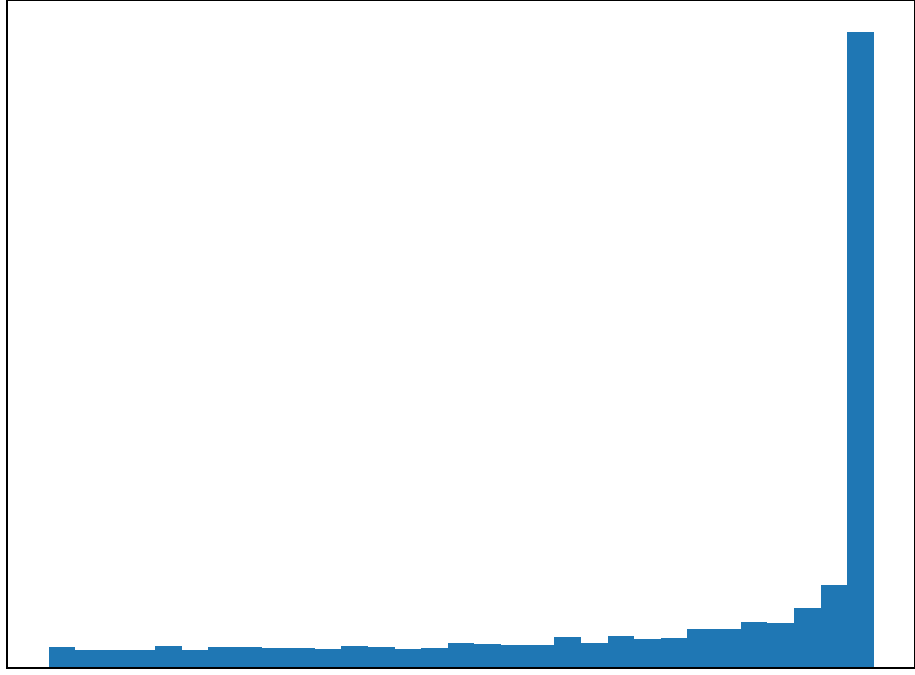}} &
    \makecell{\includegraphics[width=0.05\textwidth]{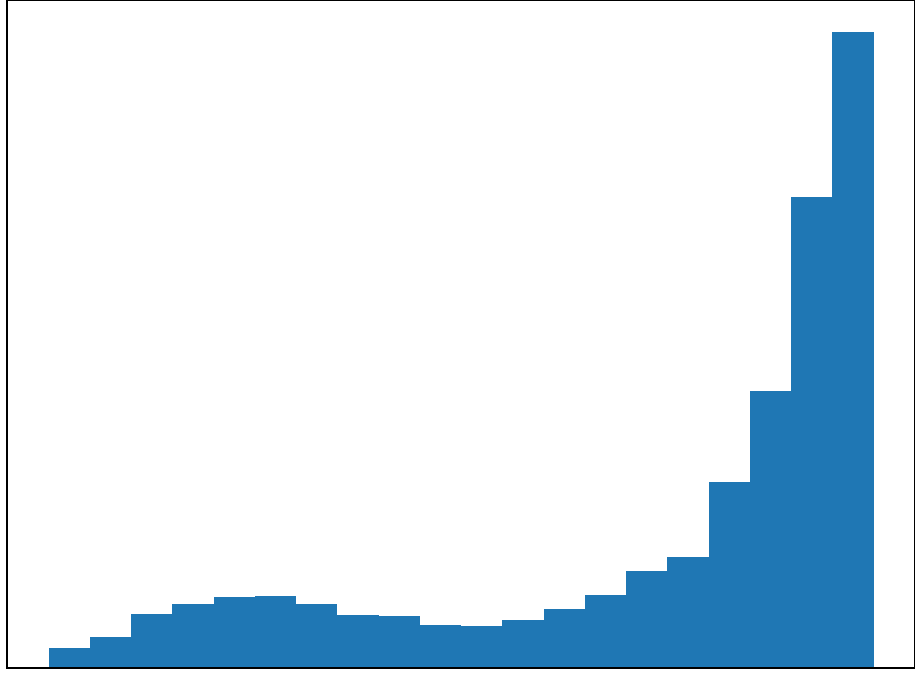}} &
    \makecell{\includegraphics[width=0.05\textwidth]{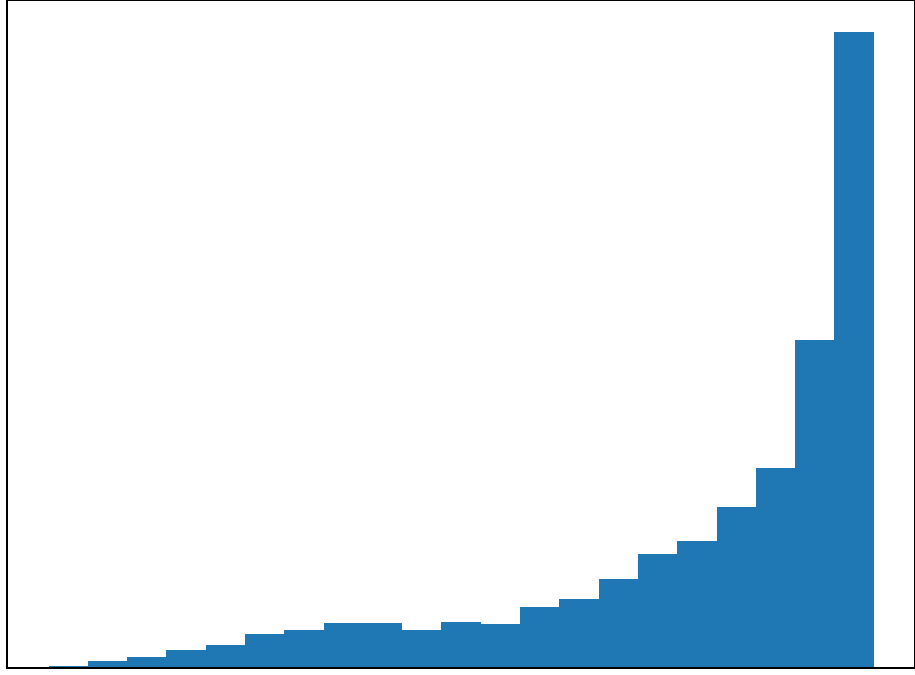}}\\
    & \makecell{Sill} &
    \makecell{\includegraphics[width=0.05\textwidth]{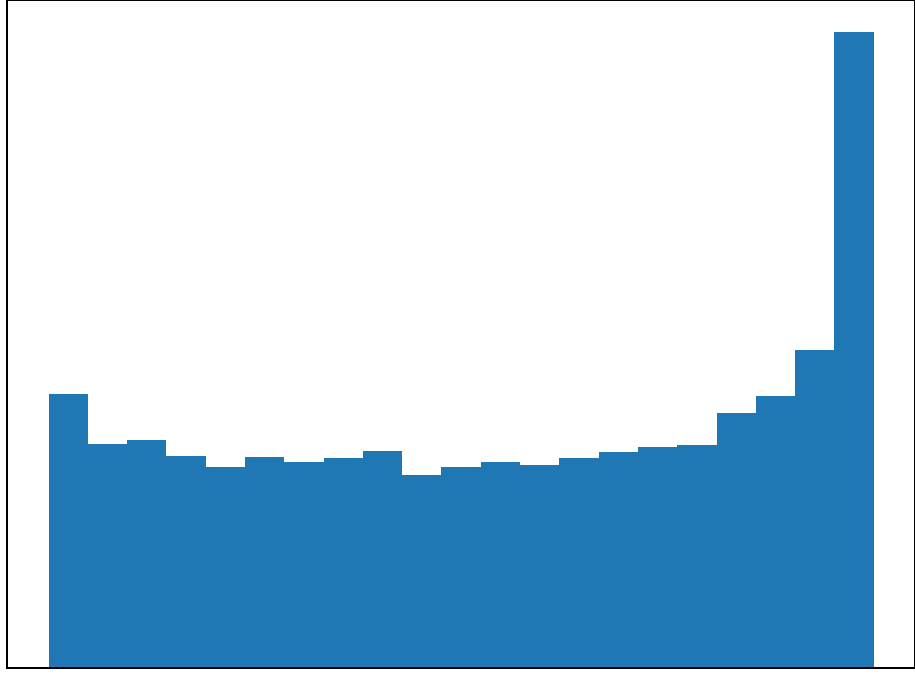}} &
    \makecell{\includegraphics[width=0.05\textwidth]{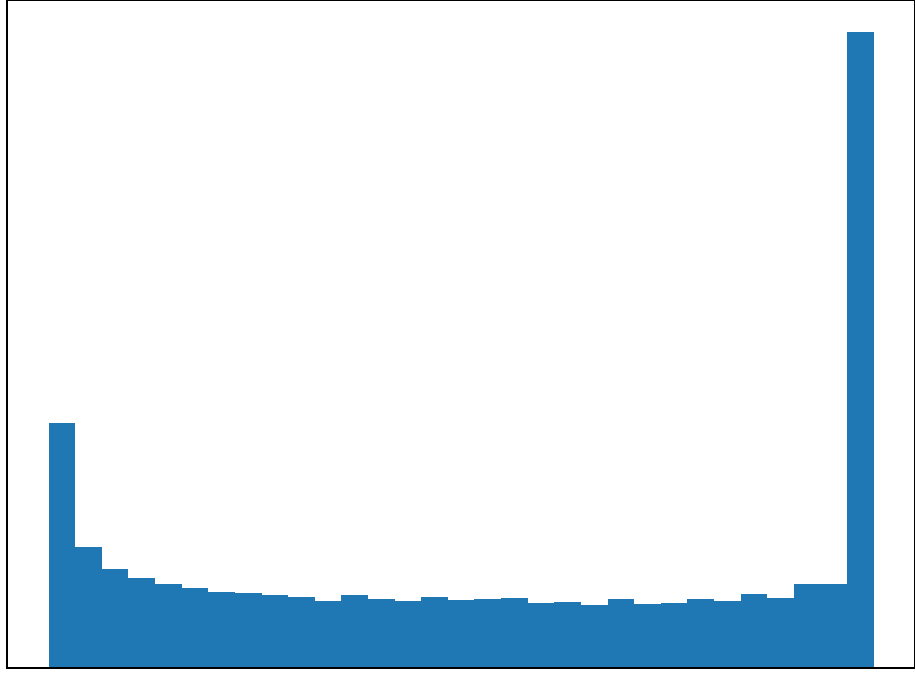}} &
    \makecell{\includegraphics[width=0.05\textwidth]{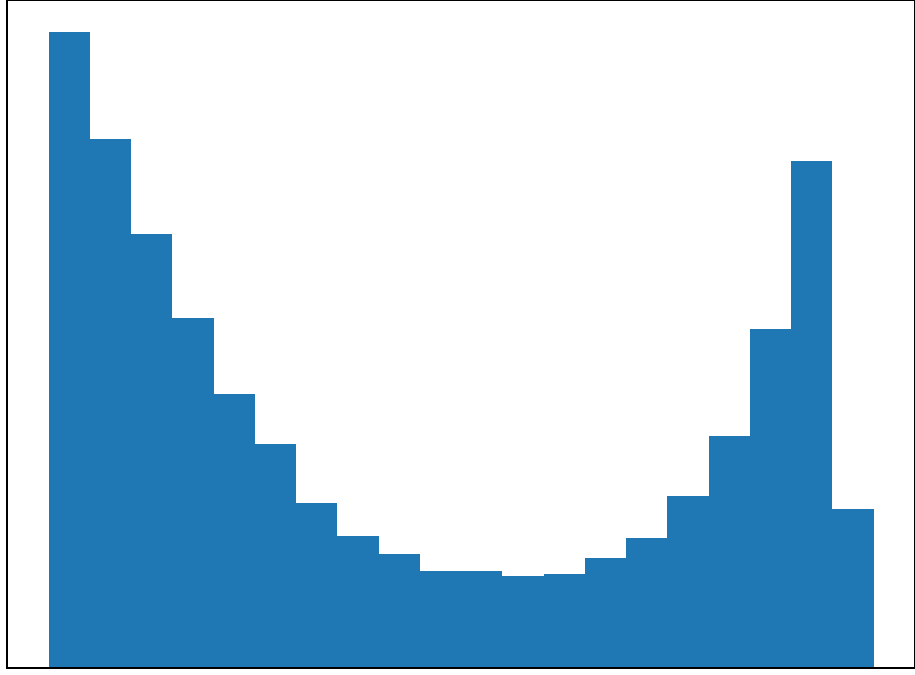}} &
    \makecell{\includegraphics[width=0.05\textwidth]{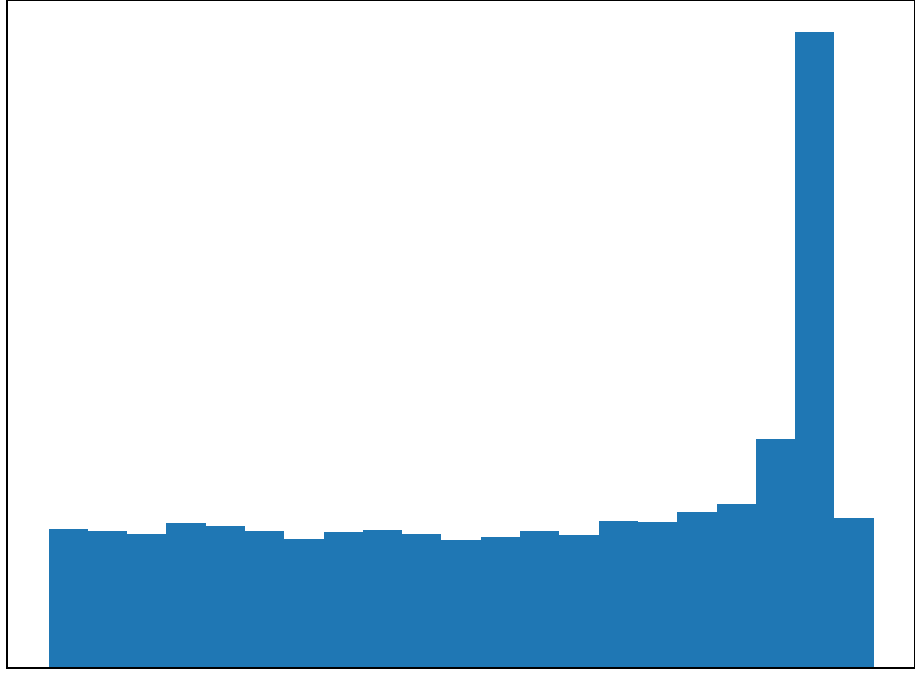}} &
    \makecell{\includegraphics[width=0.05\textwidth]{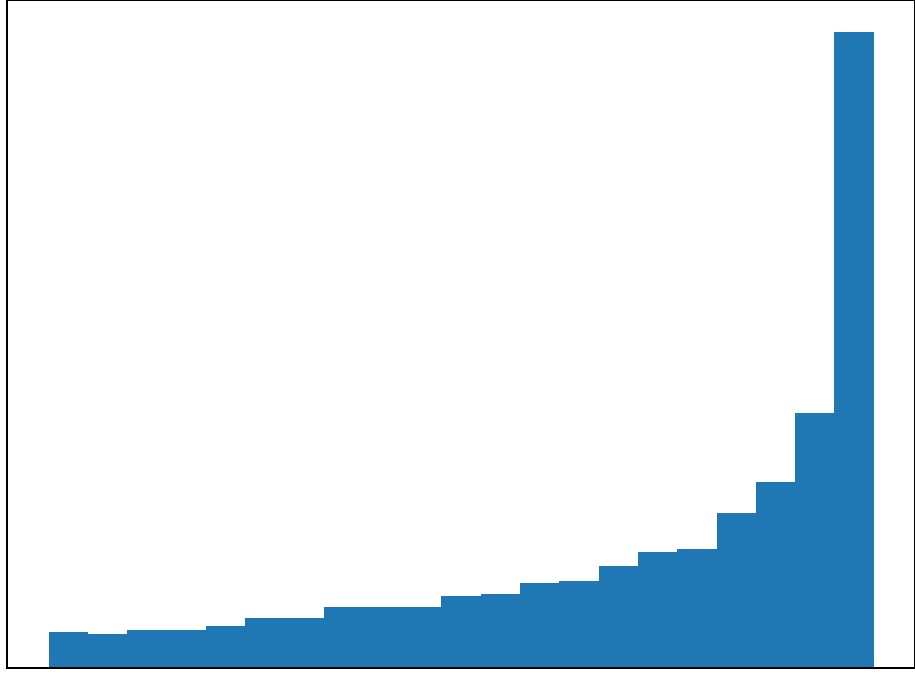}} &
    \makecell{\includegraphics[width=0.05\textwidth]{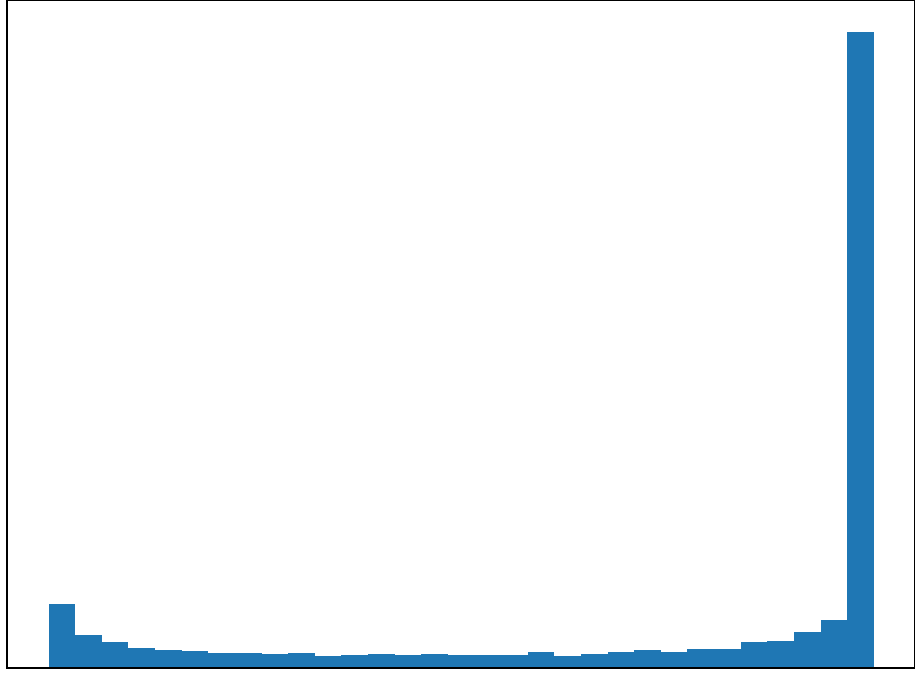}} &
    \makecell{\includegraphics[width=0.05\textwidth]{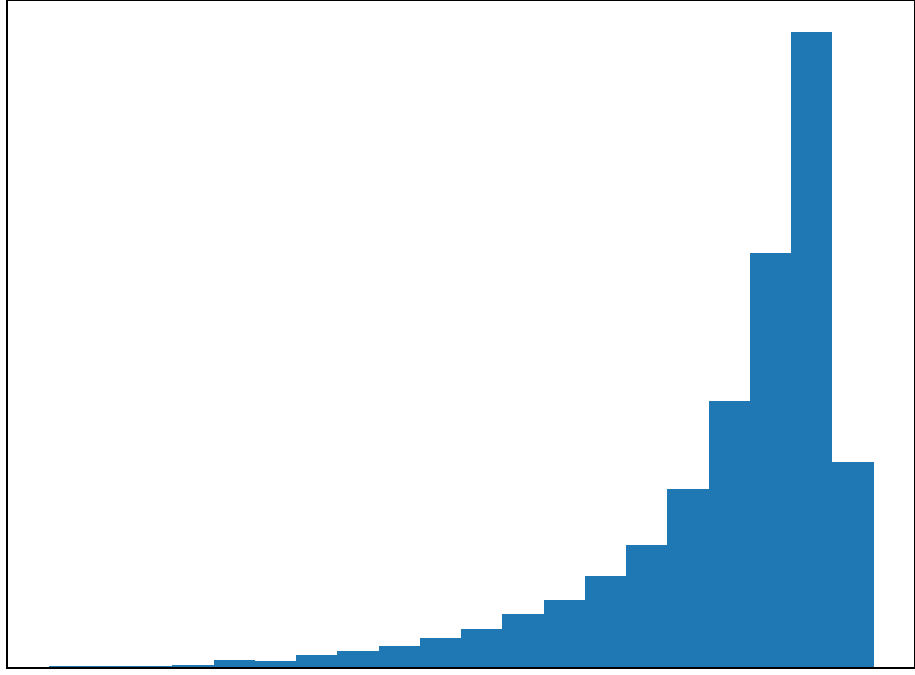}} &
    \makecell{\includegraphics[width=0.05\textwidth]{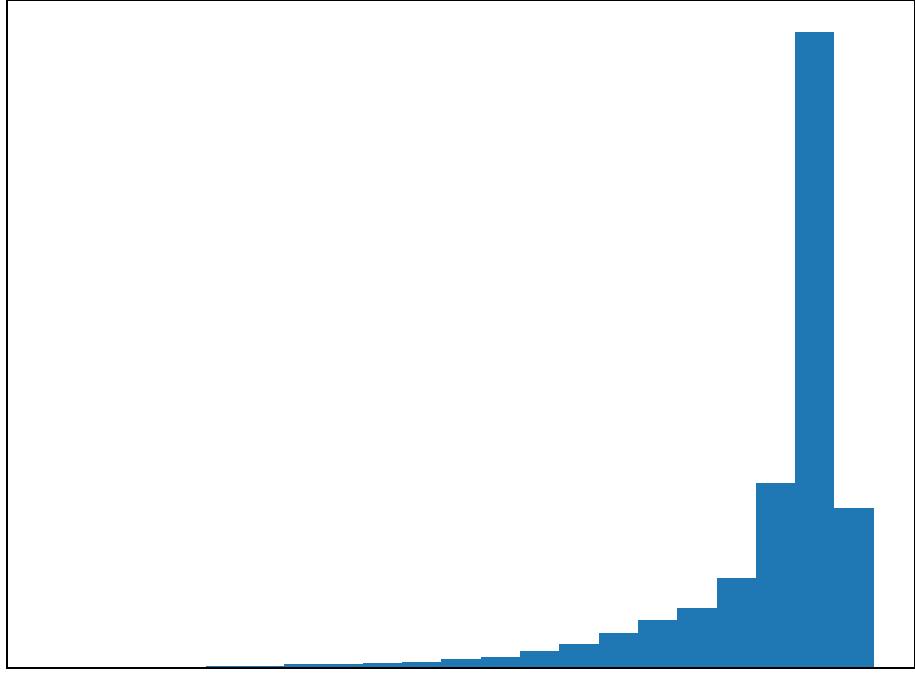}}\\
    \end{tabular}
    \caption{PIT diagrams as in Figure~\ref{fig.pit_precipitation} and Figure~\ref{fig.pit_obj} but for various variogram parameters of an exponential variogram model, which are the nugget, effective range, and sill. Results are shown for STEPS, ICON, COPULA and COPULA sorted for the lead times of $+1$\,h, $+3$\,h, and $+6$\,h. The four columns on the left show the results for the whole dataset, while the four columns on the right are restricted to instances where precipitation was observed within the considered $9 \times 9$ sub-region $V$.}
    \label{fig.pit_var}
\end{figure}

As an additional technique to assess the spatial structure of synthetic rainfall fields, we fit an exponential variogram model (e.g. \citep{Journel1976}) and compare the estimated fitting parameters. These parameters, namely, nugget, sill, and effective range, are shown in Figure~\ref{fig.pit_var}  for lead times of $+1$\,h, $+3$\,h, and $+6$\,h. Again, we consider the whole dataset (four columns on the left-hand side) and the precipitation subset (four columns on the right-hand side). Nugget describes the portion of non-spatial variance, whereas sill represents the limit of the variogram. The effective range is the distance where 95\,\% of the sill is exceeded. The forecasts of ICON and STEPS reveal an underdispersive behaviour in all three metrics for the whole dataset, whereas, for the precipitation subset, only an underestimation is visible. As for the set of object-based metrics, many cases are removed in which precipitation is neither observed nor forecast. However, in cases where precipitation is forecast in all forecast members but not observed, the ensemble members show higher variability than the object-based metrics. This may explain the number of observations in the first bin. Differences with increasing lead times are merely visible for the results of the whole dataset. However, all three metrics decrease with increasing lead time for ICON and STEPS for the precipitation subset.

COPULA sorted shows a nearly uniformly distributed histogram for lower percentiles; however, it underestimates all three metrics for many cases in the whole dataset. This peak is more distinct in the precipitation subset. In addition, observed nugget values are centred around the median. It could also be seen that nugget and effective range decrease with increasing lead time. Compared to this, COPULA shows (for the whole dataset) most of the observed values at lower percentiles, with only a few events in the last bins. For the precipitation subset, these peaks are shifted toward upper percentiles. Furthermore, all values decrease with increasing lead times. While with respect to the traditional scores considered above, COPULA sorted showed a similar validation behavior compared to COPULA, the results presented  in Figure~\ref{fig.pit_var}  indicate that the forecasts of COPULA exhibit a more realistic spatial correlation.

\section{Conclusion}
\label{sec.conclusion}
Many post-processing techniques for weather forecasts apply statistical corrections to a forecast for each individual location considered, i.e., the spatial correlation between locations contained in the input forecast is lost, and the post-processed output contains only marginal distributions. This necessitates the development of models to reintroduce spatial correlation into such post-processed forecasts, as the spatial correlation between locations is required to make predictions for larger areas like, e.g., the amount of rainfall within a river basin.

In this paper, we propose and apply such a model to the output of a neural network that combines two input forecasts and outputs calibrated marginal distributions. The proposed model uses an R-vine copula fitted to historical precipitation observations to model the joint distribution of precipitation amounts at adjacent locations. Similar to existing approaches such as the Schaake shuffle \citep{clark2004schaake} and ensemble copula coupling \citep{schefzik2013uncertainty}, the predicted precipitation amounts for each location are iteratively arranged into synthetic ensemble members. The density of the fitted copula is used to determine how well a synthetic ensemble member compares to the joint distribution of historical precipitation amounts. The output of the proposed model is a synthetic ensemble forecast with the same calibrated marginal distribution as the input forecast and, in addition, exhibits realistic spatial correlation patterns.

Methods discussed in the literature, such as the Schaake shuffle and ensemble copula coupling, derive their spatial information directly from a suitably selected set of historical observations or an ensemble forecast. However, this requires a close relationship between the source of spatial correlation and the post-processed forecast to obtain realistic results. The  model proposed in the present paper does not require such a close relationship and can be applied in cases, such as the combination model, where the marginal output distributions differ significantly from each input ensemble forecast, such that neither of them is a suitable source of spatial information. In addition, inferring spatial correlations from a set of historical observations or another ensemble forecast requires a ranking without ties of the predicted events for each location, which is not always given for weather variables such as precipitation, whose probability distribution has an atom at $0$mm. Deriving spatial correlations from a joint probability distribution avoids this problem.

In order to evaluate the performance of the proposed model, we considered several validation metrics such as bias, Brier skill score and reliability diagrams for precipitation amounts within sub-regions of $9 \times 9$ grid points. The obtained results were compared to both input forecasts and an additional arrangement of precipitation values with maximum spatial correlation. As a result, it could be shown that the calibration of the marginal distributions by the previously applied combination model carries over to the area predictions of the proposed copula model, which shows a clear improvement in forecast quality compared to both input forecasts.

Furthermore, the spatial correlation is directly evaluated using object-based metrics such as the number of connected area components, distance metrics between precipitation clusters and variogram parameters. These metrics indicate that the output of the proposed copula model has the most realistic precipitation patterns compared to all other forecast models considered in this paper.

The present paper evaluates the performance of the copula model for areas consisting of $9 \times 9$ grid points. In order to apply the model to more general cases, such as river catchments, the model requires some modifications, which will be investigated in a forthcoming study. Such a modification could be the extension of the R-vine copula model to higher dimensions, in order to include more grid points. This, however, would lead to  an increased algorithmic complexity and might not directly be feasible, depending on the size of the area. To take this into account, another modification could be the adaption of the hill climbing algorithm such that the existing copula for $9 \times 9$ areas is used to determine the permutation of values at a given grid point, i.e., for determining a permutations $\sigma_i$ only the local neighbourhood is considered. Besides considering more general areas, including additional information into the R-vine copula fitting procedure would be interesting, like, e.g., the local orography or the distinction of convective and stratiform precipitation patterns,  which could further improve the reconstruction of spatial correlation.

\bibliographystyle{apalikecustom}
\bibliography{references}

\end{document}